RUSSIAN ACADEMY OF SCIENCES
FIBER OPTICS RESEARCH CENTER
*LABORATORY OF FIBER LASERS AND AMPLIFIERS*

# OPTICAL PROPETIES OF BISMUTH-DOPED $SIO_2$- OR $GEO_2$-BASED GLASS CORE OPTICAL FIBERS

Elena G. Firstova

PhD Thesis
(in Russian)

March 2015

## Abstract

A detailed study of optical properties of bismuth-doped fibers based on $SiO_2$ and $GeO_2$ glasses containing no other dopants has been carried out. To provide important information about spectroscopic properties of IR bismuth-related active centers (BAC) the excitation-emission fluorescence spectra for a spectral region of 220 – 2000 nm have been measured. The obtained three-dimensional spectra have been presented for different host glass compositions: silicate, germanate, aluminosilicate and phosphosilicate. Energy-level configuration and main radiative transitions associated with BACs in $GeO_2$ and $SiO_2$ glasses have been revealed. Fluorescence lifetime analyses of the basic radiative transitions of BAC have been carried out. It has been shown that the energy-level schemes of BAC-Si and BAC-Ge (BAC associated with silicon and germanium, respectively) are similar, the corresponding BAC-Ge energy levels lying 10-16% lower than those of BAC-Si. It has been determined that BAC-Si, BAC-Ge and BAC-Si, BAC-P can exist simultaneously in bismuth-doped germanosilicate and phosphosilicate fibers, respectively. Optical amplification in the wavelength range 1410-1470 nm and lasing at 1460 nm in $SiO_2$-glass fiber doped with bismuth have been demonstrated for the first time. The optical gain and lasing in this fiber were based on the radiative transition between the first excited and ground energy levels belonging to BAC-Si. Anti-stokes luminescence spectra of BAC-Si and BAC-Ge in optical fibers under the two-step excitation have been measured and analyzed in detail.

# ОГЛАВЛЕНИЕ

**БЛАГОДАРНОСТИ**

Автор благодарит своего научного руководителя члена-корреспондента РАН, доктора физико-математических наук И.А. Буфетова за ценные советы при обсуждении экспериментальных результатов, помощь и поддержку при подготовке докладов на научных семинарах, а также написании научной работы. Данная работа была бы невозможна без постоянного интереса и внимания со стороны директора НЦВО РАН академика Е.М. Дианова.

Автор выражает искреннюю признательность своим коллегам по лаборатории волоконных лазеров и усилителей зав. лаб. М.А. Мелькумову, с.н.с. С.В. Фирстову и н.с. А.В. Шубину за помощь в проведении экспериментов, ценные обсуждения и замечания. Кроме того, автор работы выражает благодарность в.н.с. ИХВВ РАН В.Ф. Хопину и н.с. НЦВО РАН В.В. Вельмискину за создание заготовок волоконных световодов.

## СПИСОК ИСПОЛЬЗУЕМЫХ СОКРАЩЕНИЙ

**БР –** брэгговская решетка

**ВАЦ** – висмутовый активный центр

**ВАЦ-Si(Ge, P, Al)** - висмутовый активный центр, ассоциированный с кремнием (германием, фосфором, алюминием)

**ВВЛ** – висмутовый волоконный лазер

**ВКР** – вынужденное комбинационное рассеяние

**ГКДЦ** – германиевый кислородно-дефицитный центр

**ИН** – источник накачки

**ККДЦ** – кремниевый кислородно-дефицитный центр

**КПД** – коэффициент полезного действия

**УФ, ИК** – ультрафиолетовый, инфракрасный

**ЭПР** – электронный парамагнитный резонанс

**ЯМР** – ядерный магнитный резонанс

**FCVD (Furnace Chemical Vapor Deposition)** – метод химического осаждения из газовой фазы с использованием печи

**FWHM (Full Width of Half Maximum)** – полная ширина на уровне половинной амплитуды

**EXAFS (Extended X-Ray Absorption Fine Structure)** – протяженная тонкая структура рентгеновских спектров поглощения

**MCVD (Modified Chemical Vapor Deposition)** – модифицированный метод химического осаждения из газовой фазы

**v-$SiO_2$** / **v-$GeO_2$** – vitreous $SiO_2$/$GeO_2$ – стеклообразный оксид $SiO_2$ / $GeO_2$ без дополнительных примесей

# ВВЕДЕНИЕ

Прогресс в области лазерных технологий привел к тому, что лазеры стали активно применяться в самых различных отраслях науки и техники (наукоемкие и военные технологии, телекоммуникации, медицина, промышленность). Перспективы развития лазерных технологий связаны в первую очередь с проведением непрерывных исследований в области лазерной физики, фотоники, прикладной оптики, материаловедения. Одним из важнейших этапов таких исследований является поиск новых типов активных сред с уникальными свойствами.

В зависимости от специфики области применения возникает необходимость получения лазеров с определенными параметрами: высокой стабильностью, компактностью, качеством пучка и высокой мощностью выходного излучения и т.п. Именно это послужило причиной, особенно в последнее время, активного внедрения волоконных лазеров, которые находят новые области применения, постепенно вытесняя другие виды лазеров. Волоконные лазеры получили свое название из-за особенностей лазерной среды, которая представляет собой оптическое волокно, состоящее из стеклянной оболочки и светововедущей сердцевины с активными ионами. До настоящего времени для создания волоконных лазеров, также как и для объемных твердотельных лазеров, использовались, главным образом, редкоземельные ионы. На Рис. 1 представлены области генерации эффективных волоконных лазеров на редкоземельных ионах. Видно, что, несмотря на обширную область возможной генерации редкоземельных лазеров, существуют не доступные для их использования диапазоны 1150-1530 нм и 1625-1775 нм. В 2005 году был создан первый волоконный лазер (висмутовый лазер), не имеющей аналога среди объемных твердотельных лазеров [1, 2]. Полученный лазер излучал в области 1140-1215 нм и был реализован с использованием волоконного световода с сердцевиной из алюмосиликатного стекла, легированного висмутом.

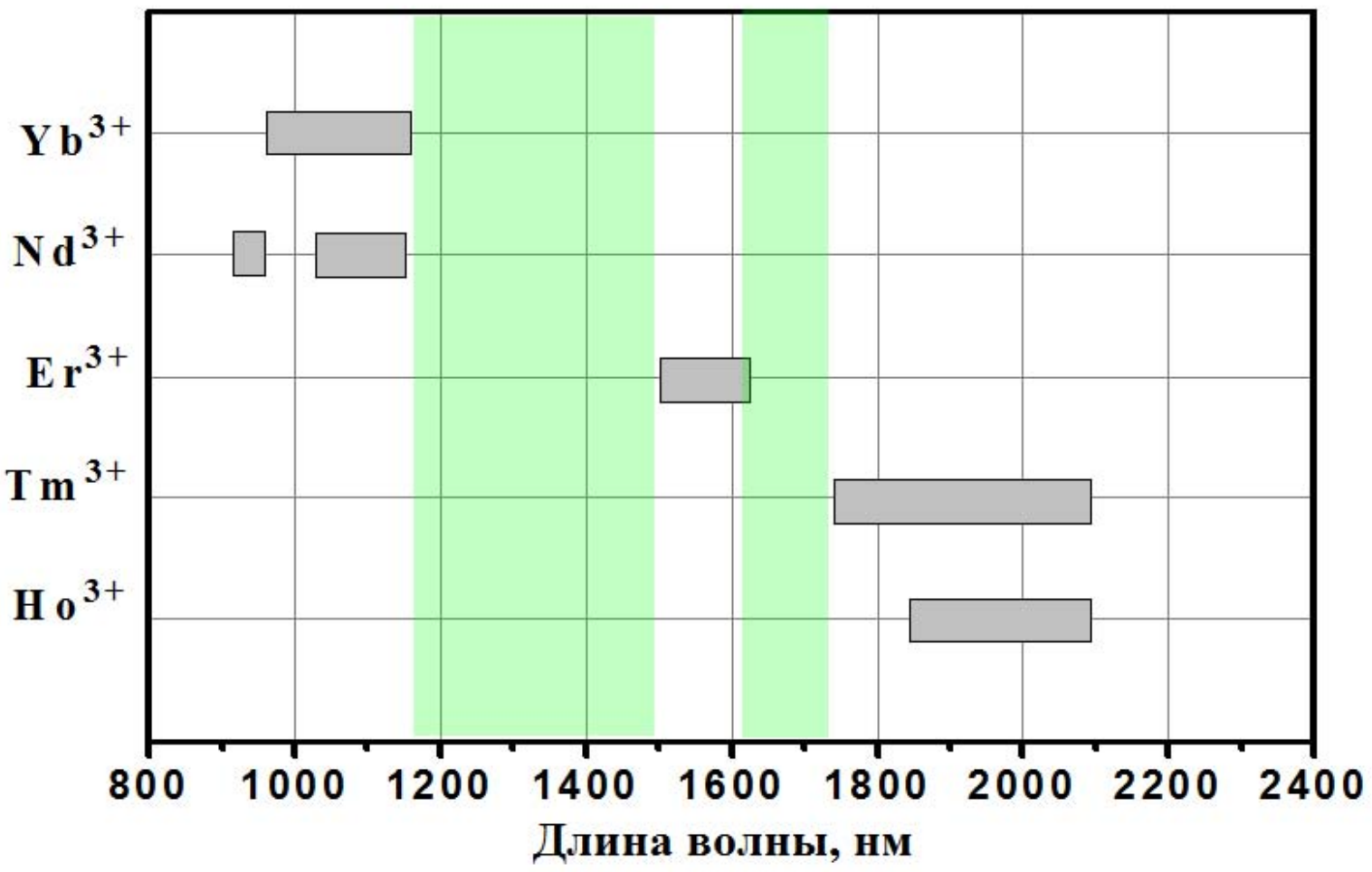

Рис. 1 Области генерации волоконных лазеров на световодах из кварцевого стекла, содержащего редкоземельные ионы [3].

Это стало первым шагом к освоению волоконными лазерами новых спектральных областей, в которых не работают лазеры на редкоземельных ионах. Позже было показано, что изменением химического состава стеклянной матрицы, можно сдвигать полосу усиления висмутовых световодов. Это позволило разработать волоконные висмутовые лазеры, которые могут излучать в спектральной области от 1140 до 1550 нм [4-11]. Недавно на висмутовых световодах была получена лазерная генерация в новой области 1625 – 1775 нм [12]. КПД висмутовых лазеров обычно находится в пределах 10-50 % в зависимости от спектральной области генерации. Столь широкий спектральный диапазон, в котором могут работать устройства на основе световодов с висмутом, открывает большие перспективы применения.

В качестве потенциально возможных применений висмутовых лазеров можно выделить следующие: в телекоммуникациях – широкополосные усилители; в астрономии - генерация желтого света путем удвоения частоты висмутовых лазеров [13-18]; в медицине – источники излучения с определенной длиной волны для дерматологии, косметологии, офтальмологии и др. [19, 20]; в нефтегазовой промышленности – для создания газоанализаторов (в частности, обнаружения метана $CH_4$ в воздухе).

Основным недостатком существующих в настоящее время содержащих висмут эффективных лазерных сред является то, что рабочие концентрации висмута не превышают 0.02 ат.% (повышение концентрации висмута приводит к снижению КПД лазеров вплоть до невозможности получения генерации), а следовательно коэффициент поглощения накачки в сердцевине эффективного световода находится на уровне от десятых долей до ~ 1 дБ/м (по этой причине висмутовые лазеры с высоким КПД являются чрезмерно длинными – до 200 м). Это не позволяет использовать схему ввода излучения накачки в световод через первую оболочку, и для эффективного поглощения (получения высокого КПД) вынуждает использовать ввод накачки в сердцевину, с соответствующими высокими требованиями к качеству излучения накачки. Отсутствие одномодовых полупроводниковых источников излучения с высокой выходной мощностью (более 2 Вт) приводит к значительному усложнению систем накачки висмутовых лазеров. Схема накачки экспериментальных образцов обычно имеет вид: многомодовый полупроводниковый лазерный диод → иттербиевый лазер → ВКР лазер → висмутовый лазер.

Основным препятствием оптимизации конструкции висмутовых лазеров является недостаточно изученная физическая природа ИК висмутовых активных центров (ВАЦ). Ситуация осложняется тем, что висмут может в стекле иметь разные степени окисления, а спектры поглощения, люминесценции в неоднородных стеклообразных средах имеют большую ширину полос и взаимно перекрываются. Поэтому изучение оптических свойств ВАЦ в ограниченных спектральных областях (для дискретного набора длин волн возбуждения и люминесценции), которое до сих пор проводилось, предоставляло заведомо недостаточный объем экспериментальных данных даже для определения схемы энергетических уровней ВАЦ, не говоря уже о построении его адекватной физической модели.

# ГЛАВА I. ОБЗОР ЛИТЕРАТУРЫ И ПОСТАНОВКА ЗАДАЧ ИССЛЕДОВАНИЯ

Ви́смут — химический элемент 5-ой группы шестого периода периодической системы химических элементов Д. И. Менделеева, имеет атомный номер 83. Обозначается символом Bi. Это самый тяжелый элемент, который имеет устойчивый изотоп. Электронная конфигурация нейтрального атома Bi имеет вид: [Xe] $4f^{14}$ $5d^{10}$ $6s^{2}$ $6p^{3}$. В соединениях висмут может проявлять несколько степеней окисления: +1, +2, +3, +5.

Висмут – элемент, имеющий способность наделять стекло особенными оптическими свойствами. Легирование стекол висмутом приводит к появлению полос поглощения и люминесценции в УФ, видимой и ближней ИК области. Начиная с конца XIX века начали проводиться исследования материалов с висмутом, которые считались перспективными для разработки на их основе веществ, способных излучать свет при поглощении ионизирующего излучения – сцинтилляторов [21]. Одним из наиболее распространенных сцинтилляционных материалов на основе Bi, является кристалл германата висмута $Bi_4Ge_3O_{12}$. Основным преимуществом таких кристаллов является большое сечение фотопоглощения гамма-квантов, обусловленное большим атомным номером (83) тяжелого компонента Bi и высокой плотностью кристалла. Типичный спектр люминесценции такого кристалла приведен на Рис. 1. 1.

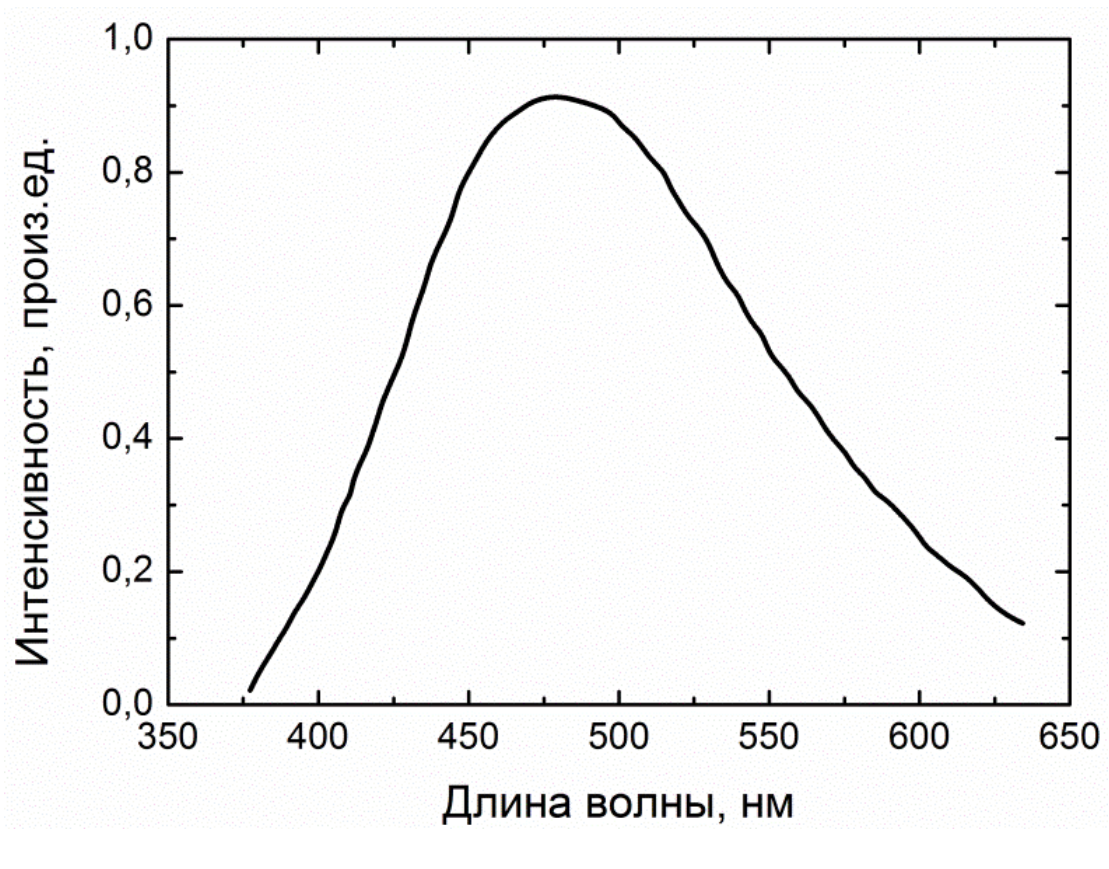

Рис. 1. 1 Спектр люминесценции кристалла $Bi_4Ge_3O_{12}$ при возбуждении γ-излучением (~660 кэВ) [22].

Последующие исследования легированных висмутом кристаллов и стекол показали, что такие материалы способны люминесцировать не только в видимой, но и в ближней ИК области. Первые сообщения, насколько нам известно, о ИК люминесценции в материалах, содержащих висмут, были еще в 80-х годах в работе В.А. Ломонова [23]. Типичные спектры ИК люминесценции кристаллов силленитов висмута, полученные в работе [23], приводятся на Рис. 1. 2. К сожалению, результаты данной работы, по-видимому, остались вне поля зрения большинства исследователей.

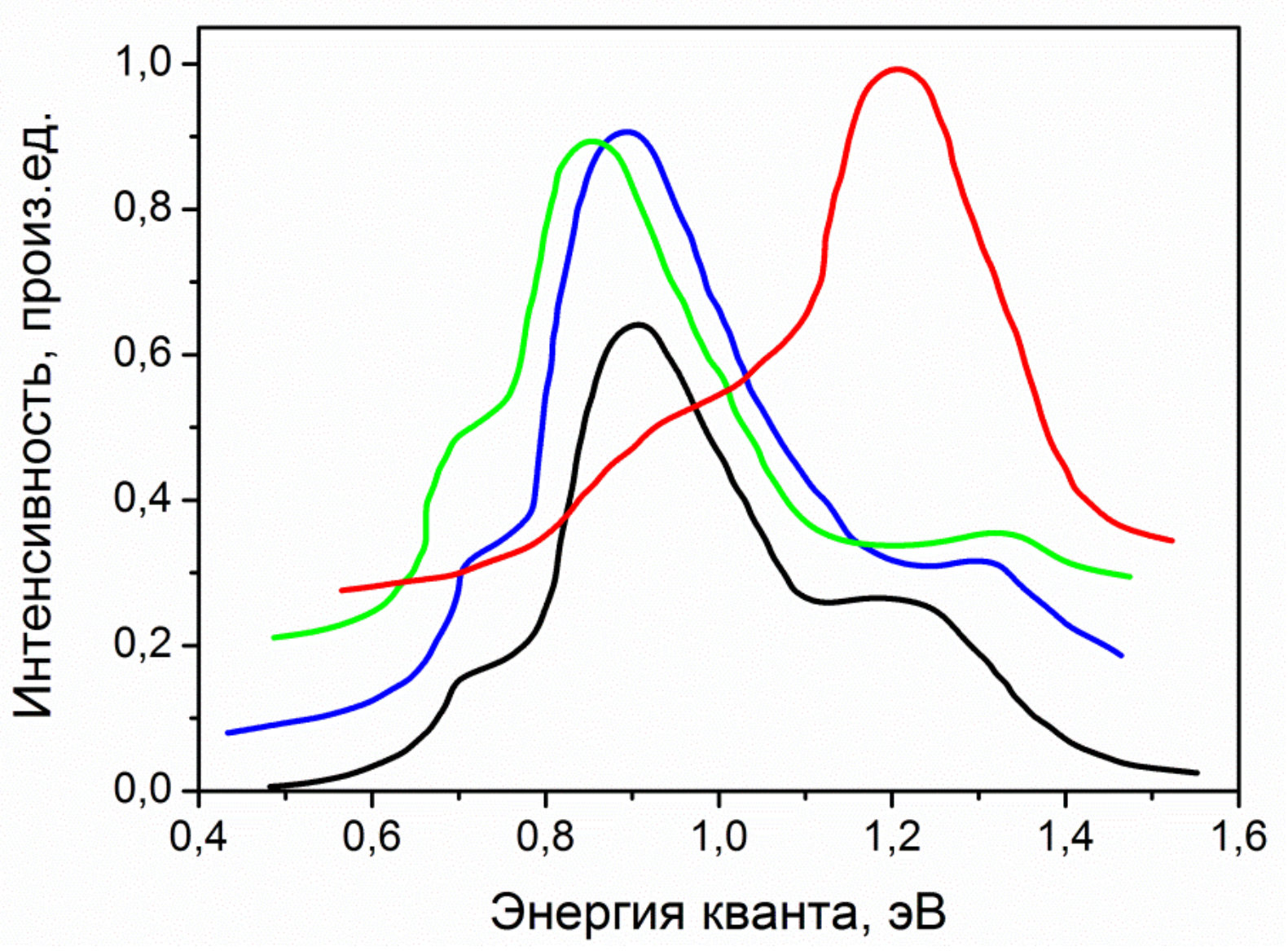

Рис. 1. 2 Спектры ИК люминесценции кристаллов силикосилленита висмута, выращенных из расплавов с различным содержанием $SiO_2$ [23].

Однако интерес к материалам с висмутом, излучающим ИК люминесценцию, возник позже – в конце 1990-х-начале 2000-х годов, после публикаций работ японских ученых K. Murata, Y. Fujimoto и M. Nakatsuka [24, 25]. В последнее время (немного более 10 лет) стекла, активированные висмутом, стали предметом интенсивных исследований. Отличительная особенность таких материалов – это наличие широкой полосы люминесценции в спектральной области от 1100 до ~1900 нм. Это позволило рассматривать их в качестве возможной активной среды для эффективных широкополосных источников когерентного и некогерентного излучения, оптических усилителей и других устройств, реализация которых была осуществлена позже (например [26], а также ссылки, приведенные во вводной части настоящей работы).

Именно данное обстоятельство послужило исходной точкой для выполнения настоящей работы, направленной на получение большего объема знаний о таких материалах и о роли висмута в них.

Данная глава будет посвящена обзору оптических свойств материалов с висмутом: кристаллов, стекол и волоконных световодов, известных на момент начала этой работы (до 2009 г). Кроме того, будут рассмотрены основные результаты по получению лазерной генерации и оптического усиления в световодах с висмутом на тот же момент времени.

## 1.1 Механизм возникновения видимой люминесценции в оптических средах, активированных висмутом

К настоящему времени было проведено множество исследований и накоплен значительный объем экспериментальных данных, касающихся оптических свойств материалов, содержащих ионы висмута. Основное внимание в публикациях 1970-1980-х годов было уделено изучению люминесцентных свойств кристаллических сред в видимом диапазоне спектра. Цель большинства работ заключалась в поиске кристаллической структуры для получения интенсивной люминесценции в видимой области спектра, в частности синей люминесценции. Данная цель была продиктована, как уже было раньше отмечено, необходимостью создания эффективных сцинтилляторов. Проведенный анализ литературы показал, что большинству кристаллических сред с висмутом присущи широкие (50-80 нм) полосы (синей (со временем жизни люминесценции $\tau \approx$ 1-2 мкс) и красной ($\tau \approx$ 5-12 мкс)) люминесценции. Даже в случае монокристаллов с висмутом можно было наблюдать достаточно широкие полосы поглощения и люминесценции, что объясняется сильным электрон-фононным взаимодействием оптического центра и решетки кристалла, несвойственным редкоземельным ионам, у которых оптические электроны экранируются от внешнего поля лигандов. Поэтому в зависимости от структуры кристалла и хи-мического состава максимум синей люминесценции может располагаться в диапазоне от 300 до 480 нм. В случае красной люминесценции данный диапазон является более ограниченным – от 580 до 640 нм. Типичные спектры синей и красной люминесценции и их спектры возбуждения представлены на Рис. 1. 3 а и б. На Рис. 1. 3 а видно, что синей люминесценции соответствует одна область возбуждения (около 220-290 нм), тогда как для красной люминесценции возможно возбуждение в 3 спектральных областях: 250-260, 410-470 и 560-620 нм (Рис. 1. 3 б).

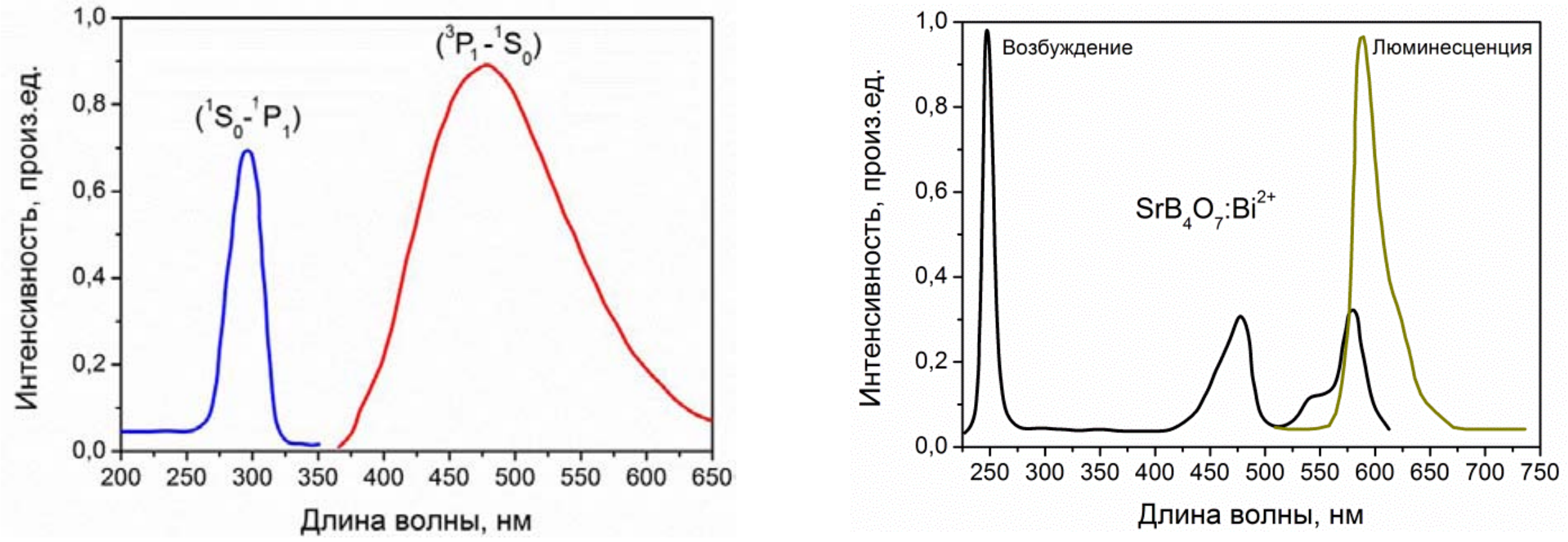

а б

Рис. 1. 3 а) Типичный спектр синей фотолюминесценции (переход $^3P_1 \rightarrow {}^1S_0$) и спектр ее возбуждения (переход $^1S_0 \rightarrow {}^1P_1$) для $GdAlO_3:Bi^{3+}$ [27]; б) типичный спектр красной фотолюминесценции и спектр ее возбуждения кристалла $SrB_4O_7:Bi^{2+}$ [28].

Из опубликованных работ было установлено, что на первоначальном этапе существовали различные гипотезы о механизме появления наблюдаемых полос люминесценции в видимой области. Данный факт стал неожиданным, поскольку значительное количество ранних работ было посвящено изучению висмута в упорядоченных средах (поликристаллам и монокристаллам), которые, как известно, являются наиболее удобными средами для изучения особенностей и структуры оптических центров. Оказалось, что в случае с висмутом, изучение люминесцентного центра даже в упорядоченной оптической среде является трудоемкой задачей из-за ряда причин, основная часть из которых перечислена ниже: 1) наличие большого многообразия форм висмута; 2) повышен-ная склонность к кластеризации – формированию поликатионных соединений; 3) наличие внешних электронов, взаимодействующих с решеткой кристалла, и способность к захвату дополнительных электронов (электронные ловушки); 4) формирование комплексов с дефектами кристалла.

Далее в настоящей работе будут рассмотрены некоторые из предлагаемых моделей механизмов возникновения полос видимой (синей и красной) люминесценции, характерной для кристаллов и стекол, содержащих висмут.

***Синяя люминесценция***

- **Экситон $Bi^{4+}$ + $e^-$.** Одна из гипотез о появлении синей люминесценции, которая использовалась для объяснения люминесцентных свойств кристаллов $Ga(Lu)BO_3$:$Bi^{3+}$, была связана с формированием комплекса $Bi^{4+}$ и электрона [29]. Отсутствие характерной (синей) люминесценции в таком кристалле, но без висмута, свидетельствовало о непосредственной взаимосвязи появления синей люминесценции и присутствия ионов висмута в кристалле. Авторами отвергалась модель "одиночного" иона трехвалентного висмута, поскольку в кристаллах $BiBO_6$ синюю люминесценцию обнаружить не удалось. В результате было установлено, что появление синей люминесценции связано с одновременным присутствием ионов Ga и Bi в кристалле. По мнению авторов работы, замещение иона галлия ионом висмута приводит к искажению симметрии расположения атомов в кристалле. Наведенное искажение может приводить к тому, что ион $Bi^{3+}$в возбужденном состоянии формирует экситон $Bi^{4+}$ + $e^-$. При объединении $Bi^{4+}$ и $e^-$ в ион $Bi^{3+}$ испускается квант синего свечения. Подобная модель также применялась для объяснения появления синей люминесценции с аномально большим стоксовым сдвигом равным 1.25 эВ (100 нм) в кристалле $InBO_3$:Bi. Подобное объяснение механизма появления синей люминесценции можно найти в работах G.Blasse и др. [30], A.M.Srivastava и др. [31].

- **Парные ионы Bi или кластеры.** Кристаллы со структурой граната, содержащие ионы $Bi^{3+}$, показывали характерную синюю люминесценцию [32]. В этом случае зависимость интенсивности люминесценции кристалла от со-держания висмута в нем была нелинейной. Полученная зависимость не могла быть описана моделью, включающей одиночный ион $Bi^{3+}$. Поэтому авторы, ссылаясь на ранние работы [29, 31, 33- 37], предложили модель парных центров (димеров или кластеров) висмута. С помощью модели парных центров, был объяснен аномально большой длинноволновый сдвиг полосы возбуждения видимой люминесценции. Следует отметить, что существует ряд работ других авторов (например, [33]), которые рассматривают межкластерные взаимодействия как причину появления аномального стоксова сдвига видимой люминесценции.

- **Ион висмута, связанный с дефектом кристалла.** Для объяснения природы наблюдаемой синей люминесценции в кристаллах, активированных ионами висмута, достаточно часто рассматривалась модель иона висмута, расположенного в окрестности дефекта решетки кристалла. В работе [38] приведен спектр поглощения кристалла $YVO_4$:Bi, в котором присутствуют полосы в УФ области, характерные не только для "одиночного" иона $Bi^{3+}$, а также комплекса, состоящего из иона $Bi^{3+}$ и дефекта кристалла. Появление таких полос поглощения авторами объяснялось процессами переноса заряда между Bi – O и V – O. В работе [39] сравнительный анализ спектров поглощения и люминесценции позволил получить косвенное подтверждение, что комплекс ион $Bi^{3+}$ + $WO_3$ является ответственным за появление синей люминесценции. Аналогичная интерпретация приводилась в другой работе, касающейся изучения оптических свойств кристалла $CaWO_4$:Bi [40].

- **Одиночный ион $Bi^{3+}$.** Получение новой информации о свойствах кристаллов с висмутом, совершенствование технологии роста кристаллов, рост производительности вычислительной техники, алгоритмов расчетов и появление новых подходов к изучению кристаллов позволили исключить основную часть предложенных моделей.

К настоящему времени лишь одна модель была неоднократно подтверждена – это модель "одиночного" иона $Bi^{3+}$ (например, [41]). Механизм появления синей люминесценции в этом случае представляет собой оптические пере-ходы между энергетическими уровнями одиночного иона $Bi^{3+}$ ($^1S_0 \rightarrow {}^1P_1$ – возбуждение, $^3P_1 \rightarrow {}^1S_0$ – люминесценция), показанные на Рис. 1. 4.

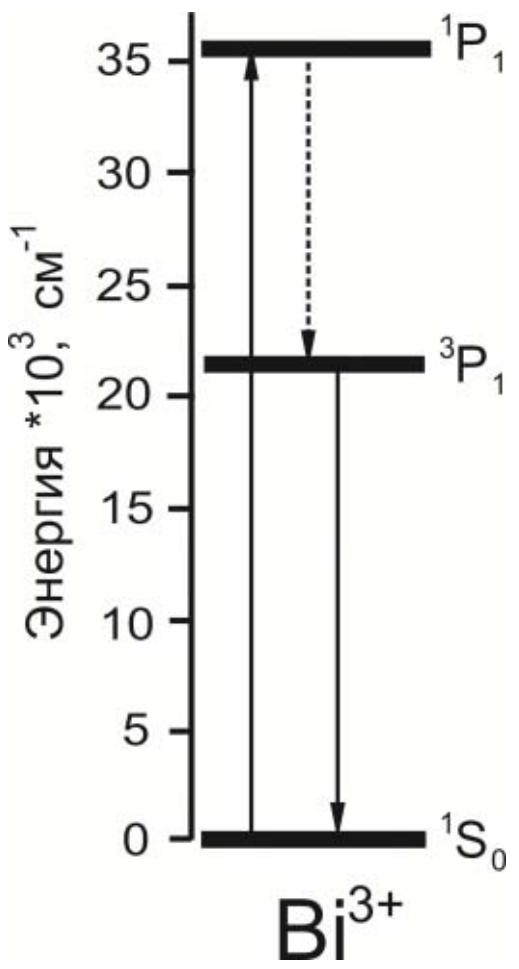

Рис. 1. 4 Схема энергетических уровней и оптических переходов иона $Bi^{3+}$ [42]. $^1P_1 \rightarrow {}^3P_1$ – безызлучательный переход, $^3P_1 \rightarrow {}^1S_0$ – синяя люминесценция, $^1S_0 \rightarrow {}^1P_1$ – поглощение (возбуждение).

Помимо изучения только спектрально-люминесцентных свойств висмутсодержащих материалов проводились работы по изучению структуры излучательного центра (пространственное положение, степень окисления висмута в решетке кристалла). В работе [43] структура центров люминесценции в кристалле $LaPO_4$:Bi изучалась с помощью метода EXAFS (Extended X-ray Absorption Fine Structure). Найдено, что висмут действительно замещает в решетке кристалла ионы $La^{3+}$, что приводит к некоторому искажению локального окружения (расположения лигандов рядом с ионом висмута). Полученные результаты сравнивались с результатами для кристалла $BiPO_4$. Результаты указывали только на присутствие ионов трехвалентного висмута (не было данных о наличии парных ионов или кластеров висмута), что полностью подтверждало гипотезу о происхождении синей люминесценции от иона $Bi^{3+}$.

***Красная люминесценция***

В отличие от полосы синей люминесценции, другая полоса люминесценции с максимумом в области 600 нм, не могла быть объяснена оптическими переходами в схеме энергетических уровней, принадлежащей иону $Bi^{3+}$. Корре-ляцию между интенсивностью красной люминесценции и содержанием $Bi^{3+}$ в кристалле обнаружить не удалось [44] – полученные данные имели гораздо более сложный вид, чем предполагалось. Изначально считалось, что красная люминесценция возникает при переходе между энергетическими уровнями комплекса, состоящего из иона висмута и дефекта кристалла. Данная модель использовалась для интерпретации спектрально-люминесцентных свойств кристаллов KCl, CaO с висмутом [45]. В таких кристаллах в результате замещения $Ca^{2+}$ ионом $Bi^{3+}$ происходило формирование дефекта и, как следствие, возникал положительный некомпенсированный заряд. Для компенсации заряда, по мнению авторов, происходило образование нового отрицательно заряженного дефекта (замещение $Ca^{2+}$ ионом $Na^{+}$). Как утверждается в работе, ионы натрия попадали в кристалл как неконтролируемая примесь [46]. Такие комплексы, по мнению авторов, стали основными центрами, ответственными за появление красной люминесценции. Но, в другой работе [47] те же авторы, в качестве механизма появлении красной люминесценции в том же кристалле, приводили другую гипотезу с участием атмосферного кислорода. Авторы утверждали, что интенсивность красной люминесценции в сильной степени зависит от кислорода, который, попадая из окружающей атмосферы в кристалл, приводил к формированию комплексов, участвовавших в образовании оптического центра, ответственного за красную люминесценцию. Стоит отметить, что во втором случае авторы не привлекали ионы $Na^{+}$. К сожалению, при объяснении экспериментальных результатов многие авторы не принимали во внимание возможности появления ионов восстановленных форм относительно $Bi^{3+}$. Не исключена вероятность наличия ионов $Bi^{2+}$ в изученных кристаллах, в частности замещая ионы $Ca^{2+}$ из-за близости ионных радиусов (ионный радиус $Ca^{2+}$ r ($Ca^{2+}$) = 0.99 – 1.14 Å и ионного радиуса $Bi^{2+}$ r ($Bi^{2+}$) = 1.14 Å [48]).

Именно дальнейшие исследования показали, что источником красной люминесценции являются ионы $Bi^{2+}$ (обсуждение основных работ по данному вопросу и их ссылки будут приведены ниже). Впоследствии было опубликова-но большое число работ по обнаружению красной люминесценции и детальному изучению механизма ее появления. В частности, в работе [49] изучались оптические свойства кристалла $BaSO_4$, в котором висмут присутствовал в качестве неконтролируемой примеси. Авторы предполагали, что висмут встраивается в решетку кристалла в двух состояниях $Bi^{3+}$ и $Bi^{2+}$. При этом наибольшая доля атомов Bi должна присутствовать в форме $Bi^{2+}$(r ($Bi^{2+}$) = 1.14Å), поскольку они замещают ионы $Ba^{2+}$, приблизительно с таким же ионным радиусом r ($Ba^{2+}$) = 1.49 Å. На Рис. 1. 5 показан типичный спектр люминесценции такого кристалла.

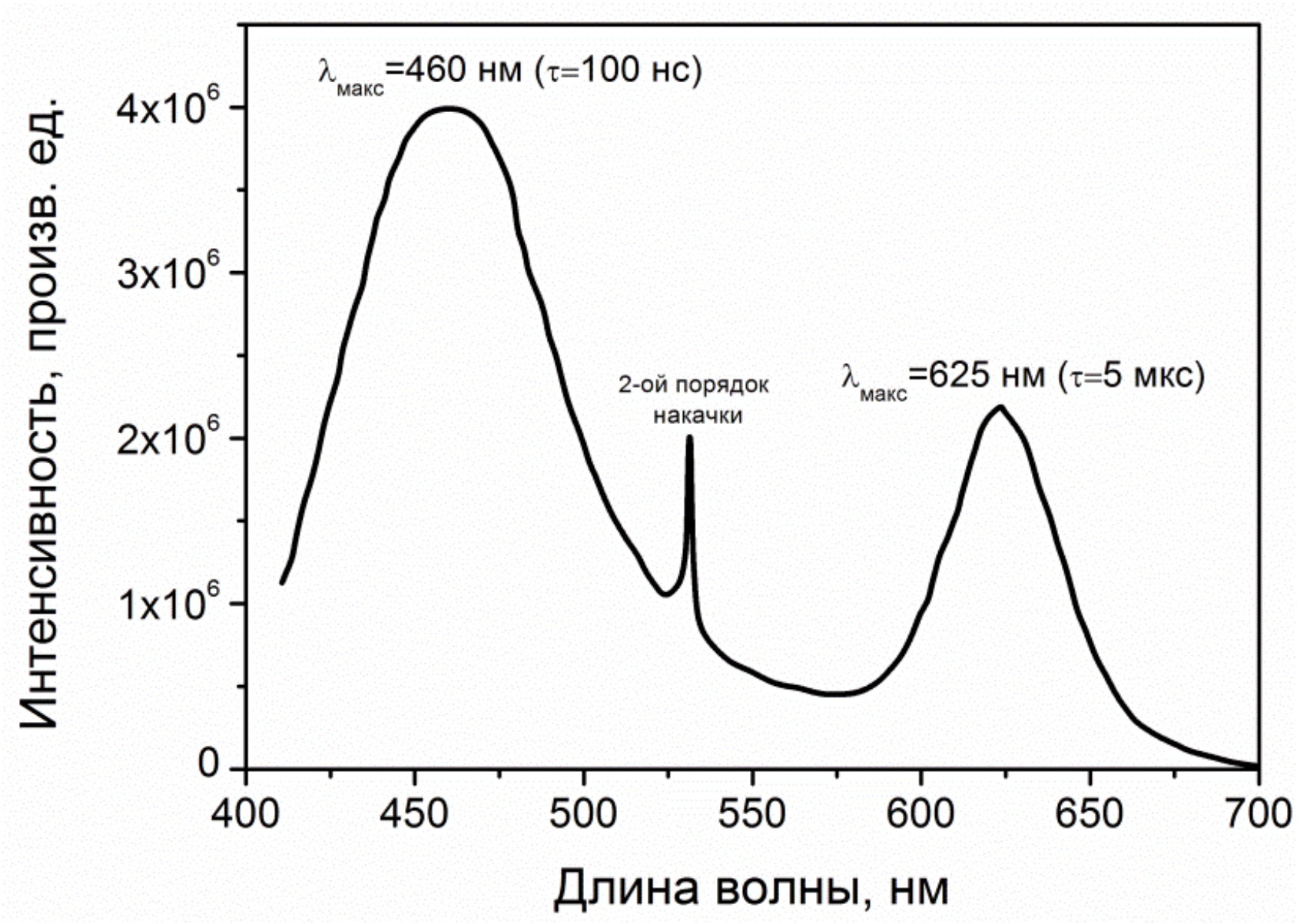

Рис. 1. 5 Спектр люминесценции кристалла $BaSO_4$ с примесью висмута при возбуждении на 266 нм (T=300 K) [49].

Полученный спектр люминесценции состоит из двух отдельных полос с максимумами на 460 (синей) и 625 нм (красной). Как уже обсуждалось ранее, ответственными за появление синей люминесценции являются ионы $Bi^{3+}$. Далее исходный кристалл подвергался термической обработке при T = 700 C на воздухе в течение небольшого промежутка времени. После термической обработки красная люминесценция кристалла полностью исчезала. Предполагалось, что отжиг на воздухе стимулирует процессы окисления ионов $Bi^{2+}$ до $Bi^{3+}$. Наблюдаемое тушение красной люминесценции указывает на причастность ионов $Bi^{2+}$ к ее появлению. Аналогичный результат был независимо получен в других работах [50, 51].

Позже подробные спектроскопические исследования, проведенные при T = 4.2 K, и теоретические расчеты структуры, свойств кристалла тетрабората стронция позволили определить структуру центра замещения $Bi_{Sr}$, рассчитать положения и форму поверхностей потенциальной энергии для основного и 3-х возбужденных состояний, принадлежащих $Bi^{2+}$[52]. Полученные результаты полностью подтверждают причастность ионов $Bi^{2+}$ к появлению красной люминесценции.

Следует также отметить, что для изучения структуры кристалла и наличия ионов $Bi^{2+}$ в них эффективной методикой является электронный парамагнитный резонанс (ЭПР), поскольку данный ион обладает наиболее выраженными парамагнитными свойствами (из-за наличия неспаренных электронов) по сравнению с остальными формами висмута. Впервые с помощью ЭПР исследовался кристалл $CdWO_4$, активированный висмутом [53], в котором была обнаружена типичная для ранее изученных кристаллов $BaSO_4$:Bi и $SrB_4O_7$:Bi красная люминесценция (Рис. 1. 6).

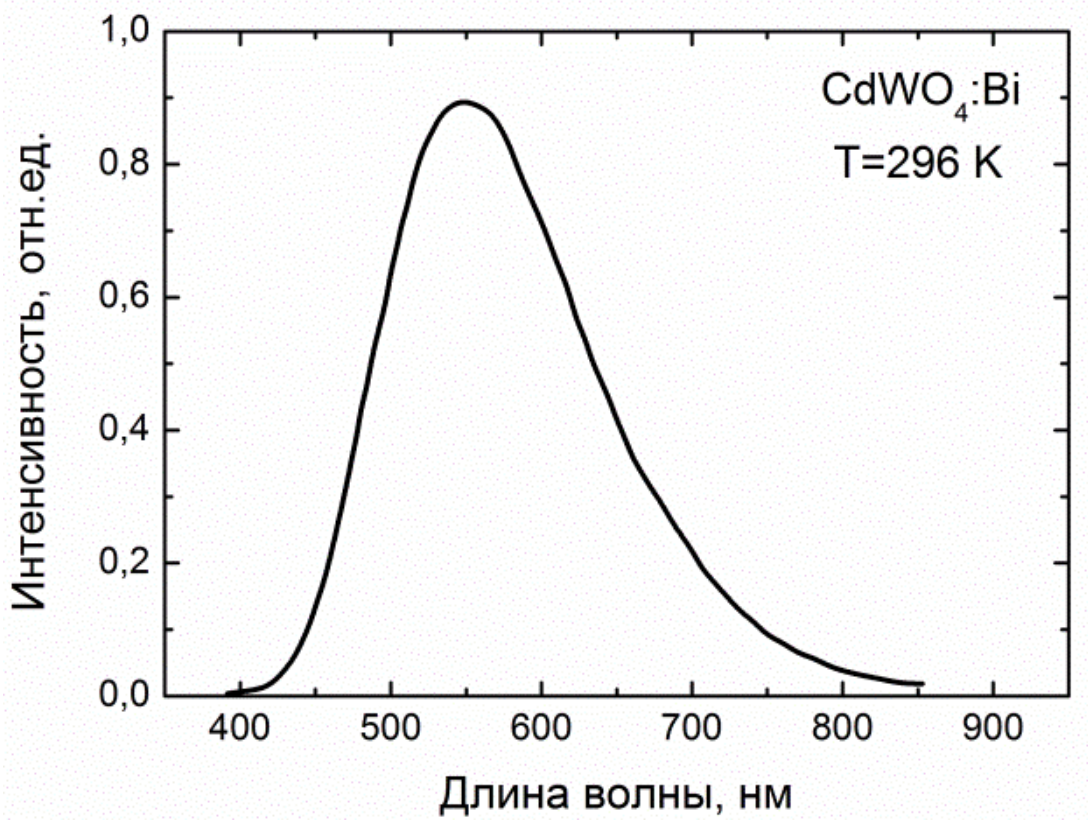

Рис. 1. 6 Спектр люминесценции кристалла $CdWO_4$:Bi при возбуждении на 351 нм [53].

Предполагалось, что ионы $Bi^{2+}$ встраивались в решетку кристалла $CdWO_4$, как и в ранее рассмотренных случаях, замещая ионы с близким радиусом. В случае кристалла $CdWO_4$ таким ионом был $Cd^{2+}$. Оказалось, что ЭПР спектр кристалла состоит из 10 эквидистантных резонансных пиков, появление которых характерно для $Bi^{2+}$ (т.к. спин ядра висмута равен I=9/2, то расщепление в постоянном магнитном поле дает 2I+1=10 компонент). Экспериментально полученный спектр ЭПР показан на Рис. 1. 7.

В более поздних работах, например, изучая кристаллы вольфрамата свинца с висмутом [54], также наблюдали характерный ЭПР сигнал от ионов $Bi^{2+}$.

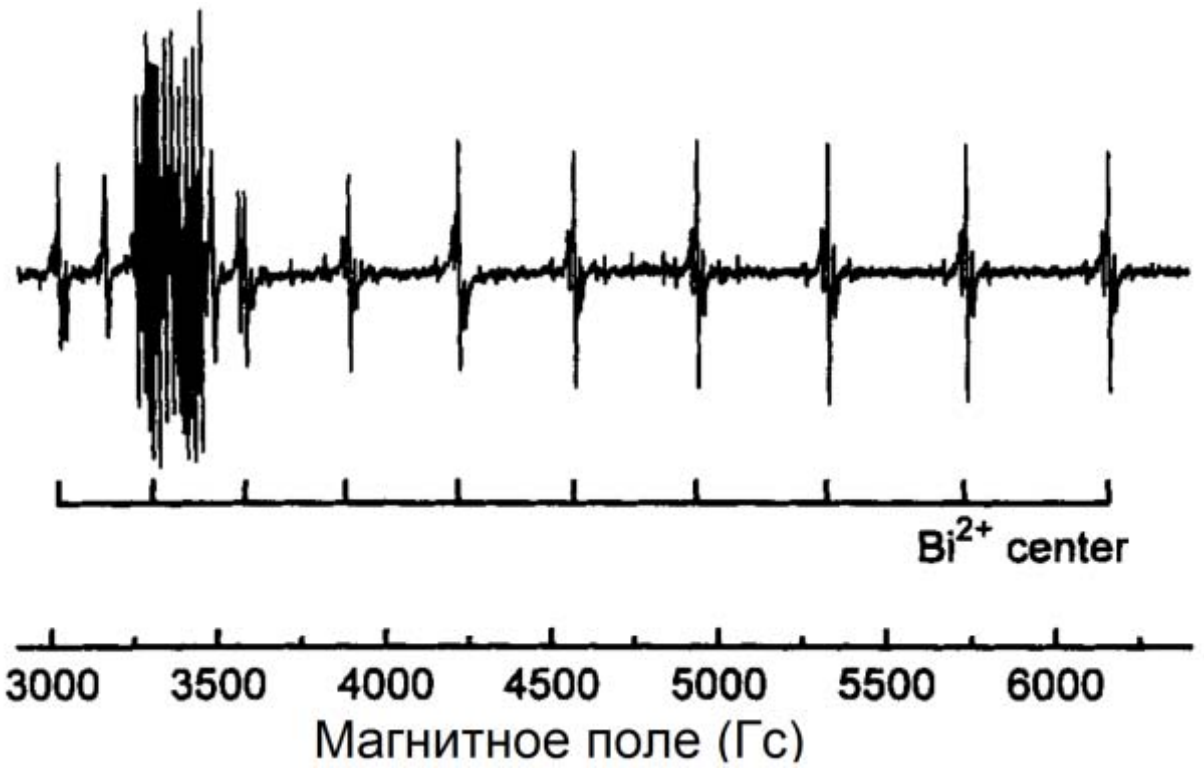

Рис. 1. 7 ЭПР спектр кристалла $CdWO_4$:Bi при T=77 K [53].

Таким образом, в настоящее время сформировалось, по-видимому, единое мнение о происхождении красной люминесценции в кристаллах, активированных висмутом. Показано, что система энергетических уровней иона $Bi^{2+}$ может полностью описать экспериментально полученные данные. На Рис. 1. 8 представлена простейшая схема уровней иона $Bi^{2+}$.

Переходы между энергетическими уровнями в представленной системе обозначены стрелками (сплошная линия со стрелкой вверх – возбуждение; сплошная (пунктирная) линия со стрелкой вниз – люминесценция (безызлучательный переход)). Наблюдаемые три полосы возбуждения красной люминесценции (переход $^2P_{3/2}(1)\rightarrow{}^2P_{1/2}$) обусловлены следующими переходами: $^2P_{1/2}\rightarrow{}^2S_{1/2}$, $^2P_{1/2}\rightarrow{}^2P_{3/2}(2)$ и $^2P_{1/2}\rightarrow{}^2P_{3/2}(1)$.

В заключение следует отметить, что попытки объяснить происхождение видимой люминесценции в кристаллах с висмутом изначально привело к большому числу гипотез и выдвигаемых моделей. Постоянное получение новой ин-формации о свойствах и особенностях таких кристаллов, совершенствование методов исследования и комплексный подход к изучению проблемы позволили

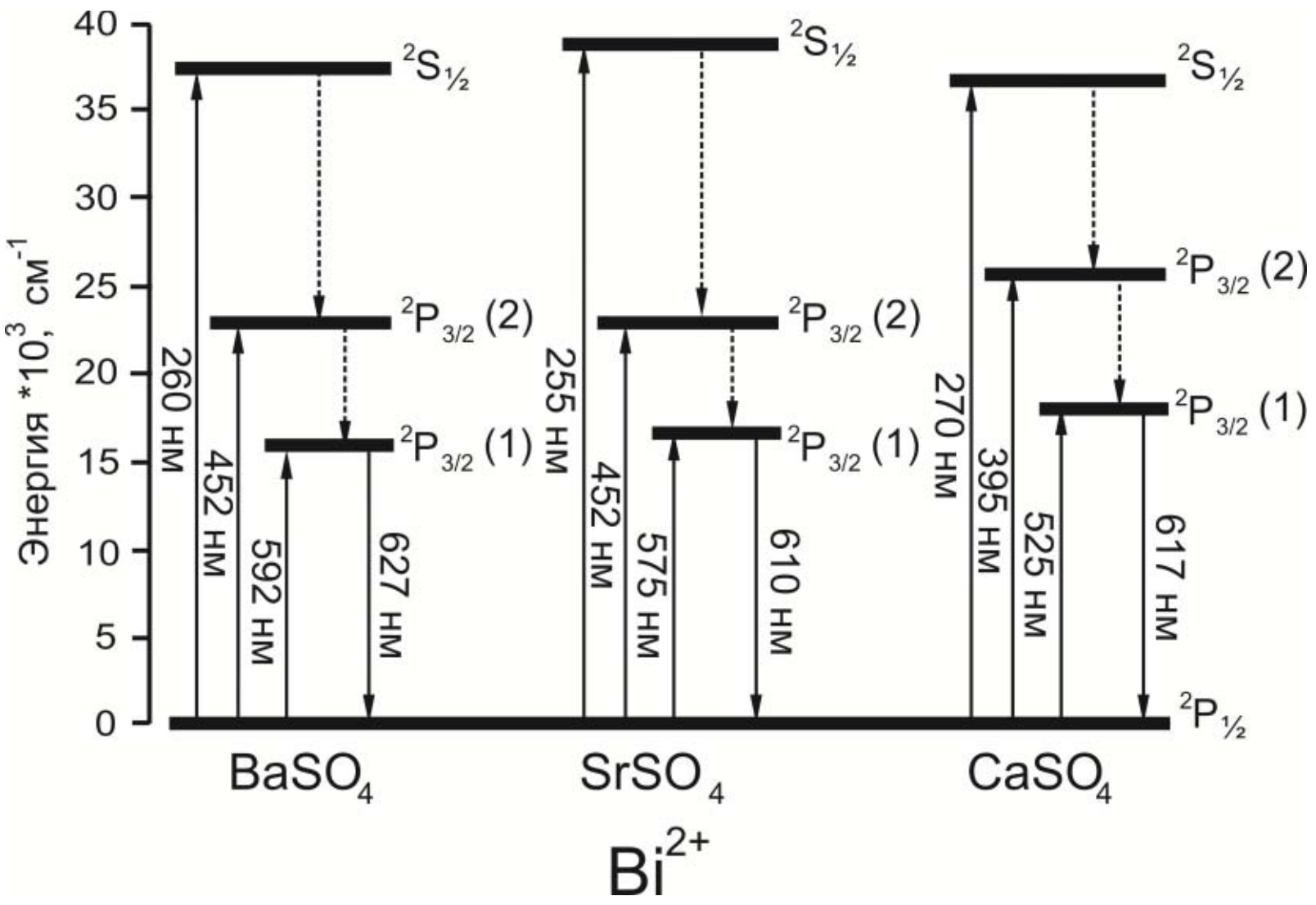

Рис. 1. 8 Схема энергетических уровней и оптических переходов иона $Bi^{2+}$ в указанных кристаллах [50]. Пунктирными (сплошными) линиями со стрелками вниз обозначены безызлучательные (излучательные) переходы. Сплошными линиями со стрелками вверх указаны характерные полосы возбуждения красной люминесценции.

определить природу таких оптических центров. Сейчас существует единая точка зрения о механизме появления синей и красной люминесценции, наблюдаемой в кристаллах с висмутом. Синяя люминесценция обусловлена оптическими переходами, характерными для иона $Bi^{3+}$, а красная люминесценция – для иона $Bi^{2+}$.

## 1.2 Люминесцентные свойства висмутовых активных центров, излучающих в ближней инфракрасной области спектра (ВАЦ). Основные существующие модели ВАЦ.

Более 10 лет назад была обнаружена ИК люминесценция в стеклах, легированных висмутом. Впервые о ней сообщалось в работе Murata и др. [24], в которой было показано, что данные среды (стекла из цеолитов с висмутом) являются перспективными материалами для реализации лазеров с высокой интенсивностью. Два года спустя, Fujimoto и Nakatsuka наблюдали ИК люминесценцию в алюмосиликатном стекле с висмутом [25]. Преимуществами таких ак-тивных сред стали широкая люминесценция в области 1100-1500 нм, и большие времена жизни – 410-650 мкс. По сути, эти публикации стали отправной точкой изучения новых лазерных материалов, легированных висмутом. Интерес к таким средам значительно усилился после получения лазерной генерации в 2005 году [2].

К настоящему времени накоплен значительный объем экспериментальных и теоретических данных, касающихся оптических свойств содержащих висмут оптических сред. Анализ имеющихся данных позволил выявить ряд закономерностей, присущих таким материалам, некоторые из которых будут подробно рассмотрены ниже.

Сначала будут обсуждаться спектрально-люминесцентные свойства оптических сред с висмутом и предлагаемые различными авторами модели активных центров.

***Кристаллические среды***

Для установления физической природы активных центров, излучающих в ближней ИК области, активнее стали изучаться кристаллические среды, легированные висмутом. Основная часть изученных кристаллов приведена ниже (Табл. 1. 1).

Следующие подходы применялись к изучению кристаллов и формирующихся в них висмутовых активных центрах:

- Изучение спектрально-люминесцентных свойств кристаллов. Сравнение результатов с ранее полученными данными.
- Применение спектроскопии комбинационного рассеяния и дифракционных методов для изучения структуры кристалла и новообразований, обусловленных включением активатора (висмута) в решетку кристалла.
- Влияние температурного отжига (в различных окислительно-восстановительных условиях) и гамма-облучения на свойства висмутовых центров.

Табл. 1. 1 Химический состав, параметры ИК люминесценции и предполагаемые модели висмутового активного центра в различных кристаллах.

| Кристалл | Содержание Bi, ат.% | Длина волны люминесценции ($\lambda_{макс}$), нм | Время жизни люминесценции, мкс | Предполагаемый ИК активный центр | Ссылка |
|---|---|---|---|---|---|
| Силикосилленит висмута | | 990, 1380 | | | [23] |
| $RbPb_2Cl_5$:Bi | | 1080 | 140 | $Bi^+$ | [55] |
| $BaF_2$:Bi | ~1 | 1070, 1500 | 2.1, 2.5 | $Bi^{2+}$ или $Bi^+$ + F центр | [56] |
| $SrB_4O_7$:Bi (поликристалл) | ~0.15 | 1250 | 180 | $Bi^+$ | [57] |
| α-$BaB_2O_4$:Bi | ~0.1 | 1139 | 526 | $Bi^+$ | [58] |
| $Ba_2P_2O_7$:Bi | ~0.03 | 1100 | ~640 | $Bi^0$ + $Ba^{2+}$ вакансия | [59] |
| $CdWO_4$:Bi | ~0.05 | 1078 | - | $Bi^{5+}$ | [60] |
| $Ba_2B_5O_9Cl$ :Bi | ~0.01 – 0.5 | 1030-1060 | ~30 | $Bi^0$ | [61] |
| CsI:Bi | ~0.2 | 1216, 1560 | - | $Bi^+$, $Bi_2^+$ | [62] |
| $KAlCl_4$:Bi | <0.1 | 980 | 525 | $Bi^+$ | [63, 64] |
| TlCl:Bi | <0.2 | 1180, ~1500 | 200-350 | $Bi^+$ + вакансия хлора | [65] |
| $KMgCl_3$:Bi | <0.1 | 950 | 400 | $Bi^+$ | [63] |

Общие закономерности, полученные при рассмотрении опубликованных результатов по оптическим свойствам кристаллов, активированных висмутом:

1) Спектр люминесценции в ближней ИК области состоит из двух широких (более 40 нм) полос с максимумами в области 980 – 1250 нм и/или 1500 – 1600 нм. Появление полосы люминесценции в коротковолновой области происходит при низких концентрациях висмута (менее 0.1 ат.%). Вторая полоса люминесценции наблюдалась только в кристаллах с содержанием висмута выше 0.1 ат.%. Из Табл. 1. 1 видно, что полоса люминесценции в длинноволновой области наблюдалась при изучении кристаллов с висмутом: $BaF_2$, TlCl, CsI. Исключением является поликристалл достаточно сложного состава $Ba_2B_5O_9Cl$:Bi.

2) Времена жизни ИК люминесценции в зависимости от кристалла может изменяться от ≈ 2 до 600 мкс.

3) Многообразие предложенных моделей активного центра свидетельствует о сложности природы центра люминесценции, а также о возможном сосуществовании нескольких различных типов люминесцирующих центров. Однако для большинства предложенных моделей имеется одно общее сходство, а именно, что в формировании активного центра принимает участие ион Bi, находящийся в низковалентном состоянии.

Приведенные выше данные отражают общее состояние дел в данной области. К сожалению, в большинстве опубликованных работ висмут рассматривался как элемент, который при включении в решетку кристалла, замещает ион другого химического элемента близкого по ионному радиусу. Такой подход может быть применен исключительно при изучении монокристаллов, но не для поликристаллов, исследования которых проводились в несколько раз чаще, чем монокристаллов. В общем понимании, поликристалл представляет собой твёрдое тело, состоящее из множества мелких кристаллитов (кристаллических зёрен), чаще всего не имеющих правильной кристаллической огранки. Граничная область между такими зернами оказывает сильное воздействие на различные свойства поликристаллов. В этой области структура материала может сильно отличаться от кристаллической. В частности, на Рис. 1. 9 приведена дифрактограмма для поликристалла $KAlCl_4$:Bi. Видно, что кроме характерных дифракционных пиков, имеется однородно рассеивающая подставка, интенсивность которой слабо зависит от угла дифракции. Это может служить доказательством наличия аморфной фазы, образовавшейся на границе кристаллических зерен.

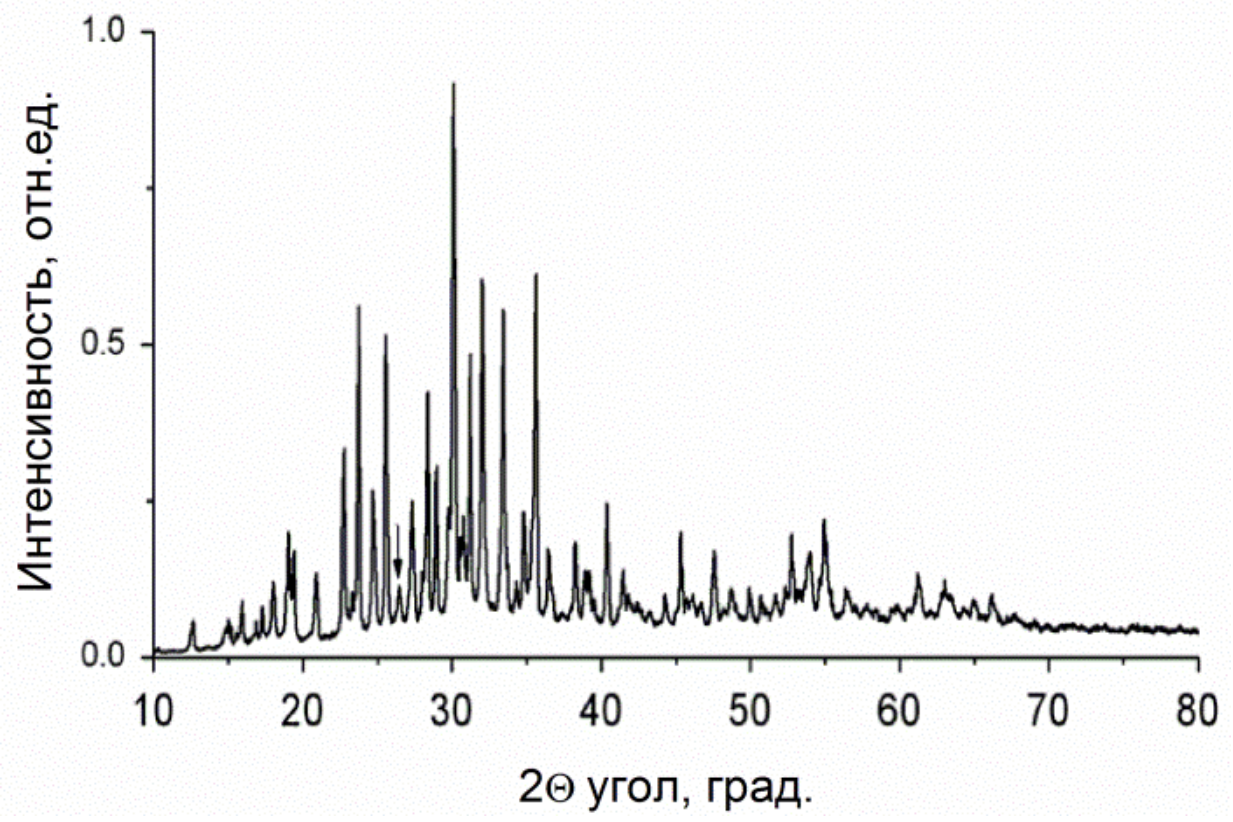

Рис. 1. 9 Дифрактограмма кристалла $KAlCl_4$:Bi [63].

Следовательно, при интерпретации полученных спектрально-люминесцентных свойств кристаллов необходимо учитывать то, что висмут может занимать положения в кристалле с различным локальным окружением, и иметь разную степень окисления. Возможно, что именно ион висмута, находящийся на границе кристаллических зерен, участвует в формировании центра ИК люминесценции. В пользу этого свидетельствует отсутствие ИК люминесценции в монокристалле $SrB_4O_7$ и ее наличие в случае поликристаллического образца [57].

Хотелось также обратить внимание на результаты по изучению влияния температурного отжига и γ-облучения на оптические свойства кристаллов (на примере α-$BaB_2O_4$:Bi). Как и ожидалось, отжиг в восстановительной атмосфере приводит к появлению ИК люминесценции (до облучения образцы не люминесцировали) [66]. Показано, что γ-облучение приводит к тому же эффекту, что и отжиг, то есть к наведению центров ИК люминесценции. Предполагается, что процесс формирования ИК люминесцентных центров при воздействии γ-облучением происходит по причине частичного восстановления ионов трехва-лентного висмута до $Bi^+$. Главными источниками свободных электронов, захват которых приводит к восстановлению ионов $Bi^{3+}$, являются существующие в кристалле дефектные центры. Примерная схема процесса восстановления может выглядеть следующим образом:

$$V''_{Ba} \rightarrow V_{Ba} + 2e,\ Bi^{3+} + e \rightarrow Bi^{2+},\ \text{и}\ Bi^{2+} + e \rightarrow Bi^{+},$$

где $V''_{Ba}$, $V_{Ba}$ – дефектные центры.

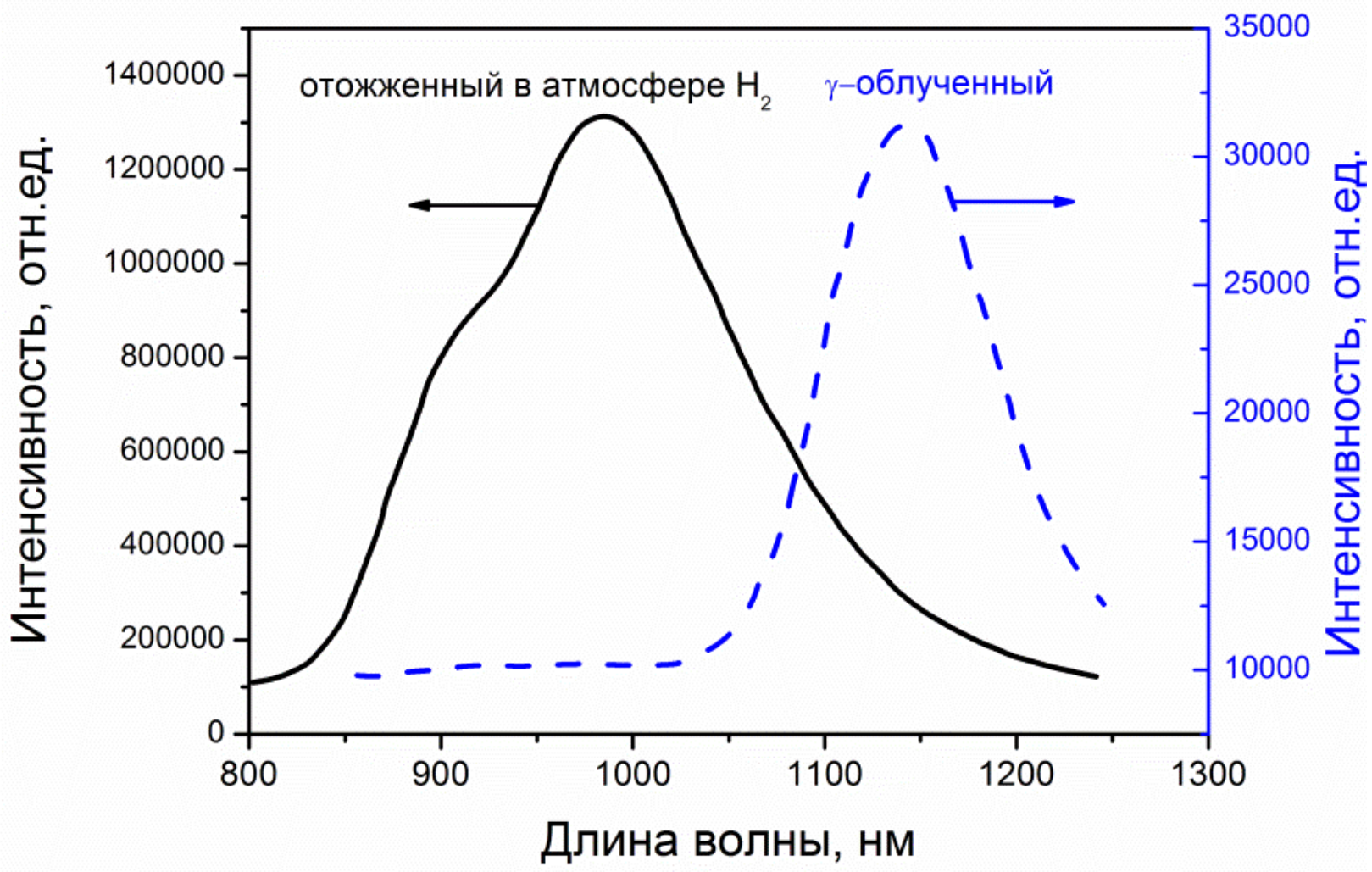

Рис. 1. 10 ИК люминесценция кристалла α-$BaB_2O_4$:Bi, отожженного в атмосфере $H_2$ (T = 800 C) и γ-облученного (доза 140 кГр) [66]. Длина волны возбуждения – 808 нм.

Необходимо отметить, что отжиг и γ-облучение приводит к одному и тому же результату – инициации процесса формирования центров ИК люминесценции, которые отсутствуют в исходном кристалле (Рис. 1. 10). Но нет доказательств, что в обоих случаях происходит формирование одних и тех же центров. На Рис. 1. 10 для сравнения показаны спектры люминесценции α-$BaB_2O_4$:Bi кристалла после отжига и аналогичного кристалла после γ-облучения. Видно, что максимум люминесценции γ-облученного образца смещен на 150 нм в длинноволновую область по сравнению с люминесценцией отожженного образца. По-видимому, наблюдаемые полосы люминесценции следует относить к различным активным центрам, а следовательно не могут быть описаны исключительно переходами, присущими системе энергетических уровней одиночного $Bi^+$, как предполагали авторы работы [66].

Подробное влияние отжига (на примере кристалла CsI:Bi) на оптические свойства активных центров было изучено в работе [67]. Авторами работы наблюдались существенные изменения в спектрах поглощения кристалла йодида цезия (Рис. 1. 11). Предполагается, что тепловая обработка кристалла приводила к последовательному восстановлению $Bi^{3+}$. После отжига при 823 К в спектре поглощения можно наблюдать заметный рост полосы около 700 нм. Авторы связали наблюдаемое изменение с формированием димероподобных структур $Bi_2^+$.

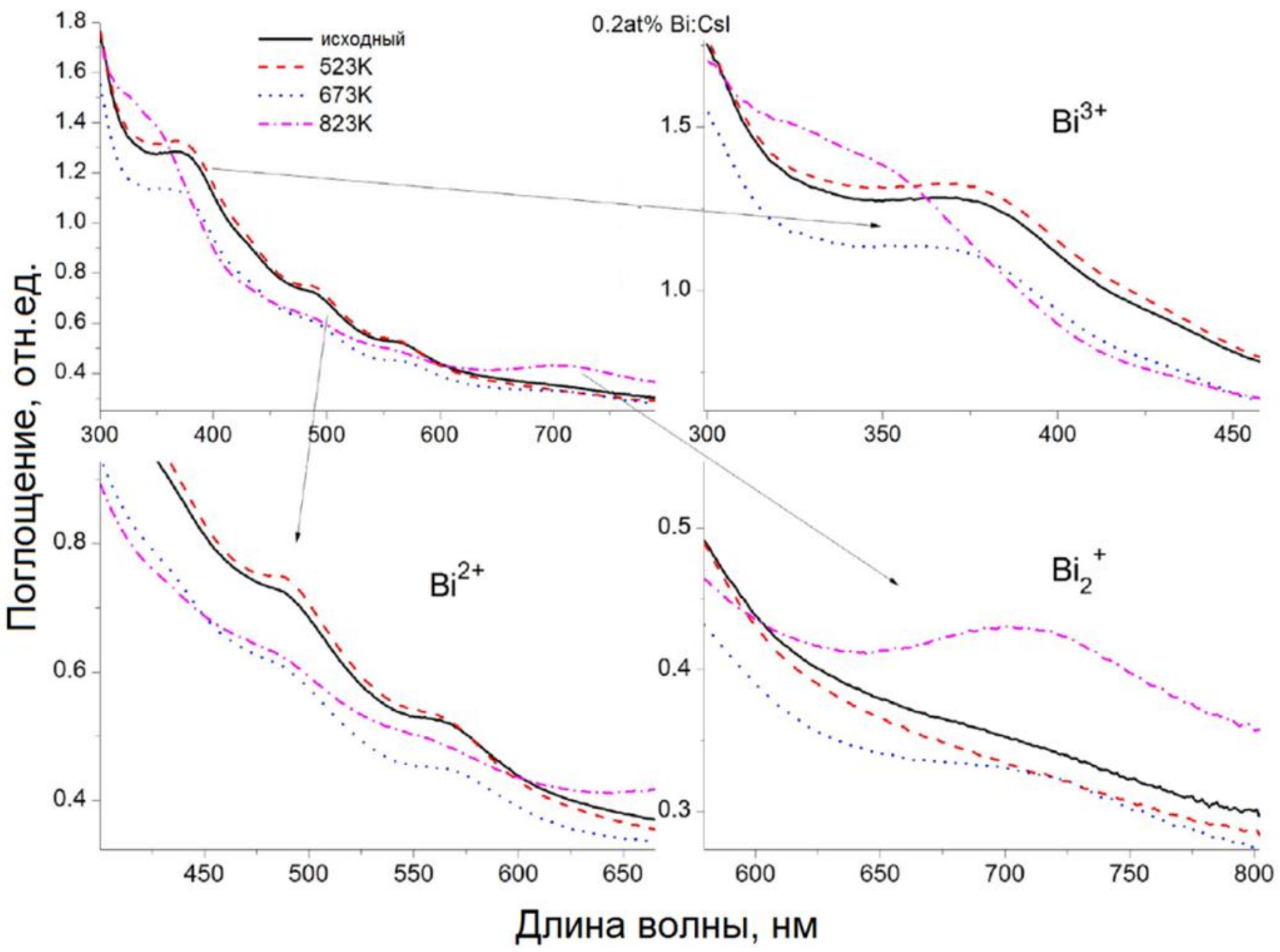

Рис. 1. 11 Спектры поглощения исходного кристалла CsI:Bi и аналогичного кристалла, отожженного при 523, 673 и 823 К [67].

На Рис. 1. 12 представлены спектры люминесценции кристалла CsI:Bi, отожженного при 523, 673 и 823 К. Видно, после отжига при температуре 673 К интенсивность полосы около 1600 нм возросла почти в 3 раза по сравнению с интенсивностью неотожженного образца.

Таким образом, совокупность проведенных исследований, касающихся кристаллических образцов, не позволяет однозначно определить висмутовый ИК центр. Это стало толчком к изучению еще более простых структур, чем монокристаллы. В частности, были изучены люминесцентные свойства соедине-ния $Bi_5(AlCl_4)_3$ [68]. Данные структуры являются примером соединений, в которых висмут существует в виде поликатионов $Bi_5^{3+}$. Такие модельные структуры обладают широкополосной ИК люминесценцией от 1 и до 4 мкм. Однако сравнительный анализ оптических свойств кристаллов с такими модельными структурами проводить нецелесообразно, поскольку для формирования последних требуется выполнение специфических условий (например, атмосфера $N_2$, очень низкая температура 150-360 С по сравнению с объемными стеклами и др.). По-видимому, вероятность формирования таких структур при высоких температурах, свойственных кристаллическим и стеклообразным средам, является достаточно низкой.

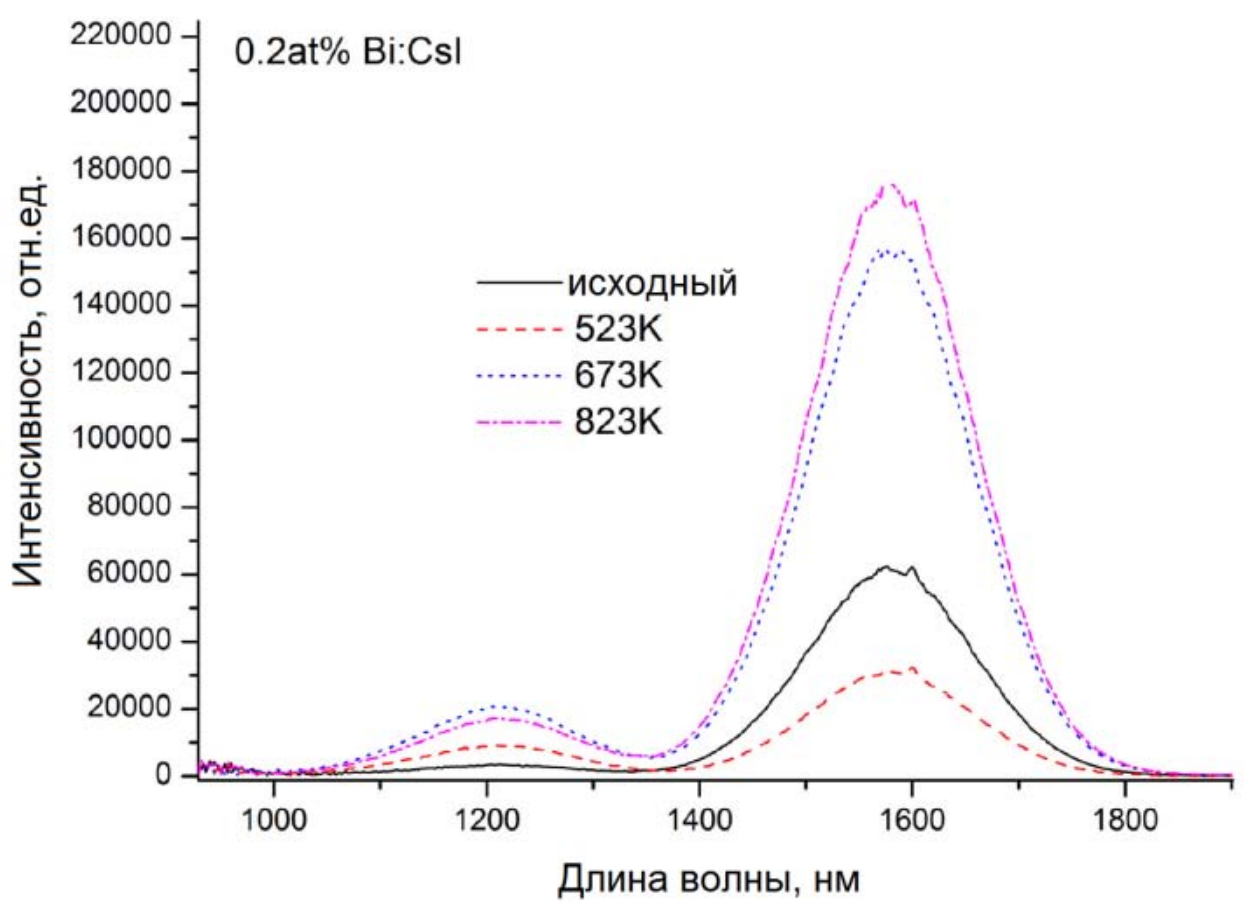

Рис. 1. 12 Спектры люминесценции кристалла CsI:Bi до и после отжига при 523, 673 и 823 К [67].

***Стеклообразные материалы***

Работы по исследованию кристаллов с висмутом занимают малую долю в общем объеме опубликованной литературы по висмутовым центрам, излучающим в ближней ИК области спектра. Основу составляют работы по исследованию стекол (оксидных и неоксидных). Висмутом легировались различные стеклянные матрицы: силикатные, германатные, боратные, фосфатные, флюоридные, халькогенидные, а также смешанные – боросиликатные, германосиликатные и другие. Поэтапное изучение различных стеклянных матриц показало, что люминесцентные свойства активных центров определяются химическим составом стекла. В настоящее время спектральная область, в которой можно получить люминесценцию висмутовых центров в стеклянной матрице, простирается от 900 до 2100 нм [69, 70]. Многими авторами (например, [71]) высказывалось предположение о том, что широкополосная люминесценция в стекле появляется в результате сосуществования нескольких разновидностей центров люминесценции. Перераспределение этих центров, то есть увеличение количества одного типа и уменьшение другого типа, в сетке стекла может зависеть от различных факторов. Из проведенного анализа литературы показано, что такими факторами могут быть: матрица стекла, температура синтеза, концентрация активатора, окислительно-восстановительные условия. Далее будет рассмотрено влияние этих факторов на формирование и оптические свойства ВАЦ в 3 типах стеклообразных материалов:

1) Стекла, синтезированные в тигле;
2) Импрегнированные висмутом пористые стекла;
3) Висмутовые волоконные световоды.

*Стекла, синтезированные в тигле*

Стеклам, полученным в тигле, посвящена самая большая доля имеющихся публикаций по теме данной работы. Химический состав такого типа стекол, как правило, является сложным, что определяется особенностями технологии приготовления. К сожалению, детальное сравнение и обобщение представленных результатов по стеклам, исследованным в разных работах, с целью поиска определенных закономерностей не представляется возможным. Это связано с рядом объективных причин. Одной из этих причин является отсутствие в работах состава полученного стекла. Как правило, приводится состав по закладке исходной шихты. Однако высокая летучесть ряда элементов (в том числе висмута) в процессе синтеза приводит к изменению соотношения исходных компонент в стекле. Другой причиной является искажение химического состава стекла, обусловленного неконтролируемой диффузией материала тигля. В частности, в работе [72] по результатам химического анализа было определено, что полученные в корундовом тигле германатные стекла, которые изначально легировались только висмутом, содержат от 1 до 7 мол. % оксида алюминия. В первой публикации авторы не предоставили такой информации и позиционировали их как германатные стекла без дополнительных примесей, кроме висмута [73].

Поэтому далее будут рассмотрены лишь некоторые результаты, которые отражают общие закономерности (без детализации и сравнений) изменения оптических свойств висмутсодержащих стекол.

Как уже было выше отмечено, синтезированные в тигле стекла имеют сложный состав. При синтезе таких стекол обычно добавляют щелочные и щелочноземельные элементы для снижения температуры плавления шихты. Используемая добавка сильно влияет на такой параметр как оптическая основность (optical basicity) стекла [74]. Чем выше значение данного параметра, тем большей основностью обладает матрица стекла. В этом случае активным ионам наиболее выгодно находится в окисленном состоянии [75]. Изучение влияния оптической основности на положение максимума полос люминесценции, относящихся к ионам висмута, проводилась, например, в работе [76]. Более подробно рассмотрим и обсудим результаты работы [77] по боросиликатным стеклам, легированных висмутом. Добавка различного содержания Na в такое стекло позволяла получать различные значения оптической основности стеклянной матрицы и оказывало влияние на валентное состояние висмута. В полученных стеклах наблюдались характерные полосы люминесценции в видимой и ближней ИК области. На Рис. 1. 13 показано изменение интенсивности синей и красной люминесценции для данного типа стекла. Монотонное снижение интенсивности синей люминесценции при снижении оптической основности происходит по причине частичного восстановления ионов $Bi^{3+}$ до $Bi^{2+}$ и более восстановленных форм. В таком случае ожидаемым был бы рост интенсивности красной люминесценции (из-за увеличения ионов $Bi^{2+}$) на начальном участке с последующим ее спадом (по причине восстановления ионов $Bi^{2+}$). Действительно, за-висимость для красной люминесценции, показанная на Рис. 1. 13 б, соответствовала ожиданиям.

| Образец | $B_2O_3$ (мол %) | Λ | $[BO_4]$ (мол %) |
|---|---|---|---|
| G1 | 5 | 0.5918 | 10.00 |
| G2 | 10 | 0.5730 | 17.94 |
| G3 | 13 | 0.5609 | 21.14 |
| G4 | 15 | 0.5533 | 23.34 |
| G5 | 16.5 | 0.5479 | 25.04 |
| G6 | 20 | 0.5261 | 20.00 |
| G7 | 23 | 0.5069 | 13.98 |
| G8 | 25 | 0.4946 | 10.00 |

а

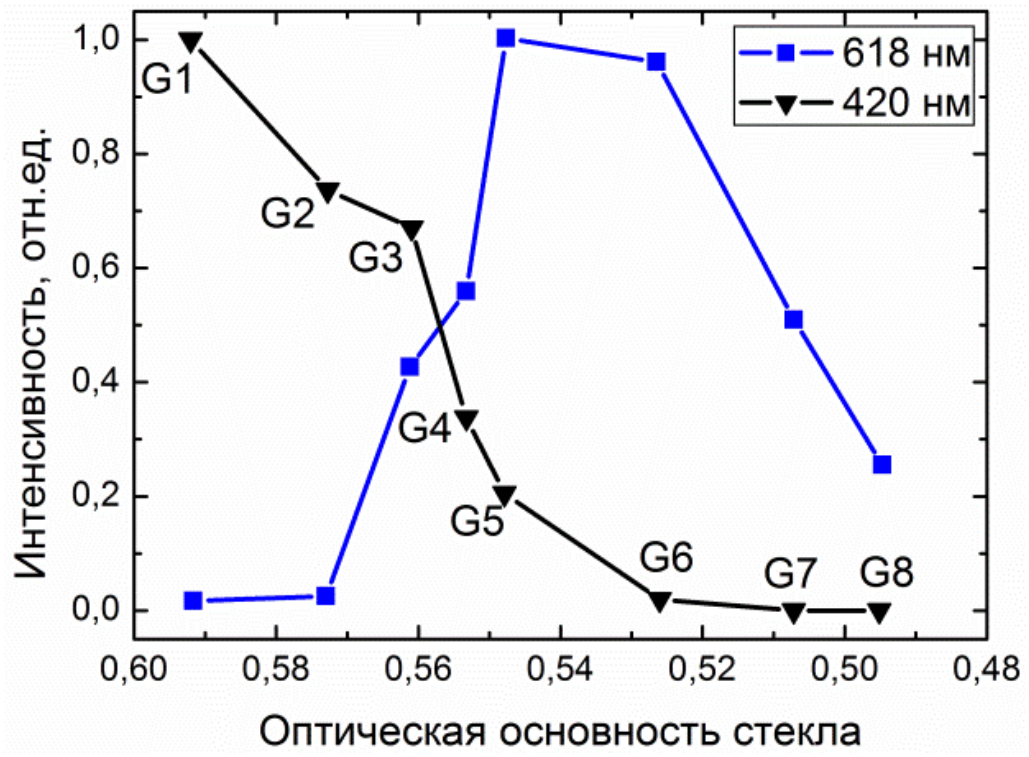

б

Рис. 1. 13 а) Концентрация $B_2O_3$, оптическая основность и процентное содержание боратных тетраэдров в различных образцах боросиликатных стекол [72]; б) зависимость интенсивности полос синей (420 нм) и красной (618 нм) люминесценции от оптической основности стекла [77].

В натриево-боросиликатных стеклах не было обнаружено корреляции между ИК люминесценцией и оптической основностью. Однако авторы работы обнаружили линейную зависимость между интенсивностью ИК люминесценции и процентным содержанием боратных тетраэдров $[BO_4]$ в сетке стекла (Рис. 1. 14). Следует отметить, что для получения стекол с интенсивной люминесценцией требуется проведение оптимизации состава стекла, поскольку высокие концентрации натрия не только не могут обеспечить желаемых характеристик, а наоборот приводят к полному отсутствию ИК люминесценции [78].

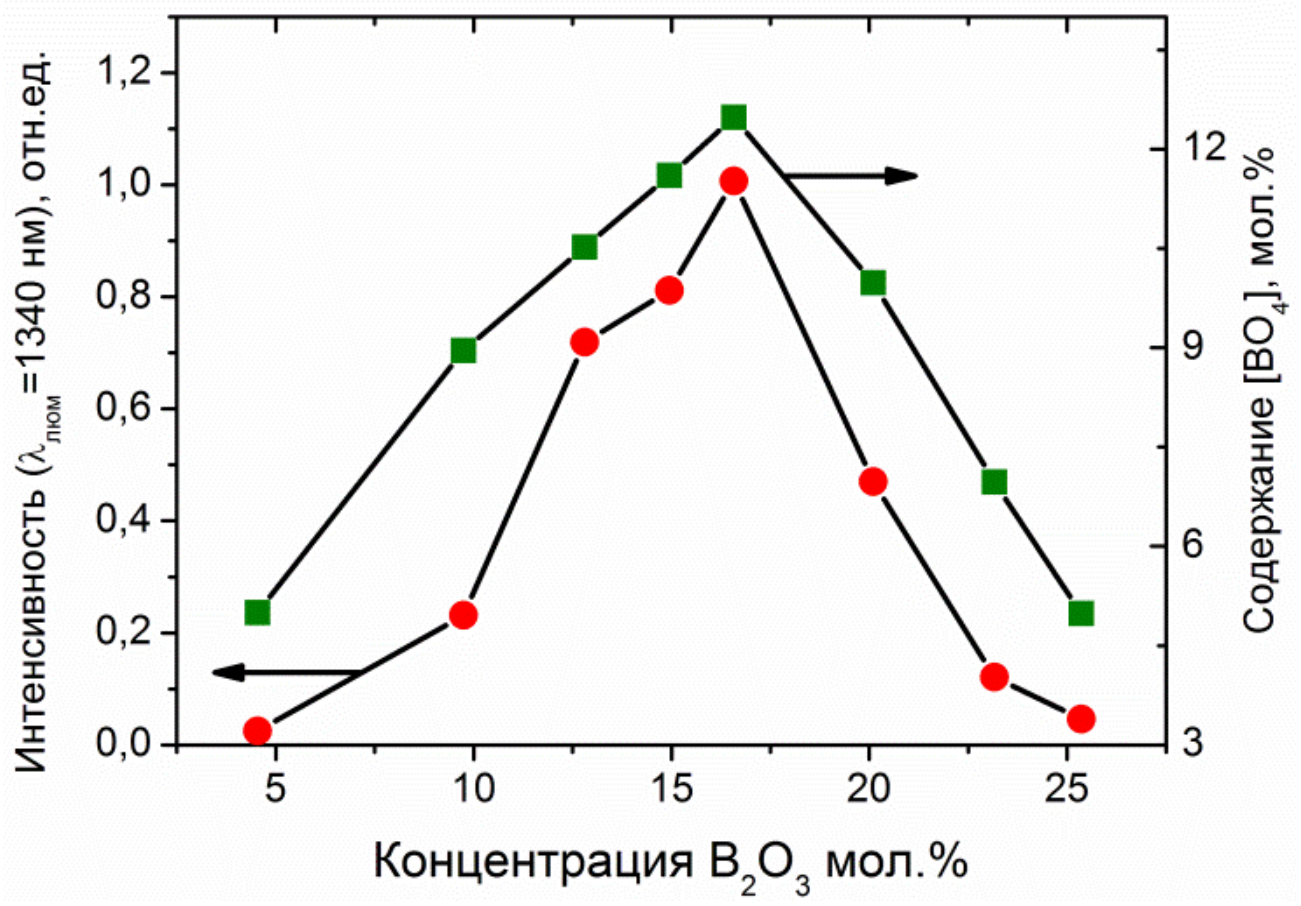

Рис. 1. 14 Зависимость интенсивности ИК люминесценции от концентрации оксида бора [78].

Следует отметить, что для некоторых видов стекол (например, оксифторидных стекол) оптическую основность можно изменять, применяя соответствующий прекурсор (соединение висмута). Данный подход был реализован в работе [79].

Таким образом, введение щелочных и щелочноземельных элементов способствует изменению степени окисления висмута в стекле, но формирование активных центров происходит без их участия. Обратная ситуация наблюдается в случае ионов алюминия, которые содержатся в большинстве исследованных стекол, в том числе в первых работах по висмуту (см. выше). Присутствие алюминия в стекле приводит к появлению ВАЦ с характерными полосами поглощения (500, 700 и 1000 нм) и люминесценции в области 1100 нм. Однако роль ионов алюминия по-прежнему остается предметом обсуждений. Одна из гипотез заключается в том, что введение алюминия является необходимым компонентом для формирования активного центра. Именно поэтому почти все первые исследования висмутсодержащих стекол ограничивались стеклами, содержащими Al. Сопутствующая Al добавка Ge создает условия (вероятно, соседствуя с активным центром) для существенного повышения интенсивности ИК люминесценции [80, 81]. Другая гипотеза возникла при сопоставлении спектрально-люминесцентных свойств ВАЦ и ЯМР спектров алюминия [72]. Было обнаружено, что в отличие от четырехкоординированного алюминия появление шестикоординированных атомов алюминия приводит к снижению интенсивности ИК люминесценции висмутовых центров. Этот результат противоречит первой гипотезе о роли алюминия. Авторы работы предполагают, что роль алюминия в сетке стекла заключается в предотвращении формирования полиатомных комплексов висмута (кластеров, димеров и др.).

Кроме того исследовались стекла, в состав которых входит другой элемент - фосфор, который, как известно, также хорошо препятствует кластеризации активных ионов (в частности, редкоземельных ионов) [3]. В работе [82] показано, что добавка в количестве до 12 мол. % $P_2O_5$ в $50SiO_2$–$25Al_2O_3$–$25CaO$– $0.5Bi_2O_3$ стекло приводит к существенному уширению спектра ИК люминесценции и одновременному снижению ее интенсивности. Уширение спектра люминесценции предположительно может быть связано с появлением дополнительной полосы люминесценции в области 1300 нм, что в целом приведет к увеличению ширины полосы люминесценции. Ответственными за данную люминесценцию могут быть активные центры, образованные вследствие структурных изменений в стекле при включении фосфора, поскольку, фосфор обладает валентностью выше, чем кремний. В других работах было показано, что введение фосфора действительно приводит к появлению полосы люминесценции на 1300 нм при возбуждении на длинах волн около ≈ 450 нм, и ≈700 нм. Аргументов о снижении доли атомов висмута, участвующих в кластеризации, при добавлении фосфора в стекло в литературе (по нашим данным) не приводилось.

Оптические свойства висмутовых центров люминесценции в германатных стеклах с висмутом также изучались (например, [83, 84]). Следует отметить, что в указанных работах стекла приготавливались в платиновых тиглях, что исключает попадание алюминия в стекло. Таким образом, в этих работах можно говорить об ВАЦ именно в германатной матрице. ВАЦ показывали ИК люминесценцию, которая имела максимум около 1200 нм. Для получения центров с такой люминесценцией требовалось создания восстановительных условий. Влияние окислительно-восстановительных условий на формирование активных центров в германатных стеклах с висмутом подробно были изучены в работе [84]. Для этого синтез стекол проводился в различных атмосферах $O_2$, $N_2$ и воздуха. В зависимости от того, в какой атмосфере было сделано стекло, наблюдалось возрастание или снижение интенсивности ИК люминесценции. Восстановительная атмосфера благоприятно сказывалась на люминесценции в области 1200 нм, интенсивность которой возрастала. Окислительные условия приводили к снижению ИК люминесценции, но было замечено, что появлялась полоса люминесценции на ≈ 750 нм.

Б.И. Денкером с сотрудниками [85, 86] достаточно подробно было изучено не только влияние атмосферы, но и других факторов, в частности, концентрации висмута, температуры синтеза на формирование висмутовых активных центров на примере магний-алюмосиликатных стекол с висмутом. Были получены следующие результаты:

- для формирования ИК активных центров наиболее благоприятной является восстановительная среда;
- рост поглощения активных центров происходит при возрастании температуры синтеза;
- обнаружена квадратичная зависимость коэффициента поглощения на 500 нм от концентрации висмута, что, по мнению авторов, может указывать на димерную природу активных центров (при концентрации активных центров в 100 раз ниже общей концентрации висмута);
- проведена оценка зарядового состояния висмута в активных центрах. Показано, что суммарный заряд равен +5. Предложена следующая модель активного центра – пара ионов висмута $Bi^{3+}$ с локализованным электроном между ними (Рис. 1. 15).

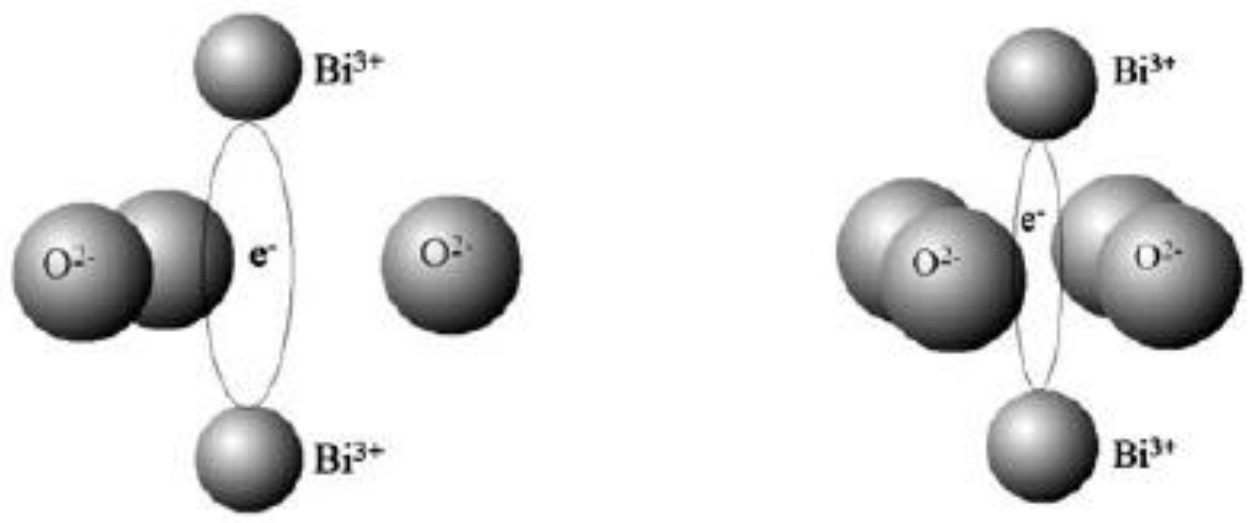

Рис. 1. 15 Предполагаемая структура заряженного димера [86].

Существует большое число публикаций, в которых получены аналогичные результаты по влиянию атмосферы (например, для германатных стекол с висмутом [87] и алюмосиликатных стекол с висмутом и лантаном [88]). Zh. Bai с коллегами обнаружил такой же эффект при изучении цеолитов [89].

Помимо атмосферы, большое количество работ было посвящено влиянию окислителей и восстановителей на формирование активных центров. Так, в работах [90, 91] было показано, что добавка оксида церия выполняет функцию окислителя висмута в стекле. На Рис. 1. 16, взятого из [90], хорошо видно, что введение оксида церия приводит к постепенному изменению окрашивания образца. Изначально образец был темный. При повышении концентрации оксида церия до 0.2 - 0.5 мол. % $CeO_2$ стекло приобретает желтовато-оранжевое окрашивание, а при более высоких концентрациях – стекло имеет бурую окраску. При высоких концентрациях церия, существенным фактором, оказывающим влияние на окраску стекла, является появление полос поглощения в УФ области, относящихся уже не к висмуту, а непосредственно к ионам церия.

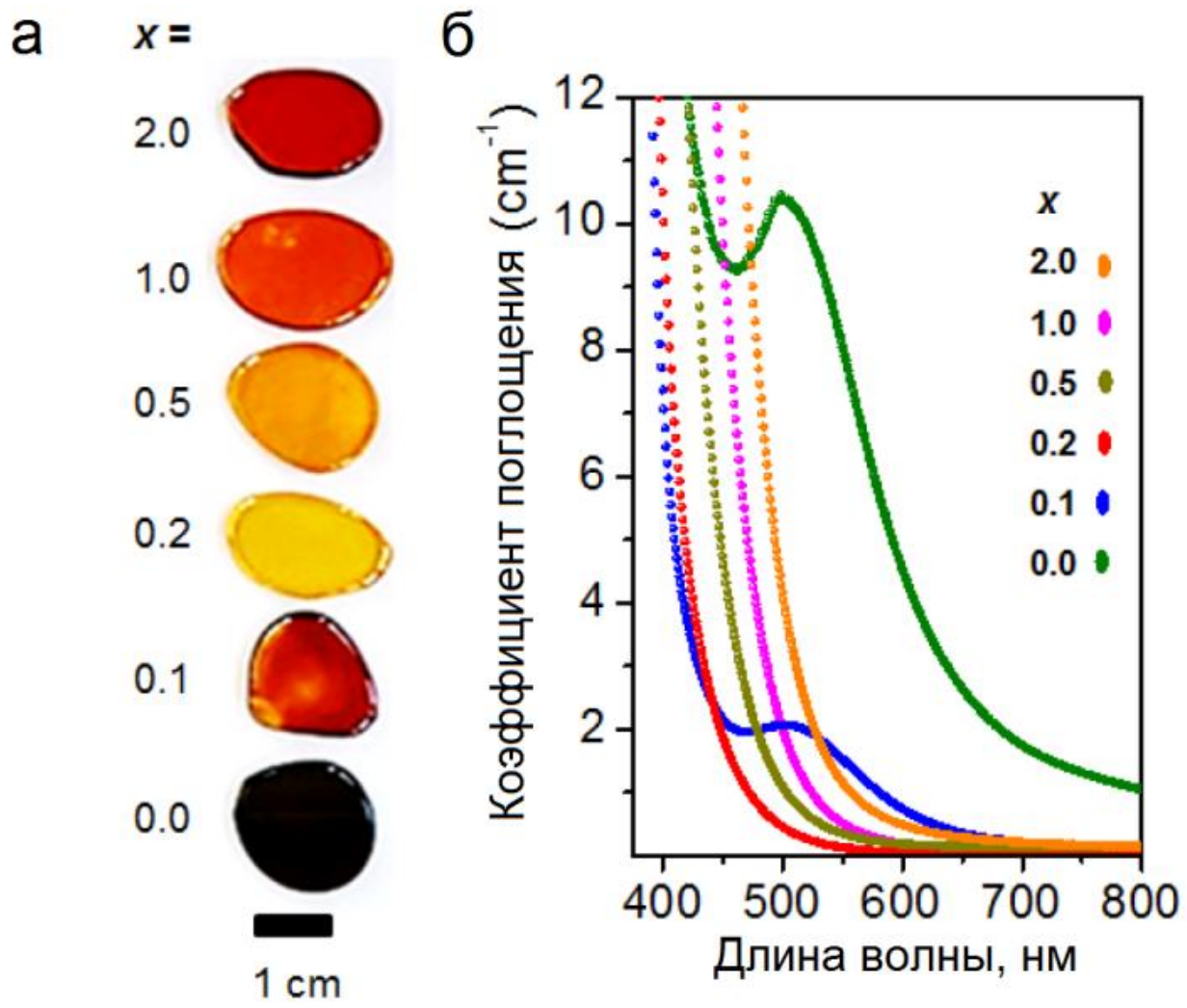

Рис. 1. 16 Фотографии (мол. %) (70.5-x) $GeO_2$ – 24.5 $Bi_2O_3$ – 5 $WO_3$: x $CeO_2$ стекол с равной концентрацией оксида церия (а) и соответствующие спектры поглощения (б) [90].

Зависимости величин интенсивностей поглощения и люминесценции представлены на Рис. 1. 17. Видно, что с возрастанием концентрации оксида церия интенсивность красной и ИК люминесценции снижаются по различным законам. Однако обе люминесценции исчезают одновременно – при одной концентрации оксида церия (в данном случае 0.3 мол. %).

В работе Khonthon и др. [92] для получения восстановительных условий использовалась добавка углерода в процессе приготовления стекол. Показано, что последовательное увеличение содержание углерода на первом этапе (до 1.5 вес. %) приводит к росту ИК люминесценции. Последующее увеличение уг-лерода (выше 1.5 вес. %) отрицательно влияет на уровень ИК люминесценции, то есть происходит ее снижение. Khonthon обнаружил, что в легированных вис-

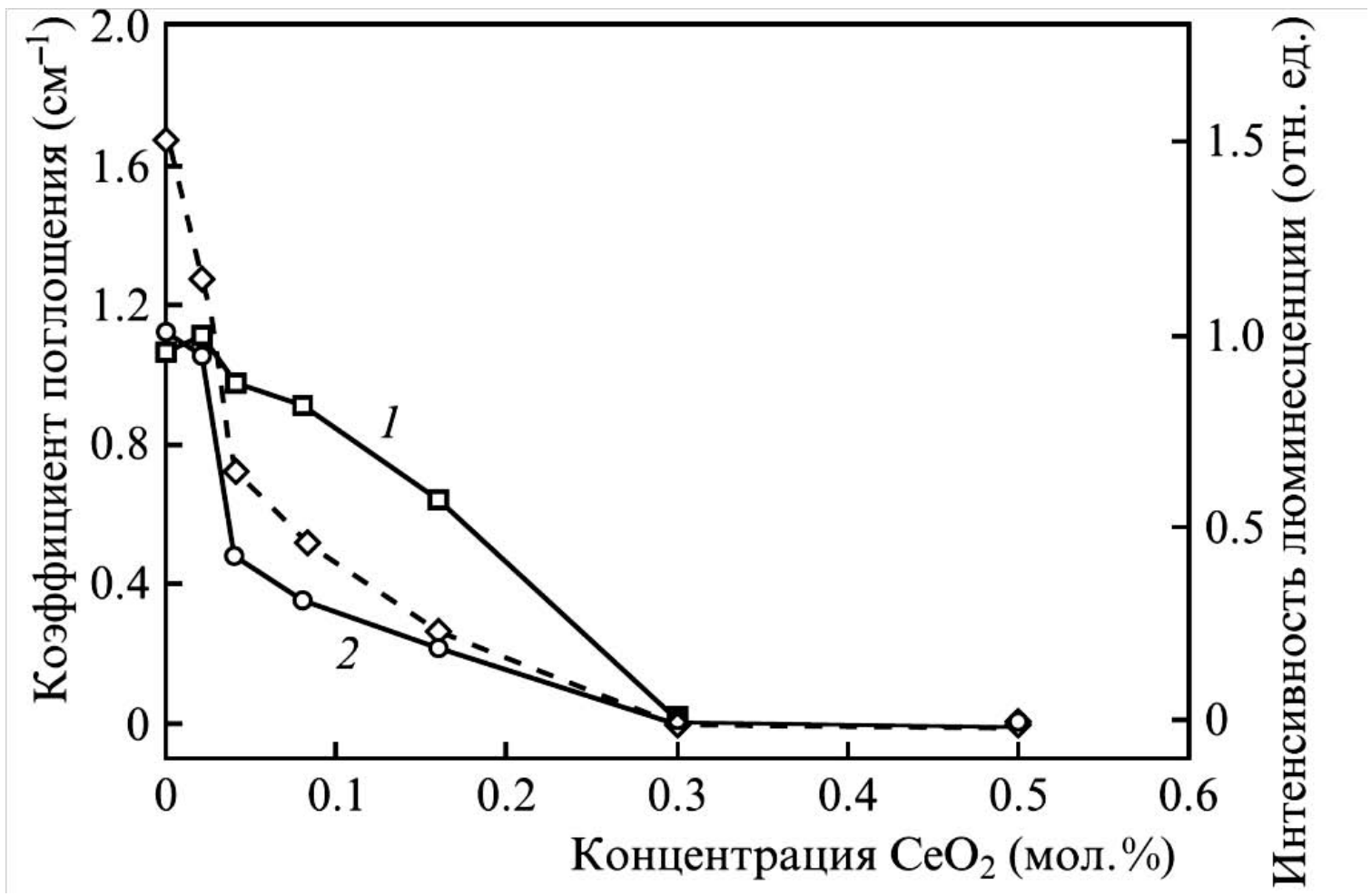

Рис. 1. 17 Зависимость поглощения на 500 нм (штриховая) и интенсивности красной (1) и ИК (2) люминесценции от концентрации оксида церия для стекол с 0.25 мол. % оксида висмута [91].

мутом образцах с содержанием углерода выше 1.5 вес.% формировались висмутовые комплексы. Этот результат показывает, что создание чрезмерных восстановительных условий может отрицательно сказываться на формировании ИК центров.

*Импрегнированные висмутом пористые стекла*

Число публикаций, касающихся импрегнированных (от impregnation - пропитывание) висмутом пористых кварцевых стекол, значительно уступает количеству работ по стеклам, синтезированным в тиглях. Тем не менее, в таких материалах получены важные результаты, которые необходимо рассмотреть подробнее.

Пористые стекла – материалы со специальной пористой структурой различной морфологии, которые позволяют эффективно контролировать химическое равновесие активных центров в наноразмерных структурах выбором температуры и атмосферы, избежать эффектов кластеризации и концентрационного тушения [93, 94]. Консолидируя наноразмерные структуры, удается сохра-нить неравновесное распределение активных центров, сформировавшихся в этих структурах. Обычно пористые стекла получаются выщелачиванием $SiO_2$ – $B_2O_3$ – $Na_2O$ – $Al_2O_3$ и $SiO_2$ – $B_2O_3$ – $Na_2O$ стекол в азотной кислоте $HNO_3$ [42, 95]. Размер пор составляет около 20 нм и может варьироваться в зависимости от условий приготовления. Полученные пористые стекла пропитываются раствором нитрата висмута в течение нескольких дней. Затем образцы высушиваются при T=150°C и подвергаются термообработке (с консолидацией стекла).

В работе [95] показано, исходные образцы кварцевого стекла с висмутом обладают исключительно синей люминесценцией. Температурный отжиг при 850 C приводит к снижению интенсивности синей люминесценции и появлению полос красной люминесценции в области 600 нм, что обусловлено восстановлением ионов $Bi^{3+}$ до $Bi^{2+}$. ИК люминесценция была обнаружена только после нагрева стекла до температур выше 1400 °C.

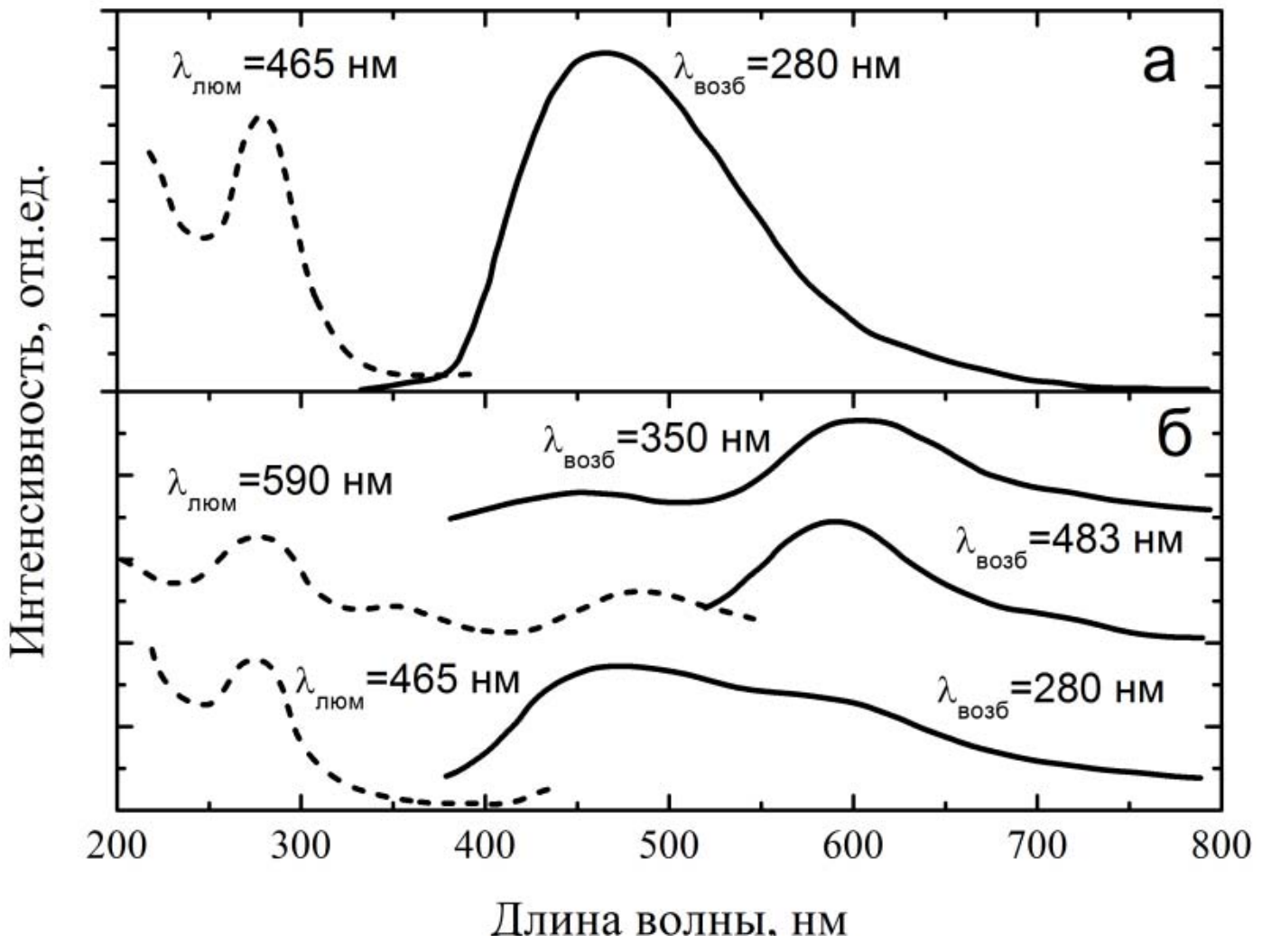

Рис. 1. 18 Спектры люминесценции (сплошная) и возбуждения люминесценции (штриховая) линия пористого стекла, термически обработанного а) на воздухе, б) в аргоне [42].

Получить ИК люминесценцию в пористых стеклах можно также при более низких температурах отжига около 1000 °C, но для этого требуется создание восстановительной атмосферы при отжиге [42]. На Рис. 1. 18 а, б. приведены спектры люминесценции образцов, термически обработанных на воздухе и в аргоне. Видно, что образец, отожженный в атмосфере воздуха, имел только си-нюю люминесценцию, характерную для $Bi^{3+}$, тогда как в атмосфере аргона синюю и красную люминесценцию. Появление красной люминесценции связано с наличием ионов $Bi^{2+}$, образование которых происходит при частичном восстановлении ионов $Bi^{3+}$. В ближней ИК области также наблюдалось появление двух широких полос люминесценции с максимумами, расположенными на 1100 и 1400 нм (Рис. 1. 19 а).

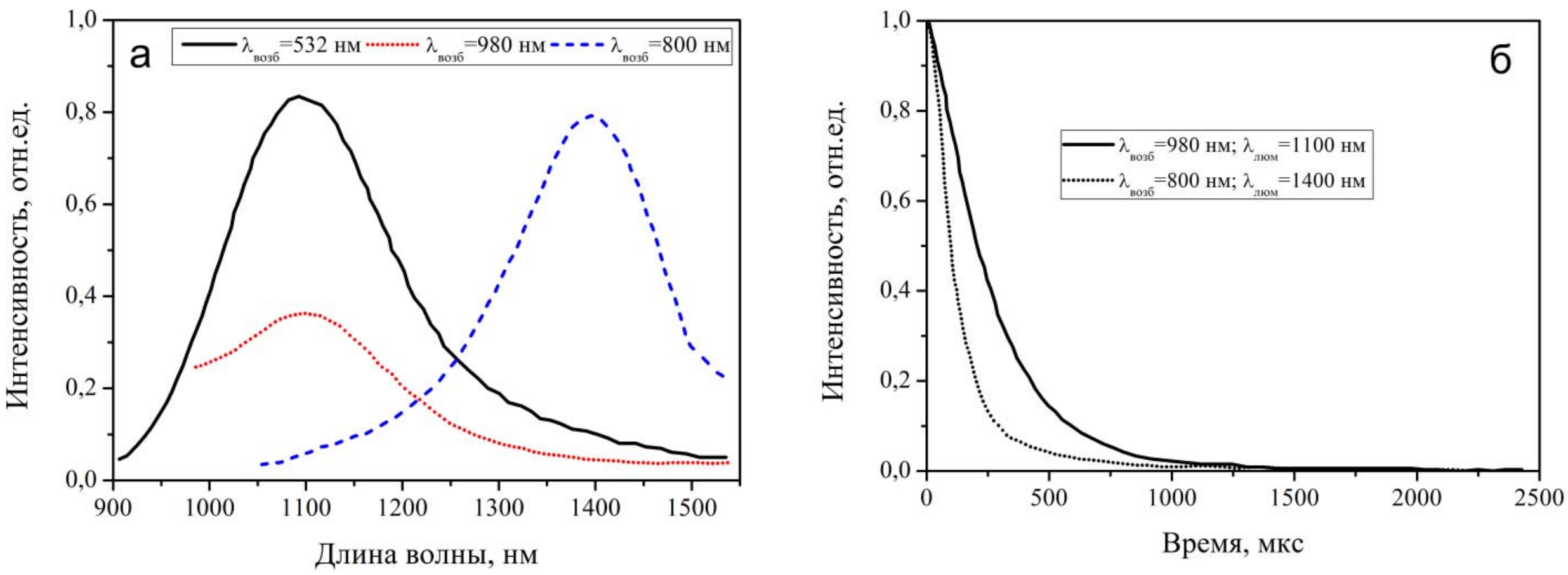

Рис. 1. 19 а) Спектры люминесценции и б) временные зависимости затухания ИК люминесценции висмут содержащего пористого стекла, отожженного в атмосфере аргона [42].

Временные зависимости затухания ИК люминесценции на 1100 нм и 1400 нм приведены на Рис. 1. 19 б. Полученные времена жизни люминесценции на 1100 нм и 1400 нм равны 108 и 254 мкс, соответственно. Предполагалось, что в формировании активного центра в пористых стеклах участвует ион $Bi^{+}$. С этой целью были проведены эксперименты по облучению пористых стекол, импрегнированных висмутом [96]. Стекла были приготовлены в окислительных условиях, а потому в них наблюдали только синюю люминесценцию. Затем данные стекла облучали излучением с длиной волны 800 нм, длительностью импульсов – 60 фс с частотой повторения – 1 кГц. На Рис. 1. 20 представлены спектры люминесценции облученных образцов стекла с висмутом в видимой и ближней ИК области. Помимо синей люминесценции, которая наблюдалась в исходных образцах, были обнаружены полосы красной и ИК люминесценции для облученных образцов. Присутствие красной люминесценции свидетельст-вует о протекании восстановительных процессов висмута при облучении стекла излучением фемтосекундного лазера.

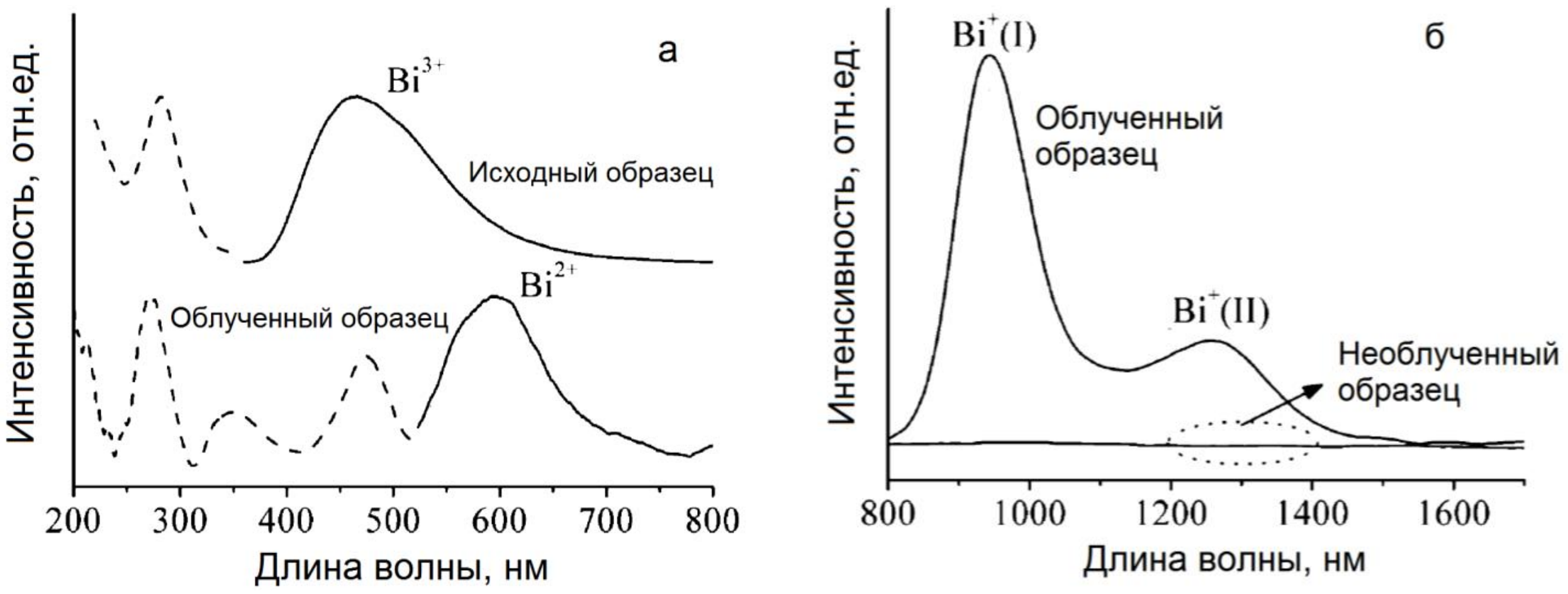

Рис. 1. 20 Спектры люминесценции облученного и необлученного стекла, приготовленного в атмосфере кислорода, в видимой (а) и ближней ИК (б) области [96]. Облучение стекол производилось излучением с λ=800 нм (длительность импульсов - 60 фс; частота повторения – 1 кГц).

В работе утверждается, что данный процесс не связан с локальным нагревом стекла под воздействием излучения высокой интенсивности. Предполагается, что трансформация $Bi^{3+}$ в $Bi^{2+}$ происходит по следующей схеме: $Bi^{3+} + e \rightarrow Bi^{2+}$. Следует отметить, что предыстория (способ приготовления, условия) стекла влияет на люминесцентные свойства образовавшихся активных центров. В частности, для облученных образцов, полученных в атмосфере кислорода, происходит появление полос в области 1000 нм и 1250 нм. Помимо этих полос люминесценции, дополнительная полоса люминесценции с максимумом на 1400 нм присутствует в спектре люминесценции облученного образца, отожженного в атмосфере аргона.

В работе [97] стекла были изготовлены по технологии “золь-гель”. Целью работы являлось изучить оптические свойства ВАЦ при изменении размера пор и выбора прекурсора. В результате было показано, что при размерах пор стекла менее 20 нм не наблюдается ИК люминесценция в области 1400 нм, то есть не происходит формирование ВАЦ в стекле. Использование в качестве прекурсоров соединений, содержащих димеры висмута, позволяло получать ха-рактерные ВАЦ полосы люминесценции в области 820 и 1400 нм. Авторами сделан вывод, что активными центрами являются димеры висмута ($Bi_2$).

*Висмутовые волоконные световоды*

В последней части литературного обзора будут изложены основные результаты, касающиеся оптических свойств ВАЦ в волоконных световодах различного состава. Отдельное рассмотрение висмутовых волоконных световодов связано с существованием отличительных особенностей по сравнению с объемными стеклами, в частности отличие температурных режимов охлаждения. Для световодов (из-за малого объема) процесс охлаждения происходит с существенно более высокими скоростями, чем для объемных образцов, что может повлиять на формирование активных центров. Кроме того, висмутовые световоды являются единственной оптически активной средой, в которой получена лазерная генерация. Данная особенность позволяет уделить особое внимание таким средам и подробно рассмотреть их оптические свойства.

Впервые волоконные световоды, легированные висмутом, были с небольшой разницей во времени изготовлены в НЦВО РАН (Россия) [1] и Sumitomo Electric Industries (Япония) [98] в 2005 году. В НЦВО РАН на данном световоде была получена впервые в мире лазерная генерация в области 1140 – 1215 нм [2]. Дальнейших публикаций от японских ученых не последовало. Первый лазер стал мощным толчком к проведению многочисленных исследований в данном направлении научными группами из различных стран.

В данном разделе будут изложены основные результаты, полученные до проведения исследований, описанных в настоящей работе.

**а) Алюмосиликатные световоды**

Изначально сердцевина волоконных световодов изготавливалась из алюмосиликатного стекла, поскольку считалось, что алюминий является неотъемлемым компонентом для формирования висмутовых центров люминесценции. По этой причине в течение нескольких лет изучались образцы с сердцеви-ной из стекла, солегированного висмутом и алюминием. Типичные спектры поглощения и люминесценции таких световодов представлены на Рис. 1. 21.

Для световода с сердцевиной из алюмосиликатного стекла, легированного висмутом, характерно наличие полос поглощения в области ≈1400, 1000, 810, 700, 500 и ≈230 нм (Рис. 1. 21 б). Спектральное положение наблюдаемых в данном случае полос слабо зависит от методов и условий (атмосфера, температура) изготовления образца, концентрации висмута. В отличие от спектрального положения, интенсивность полос сильно зависит от концентрации висмута (Рис. 1. 21 в), способа легирования и других технологических параметров [99]. Следует отметить, что введение дополнительных добавок, таких как Ge, P, Ti, также не приводит к существенному искажению люминесцентных свойств висмутовых центров в алюмосиликатном стекле [100, 101].

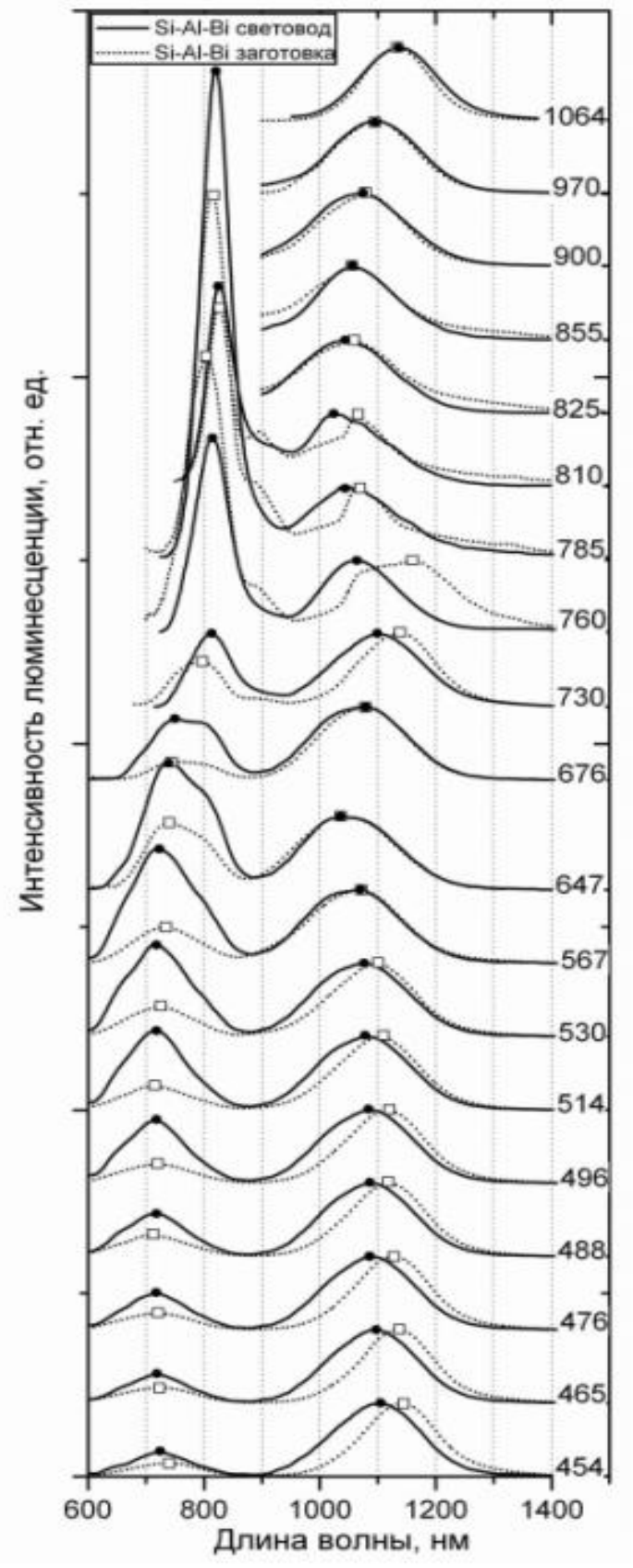

а

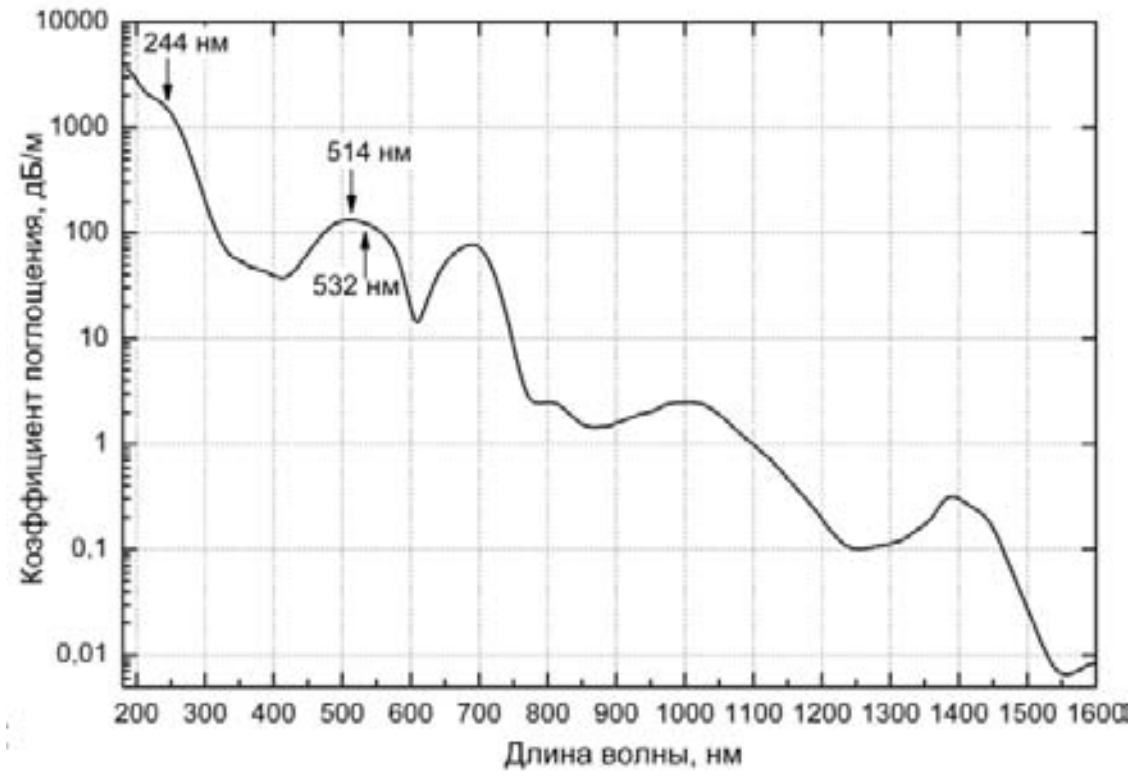

б

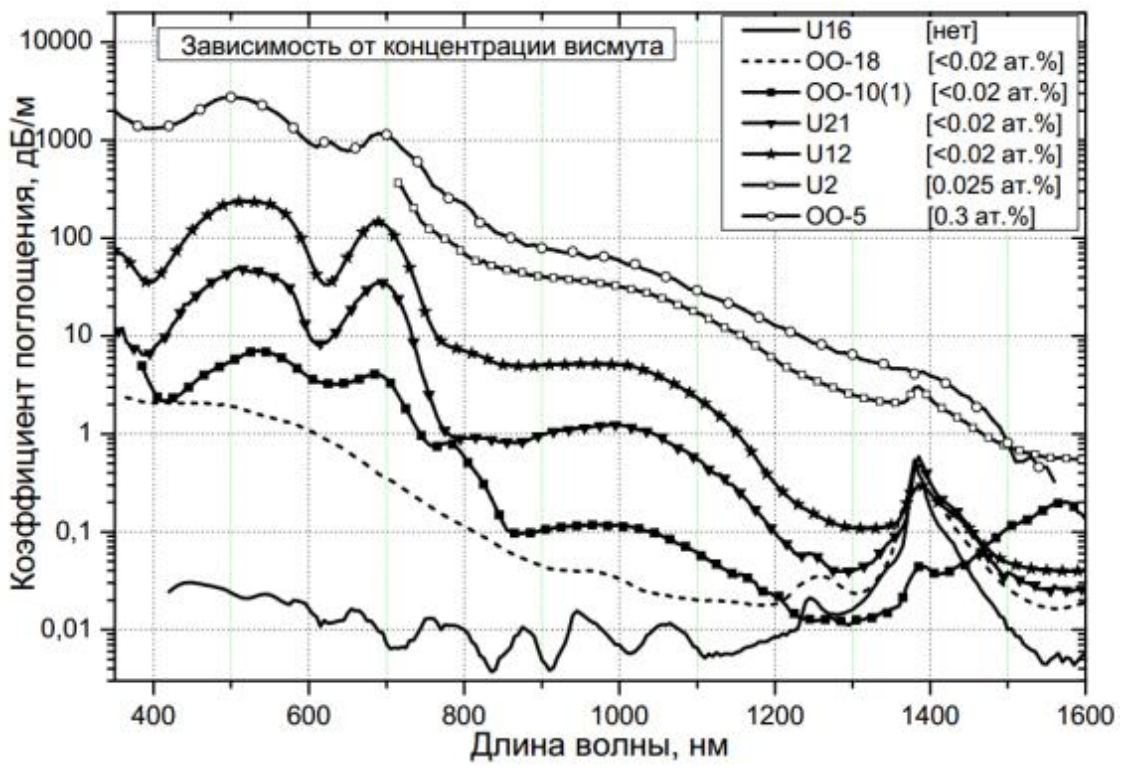

в

Рис. 1. 21 а) Спектры люминесценции преформы и волоконного световода (справа у каждого спектра указана длина волны возбуждения), б) спектр поглощения световода [102], в) спектры оптических потерь для алюмосиликатных световодов с различной концентрацией висмута [99].

При возбуждении в каждую из указанных полос поглощения наблюдается ИК люминесценция в области 1150 нм (Рис. 1. 21 а). При некоторых длинах

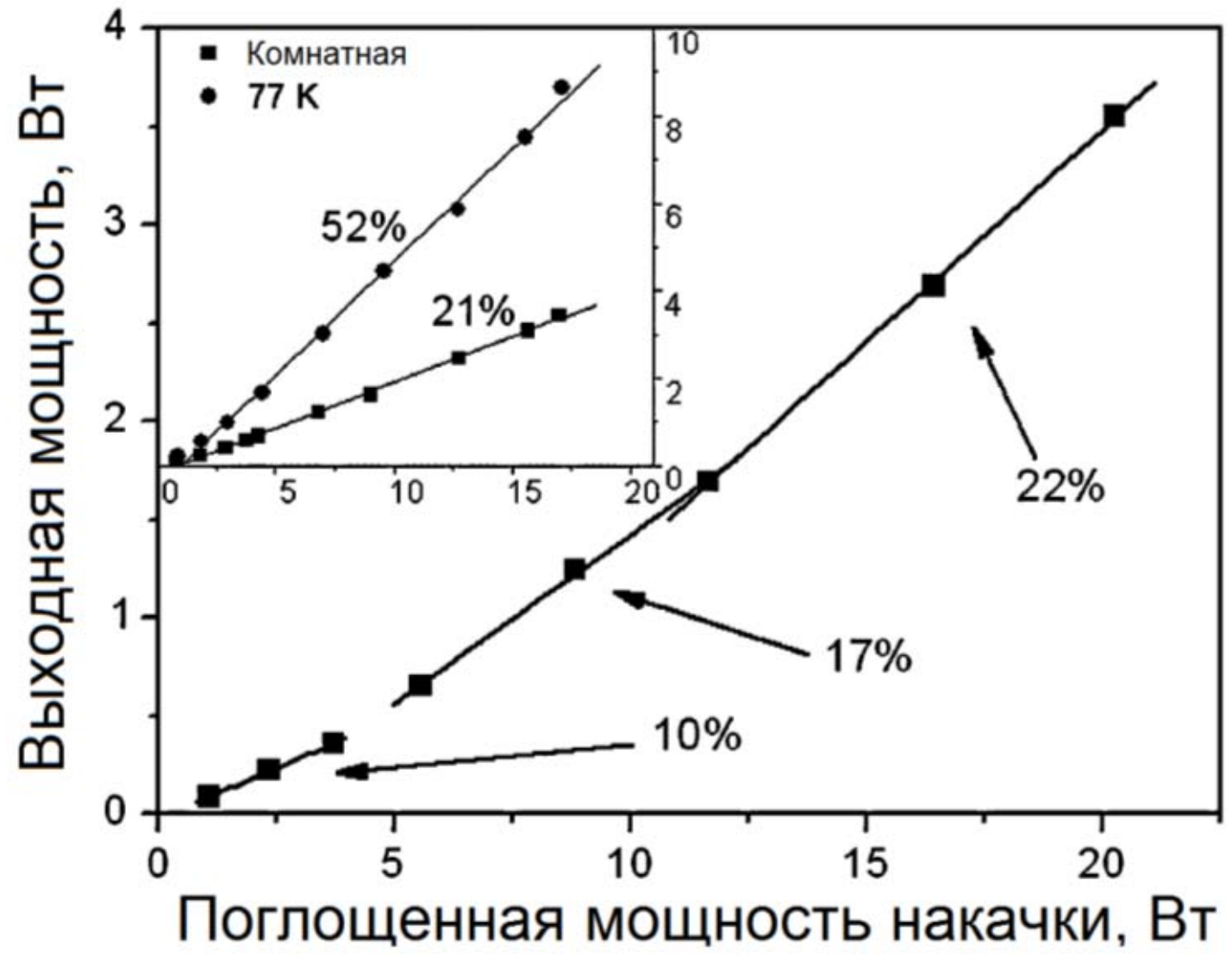

Рис. 1. 22 Эффективность висмутового лазера, генерирующего на 1160 нм, при накачке 1068 нм (и 1075 нм на вставке) для комнатной температуры и 77 К [9].

волн возбуждения (450 нм$<\lambda_{\text{возб}}<$670 нм) можно наблюдать полосу в коротковолновой области (около 750 нм). Механизм появления обеих полос до сих пор не установлен. Данный тип активных центров устойчив к термическому отжигу [103], УФ облучению [104], насыщению молекулярным водородом ($H_2$) [105]. Такие световоды можно использовать для создания лазеров непрерывного [9, 106-109] и импульсного действия [110 - 112], усилителей коротких импульсов в области 1160 – 1180 нм [113]. КПД таких лазеров достигает ≈ 20 % при комнатной температуре и возрастает до 50 % при температуре 77 К (Рис. 1. 22).

**б) Фосфоро-, германо- и фосфорогерманосиликатные световоды**

Позже было показано, что присутствие алюминия в стекле не является обязательным условием для формирования активных центров [107]. В заготовках световодов из фосфорогерманосиликатного стекла, легированного висмутом, была обнаружена широкая люминесценция, максимум которой был смещен в длинноволновую область (около 1300 нм) относительно максимума люминесценции алюмосиликатных образцов [101].

В спектре поглощения фосфорогерманосиликатного световода можно также, как и в случае алюмосиликатного световода, наблюдать достаточно широкие полосы. Типичный спектр фосфоросиликатного световода с висмутом приведен на Рис. 1. 23. Характерные полосы поглощения расположены в области 450 нм, 750-780 нм и 1200-1300 нм. При возбуждении в указанные полосы поглощения возникает ИК люминесценция в области 1280-1300 нм (Рис. 1. 23).

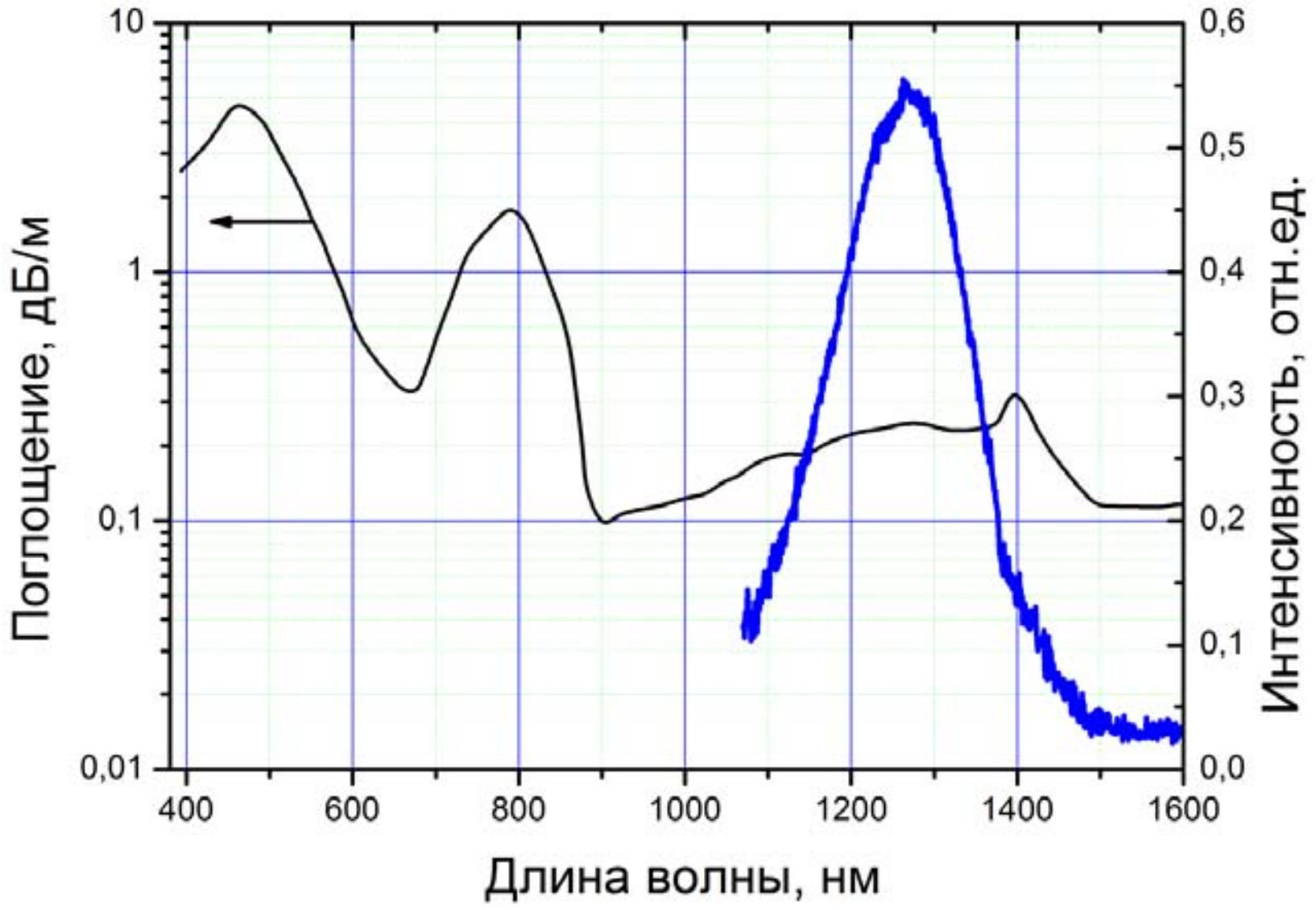

Рис. 1. 23 Типичный спектр поглощения и люминесценции фосфоросиликатного волоконного световодова, легированного висмутом [114]. Длина волны возбуждения – 1060 нм.

Световоды с сердцевиной из фосфорогерманосиликатного стекла, легированного висмутом, впервые подробно изучались в работе [115]. В ней проводилось исследование изменения оптических свойств активных центров при вариации соотношения германия и фосфора. В Табл. 1. 2 приведен состав исследованных образцов. На Рис. 1. 24 для этих образцов представлены спектры люминесценции при возбуждении на 1230 (а) и 808 нм (б). Видно, германосиликатный световод с висмутом обладает широкой люминесценцией в ИК области с максимумом на 1400 нм, тогда как фосфоросиликатный – на 1300 нм. Варьируя состав стекла сердцевины (получая световоды с различным соотношением оксида германия и оксида фосфора), авторы наблюдали плавное изменение спектра люминесценции (возрастание интенсивности полосы люминесценции на 1400 нм и снижение на 1300 нм при увеличении содержания германия и снижении фосфора). Был сделан вывод о том, что добавка фосфора способствует формированию активных центров, излучающих в области 1300 нм, а присутствие германия – центров, излучающих около 1400-1550 нм. В последнем случае не исключалась возможность влияния кремния на формирование активных центров.

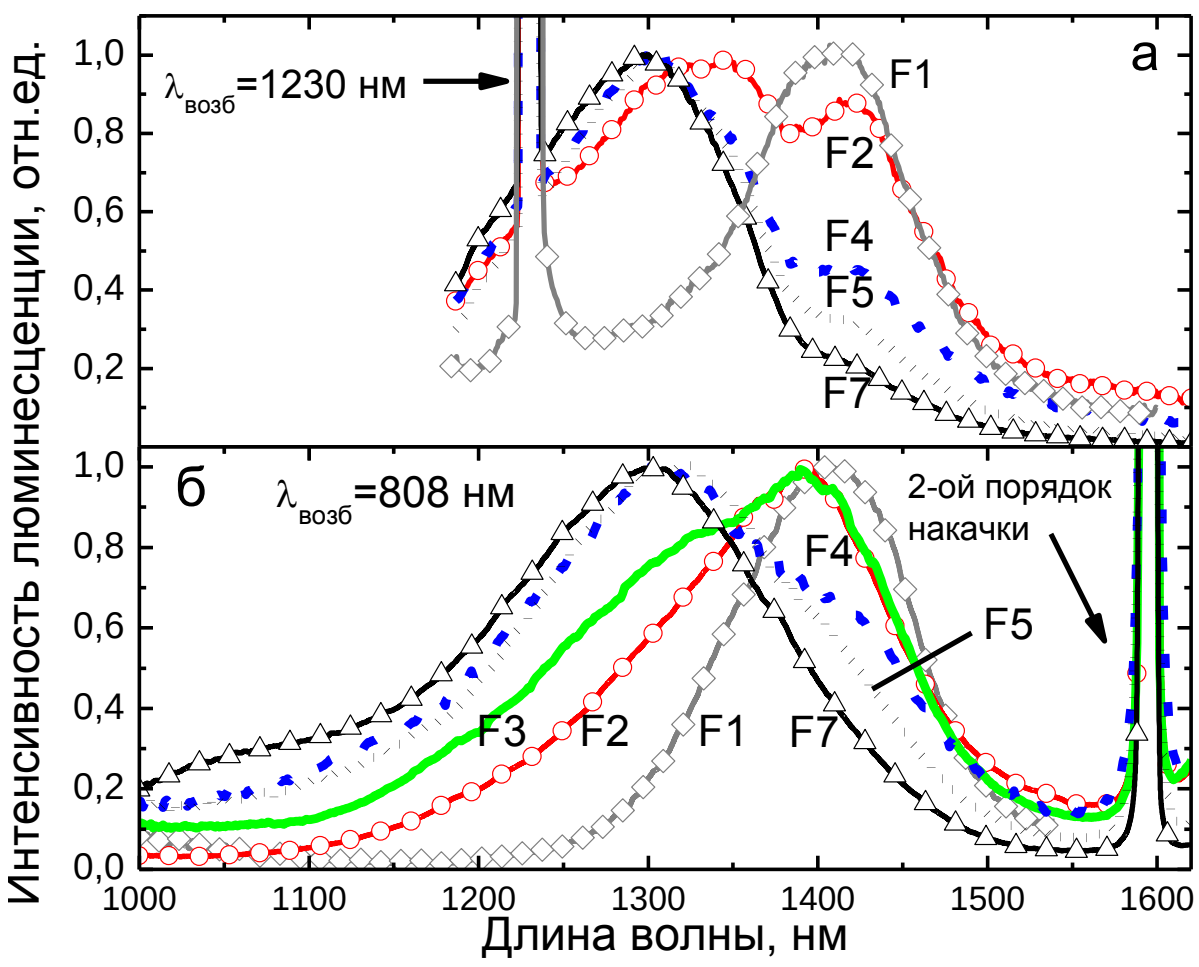

Рис. 1. 24 Спектры люминесценции фосфорогерманосиликатных световодов с висмутом при возбуждении излучением на а) 1230 нм и б) 808 нм [115].

Табл. 1. 2 Фосфорогерманосиликатные световоды, легированные висмутом [115].

| Название световода | Концентрации легирующих добавок | |
|---|---|---|
| | Ge (ат.%) | P (ат.%) |
| F1 #33 | 4.3 | - |
| F2 #28 | 4.4 | 0.5 |
| F3 #34 | 4.3 | 0.6 |
| F4 #38 | 2.9 | 1.9 |
| F5 #31 | 0.5 | 2.7 |
| F6 #32 | 0,3 | 3.9 |
| F7 #29 | - | 3.7 |

Люминесценция в области 1400 нм наблюдалась также в работе [116], в которой проводилось исследование германосиликатного световода с висмутом. Помимо люминесценции около 1400 нм, в той же работе было показано, что такой световод при возбуждении на 975 нм люминесцирует в области 1600-1650 нм. Природа происхождения данной полосы люминесценции в работе не установлена. Полученный результат независимо был подтвержден в другой работе [8].

Таким образом, люминесценция фосфоро-, фосфорогермано- и германосиликатных световодов покрывает спектральную область от 1250 до 1600 нм. Следовательно, такие световоды могут рассматриваться как кандидаты на роль активных сред для создания лазеров в спектральном диапазоне 1300 – 1550 нм. Результаты по демонстрации первой лазерной генерации в области от 1300 до 1470 нм на фосфоро-, фосфорогермано-, германосиликатных световодах с вис-мутом приведены в работах [117, 118]. Использование световода с сердцевиной из фосфорогерманосиликатного стекла, легированного висмутом, позволило продемонстрировать лазерную генерацию в диапазоне от 1300 до 1550 нм. Дальнейшие исследования показали, что наиболее подходящей стеклянной матрицей для висмута с целью получения лазерной генерации в области 1300 нм являются фосфоросиликатные световоды. Введение германия в фосфоросиликатное стекло с висмутом позволяло расширить спектральную область генерации до 1550 нм [8].

**в) Световоды из кварцевого стекла (без других легирующих добавок)**

Как выше было показано, что чем сложнее состав образца, тем труднее проводить расшифровку спектров из-за сильного перекрытия полос. С этой точки зрения интересно изучить оптические свойства наиболее простого типа световода с сердцевиной из кварцевого стекла с висмутом. Впервые чистое кварцевое стекло, легированное висмутом, был изготовлено Neff с коллегами по порошковой технологии [119]. В нем удалось измерить только спектр люминесценции при одной длине волны возбуждения (на 532 нм). Максимум люминесценции для чистого кварцевого световода с висмутом располагается на 1425 нм. Таково было состояние дел на момент выбора тематики исследований и планирования настоящей работы.

## 1.3 Выводы к Главе I

Из анализа накопленной информации по изучению висмутсодержащих материалов можно сделать следующие выводы:

1) Имеются экспериментальные доказательства о принадлежности полос синей и красной люминесценции, наблюдаемых в кристаллах, активированных висмутом: синяя люминесценция (≈400-500 нм) возникает в результате электронного перехода $^3P_1 \rightarrow {}^1S_0$ между энергетическими уровнями иона $Bi^{3+}$; появление красной люминесценции (≈590-640 нм) обусловлено переходом $^2P_{3/2}(1)\rightarrow{}^2P_{1/2}$, связанным с ионом $Bi^{2+}$, наличие которого подтверждается результатами ЭПР измерений.

2) Изменением химического состава стекла можно получить ВАЦ с люминесценцией в ИК диапазоне от 800 до 2100 нм. Следует подчеркнуть, что оптические свойства ВАЦ зависят от выбора стеклянной матрицы, а также от легирующих примесей (Al, P, Ge и др.), вводимых совместно с висмутом в стекло.

3) В опубликованных работах отсутствует подробное исследование спектрально-люминесцентных характеристик большинства стекол и световодов (за исключением алюмосиликатных световодов). Как правило, изучение оптических свойств материалов с висмутом ограничивалось измерением спектров люминесценции на дискретном наборе длин волн возбуждения (см. Рис. 1. 21 а). Очевидно, что при таком подходе интерпретация экспериментальных данных была сильно затруднена, что не позволяло построить модель ВАЦ.

4) Остается открытым вопрос о природе и схеме энергетических уровней ВАЦ, ответственных за усиление оптических сигналов и лазерную генерацию. Получение новой информации об активных центрах необходимо для продвижения в понимании их физической природы.

# ГЛАВА II. ЭКСПЕРИМЕНТАЛЬНЫЕ ОБРАЗЦЫ, ТЕХНОЛОГИЯ ИЗГОТОВЛЕНИЯ И МЕТОДИКИ ИССЛЕДОВАНИЯ

Данная глава посвящена описанию экспериментальных образцов, технологии их изготовления, основных схем проведения экспериментов, в частности, измерения спектров поглощения, трехмерных спектров возбуждения-эмиссии люминесценции висмутовых активных центров, времен затухания люминесценции и спектров антистоксовой люминесценции при ступенчатом возбуждении.

## 2.1 Экспериментальные образцы. Технология изготовления.

В настоящей диссертационной работе в качестве экспериментальных образцов были выбраны преформы и вытянутые из них волоконные световоды, легированные висмутом. Данный выбор был сделан по следующим причинам:

1) наличие в них ВАЦ, способных усиливать световые сигналы. Именно это позволит с полной уверенностью говорить о том, что исследованы оптические свойства именно активных центров;

2) простота проведения измерений при различных температурах в широкой спектральной области (от УФ до ближнего ИК);

3) технологии волоконных световодов позволяют получать световоды с простым химическим составом стекла сердцевины, что должно упростить интерпретацию получаемых данных.

Обозначение, состав стекла образцов, исследованных в данной диссертационной работе, представлены в Табл. 2. 1.

Помимо, висмутовых световодов с сердцевиной из v-$SiO_2$ (SBi образец) и v-$GeO_2$ (GBi образец) проводились исследования световодов с сердцевиной из v-$SiO_2$ с добавкой $Al_2O_3$, $P_2O_5$ или $GeO_2$.

Табл. 2. 1. Обозначения, состав сердцевины и метод изготовления волоконных световодов

| **Обозначение** | **Состав стекла сердцевины** | **Метод изготовления** |
|---|---|---|
| SBi | $100SiO_2$+Bi | Powder-in tube [120] |
| GBi | $100GeO_2$+Bi | MCVD |
| GSBi | $5GeO_2+95SiO_2$+Bi | MCVD |
| ASBi | $3Al_2O_3+97SiO_2$+Bi | MCVD |
| PSBi | $10P_2O_5+90SiO_2$+Bi | MCVD |

Концентрация висмута в сердцевинах световодов не превышала порога чувствительности нашей измерительной аппаратуры по этому металлу (0.02 ат.%). При отсутствии данных по прямым измерениям концентрации величину легирования сердцевины висмутом можно было качественно оценить по уровню оптического поглощения излучения в максимумах полос соответствующих ВАЦ в ближней ИК области (~ 1 дБ/м). Образцы с низкими концентрациями висмута были выбраны по причине того, что они обладают наилучшими генерационными характеристиками. Использование световодов с высоким содержанием висмута приводит к заметному снижению кпд таких лазеров вплоть до исчезновения генерации.

Все световоды имели внешний диаметр 125 мкм. Исследовались как одномодовые (на длине волны 1,2 мкм), так и многомодовые световоды. В некоторых измерениях (поглощение и люминесценция в УФ области) использовались отрезки заготовок волоконных световодов. В условиях проведенных измерений, различий между ними не обнаружено.

*Технология изготовления экспериментальных образцов*

В Табл. 2. 1 для каждого образца указана технология изготовления. Изготовление образца осуществляется, как правило, в два этапа: создание заготовки и последующая вытяжка световода. Первый этап является самым важным, поскольку происходит формирование состава стеклянной матрицы с висмутом. Из Табл. 2. 1 видно, что GBi, GSBi, ASBi, PSBi световоды изготавливались с помощью одной технологии MCVD (Modified chemical vapor deposition) [121], а SBi – по технологии Powder-in-Tube [119, 122], между которыми существуют значительные различия. Технология MCVD основана на послойном осаждении на опорную кварцевую трубу стеклообразующих компонентов ($SiO_2$, $GeO_2$, $Al_2O_3$, $P_2O_5$ и др.). Все легирующие добавки, включая висмут, вводятся из газовой фазы. Данный процесс происходит в две стадии: 1) на внутреннюю поверхность опорной трубы (Heraeus F300) сначала наносится дополнительный слой высокочистого кварцевого стекла, легированного фтором и фосфором (с концентрацией оксида фосфора $C_{P2O5} \leq 1$ мол.%, концентрация фтора выбирается из условия компенсации возрастания показателя преломления кварцевого стекла за счет легирования фосфором) для снижения потерь излучения; 2) после этого осаждаются стеклообразующие и легирующие оксиды, которые участвуют в формировании сердцевины световода. Температура, при которой происходят эти процессы, равна ~ 2000˚С. После осаждения и проплавления всех слоев температура кислородно-водородной горелки еще повышается, и происходит схлопывание трубы под действием сил поверхностного натяжения. Получившаяся заготовка медленно (более 1 часа) остывает до комнатной температуры.

Основное отличие технологии powder-in-tube, с помощью которой был изготовлен световод SBi, от технологии MCVD заключается в том, что там не происходит послойного осаждения стеклообразующих компонентов, формирующих сердцевину. В технологии powder-in-tube сердцевина формируется при спекании и проплавлении порошков (оксида висмута и оксида кремния) при высокой температуре (около 1900 ˚С) в кварцевой трубе под вакуумом.

Введение оксидов алюминия, фосфора или германия (ASBi, PSBi, GSBi или GBi) повышает показатель преломления сердцевины относительно кварцевой оболочки, что позволяет добиваться распространения света по сердцевине за счет полного внутреннего отражения. Образец SBi был изготовлен из чистого кварцевого стекла без каких-либо добавок (кроме висмута), поэтому показатель преломления сердцевины такого световода был близок к показателю пре-ломления чистого кварцевого стекла. Чтобы получить световедущую структуру, а впоследствии и волоконный световод, в качестве опорной использовалась не труба из чистого кварцевого стекла типа Heraeus F300, а труба из кварцевого стекла, легированного фтором, с пониженным показателем преломления. Для получения одномодового световода на заготовку нахлопывалась дополнительно кварцевая труба из чистого $SiO_2$ необходимой толщины. В результате профиль показателя SBi световода выглядел следующим образом (см. Рис. 2. 1).

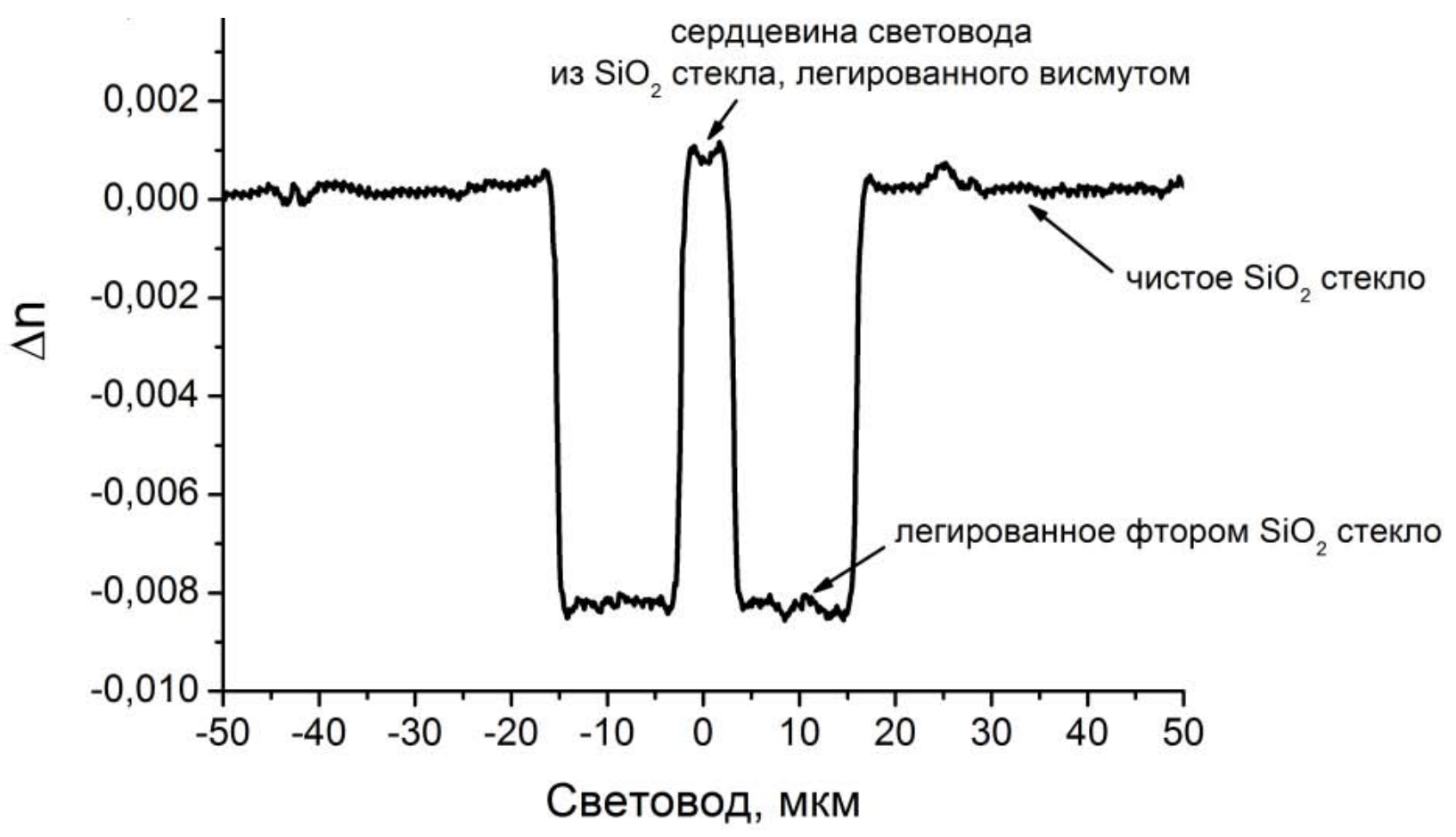

Рис. 2. 1 Профиль показателя преломления SBi световода.

Последним этапом создания исследуемых образцов является вытяжка световодов из заготовок. Данный процесс одинаков для всех световодов из Табл. 2. 1, который осуществляется при температуре, близкой к 2000 ºС. Следует отметить, что в процессе вытяжки волоконного световода его температура снижается от ~ 2000 ºС до температуры ниже температуры стеклования кварцевого стекла (1200 ºС) очень быстро, за время ~ 1 мс, что существенно отличается от режима охлаждения объемных образцов стекла. Кроме того, процесс вытяжки включает в себя нанесение на световод полимерного защитного покрытия.

## 2.2 Схема измерения спектров поглощения

Спектры поглощения экспериментальных образцов (световодов) неоднократно приводились в ряде работ (например [123]). Для получения спектров поглощения “слабого” сигнала, как и в других случаях, использовалась стан-дартная методика “облома”, заключающаяся в сравнении спектров пропускания света через длинный и короткий (после «обламывания» длинного световода) отрезки световодов при неизменном вводе сигнала. В качестве источника сигнала использовалась галогенная лампа для видимой и ИК области, дейтериевая лампа – для УФ области (DH 2000 Mikropack). Изменением населенности ВАЦ от излучения ламп можно пренебречь. Использование “слабого” сигнала не приводило к искажению спектров поглощения. Спектр пропускания регистрировался с помощью оптических спектроанализаторов: Ocean Optics 2000, HP 70950B для разных спектральных областей. Для расчета величины оптических потерь использовалась следующая формула:

$$\alpha(\lambda) = \frac{10}{L_{длин} - L_{кор}} \cdot \lg \frac{I_{кор}(\lambda)}{I_{длин}(\lambda)} \quad [\text{дБ/м}] \qquad (2.\ 1)$$

где $L_{кор}, L_{длин}$ – длины короткого и длинного отрезка световода, соответственно; $I_{кор}, I_{длин}$ – интенсивности света после прохождения через короткий и длинный отрезки световода, соответственно, $\lambda$ – длина волны.

Измерения спектров поглощения в УФ области (190-350 нм) проводились на полированных отрезках заготовок волоконных световодов различной толщины (от 0.2 до 3 мм и ограничивалась сверху их оптической неоднородностью), выбор которой зависел от уровня поглощения. Измерения проводились на спектрофотометре Shimadzu UV-3101PC. Спектральное разрешение прибора составляло 1 нм.

## 2.3 Методика измерения и построения трехмерных спектров возбуждения-эмиссии люминесценции

Особенностью стекол, легированных висмутом, является существенная зависимость их люминесценции, в частности спектрального положения максимума, ширины спектра, от длины волны возбуждающего излучения (Рис. 1. 21 а). Поэтому для решения поставленных задач в рамках данной работы необходимым являлось проведение измерений спектров люминесценции в широком спектральном диапазоне при сканировании длины волны возбуждающего излучения.

Измерения спектров люминесценции SBi, GBi, GSBi, PSBi, ASBi образцов проводились в нескольких спектральных областях:

- Область I - ($450\ \text{нм} < \lambda_{ex} < 1600\ \text{нм}$; $875\ \text{нм} < \lambda_{em} < 1700\ \text{нм}$).
- Область II - ($450\ \text{нм} < \lambda_{ex} < 1600\ \text{нм}$; $400\ \text{нм} < \lambda_{em} < 875\ \text{нм}$).
- Область III – ($220\ \text{нм} < \lambda_{ex} < 500\ \text{нм}$; $240\ \text{нм} < \lambda_{em} < 1600\ \text{нм}$).

В Областях I и II измерения проводились на волоконных световодах. Схема регистрации люминесценции представлена на Рис. 2. 2. Данная схема состоит из волоконного широкополосного спектрально-неселективного Х-образного делителя мощности (50/50). С помощью акустооптического фильтра

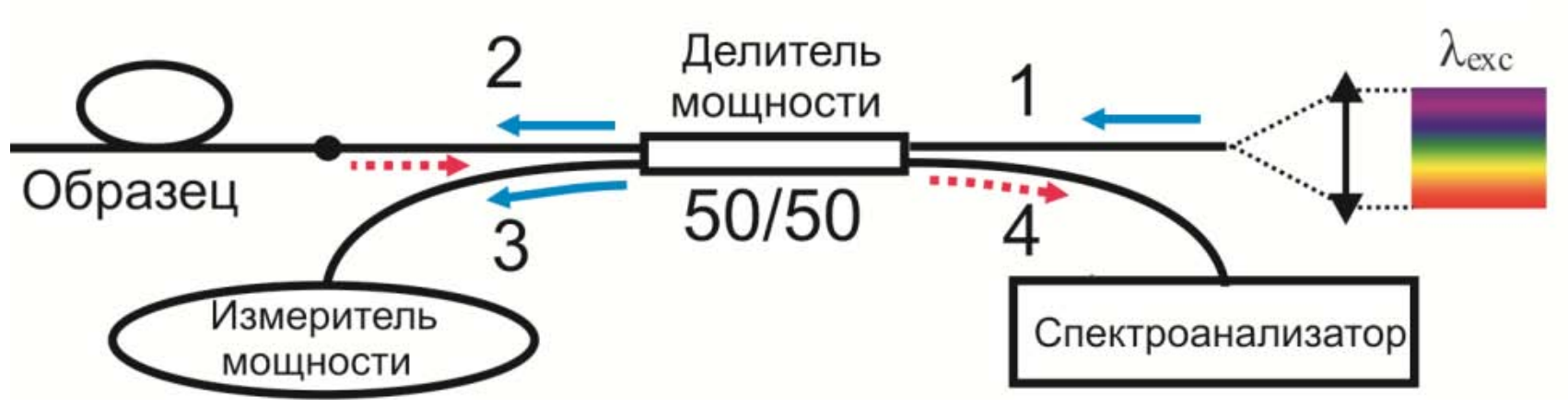

Рис. 2. 2 Экспериментальная схема регистрации люминесценции. Сплошными стрелками указано направление распространения возбуждающего излучения, пунктирными – излучения люминесценции. Точкой обозначено сварное соединение исследуемого световода с подводящим возбуждающее излучение световодом.

из непрерывного спектра излучения выделялась полоса шириной 3 нм, которая вводилась в сердцевину исследуемого световода. Далее это излучение поровну делилось между каналами 2 и 3. Через канал 2 излучение накачки проходило к исследуемому висмутовому световоду. Канал 3 волоконного делителя мощности использовался для мониторинга мощности возбуждающего излучения (не превышала 1 мВт во всем диапазоне возбуждения). Излучение люминесценции образца распространялось по каналу 2 в обратном направлении относительно направления излучения накачки. Регистрация осуществлялась с помощью Ocean Optics 2000 и HP 70950B спектроанализаторов с разрешением 10 нм через выход 4. Полученные спектры люминесценции были исправлены на спек-тральную чувствительность канала регистрации и нормировались на введенную в световод мощность излучения накачки. Измерения в областях I и II проводились при комнатной температуре и при температуре кипения жидкого азота.

В Области III использовать световоды было невозможно из-за отсутствия соответствующих источников накачки, поэтому измерения проводились на отрезках заготовок световодов с помощью спектрофлюориметров LS55 фирмы Perkin Elmer и FLSP920 фирмы Edinburgh Instruments. Во всех измерениях величина спектрального разрешения спектрофлюориметров как по каналу возбуждения, так и по каналу регистрации люминесценции составляла 10 нм (большая величина щелей монохроматоров была выбрана для повышения чувствительности схемы). Источником возбуждения были ксеноновые лампы с суммарной мощностью 200 Вт. Все полученные спектры люминесценции были исправлены на калибровочную кривую приборов. Для получения единой картины полученные спектры люминесценции сшивались.

В результате для каждого образца были измерены спектры люминесценции от 240 до 1600 нм при перестройке длины волны возбуждения в диапазоне 220-1600 нм с шагом 10 нм. Для удобства представления такого объема экспериментальных данных и проведения их анализа (интерпретации) формировались трехмерные спектры возбуждения-эмиссии люминесценции. В настоящее время такое представление часто используется при исследовании органических веществ. Обработка и построение спектров возбуждения-эмиссии люминесценции проводилась в аналитическом пакете Origin 7.0. Полученные трехмерные спектры возбуждения-эмиссии люминесценции представляют собой сочетание спектров люминесценции (вдоль линии $\lambda_{ex}$ = const) и спектров возбуждения люминесценции (вдоль линии $\lambda_{em}$ = const). В этом случае положение каждого пика люминесценции, наблюдаемого на такой диаграмме, характеризуется двумя параметрами: $\lambda_{ex}$ – длина волны возбуждения и $\lambda_{em}$ – длина волны люминесценции.

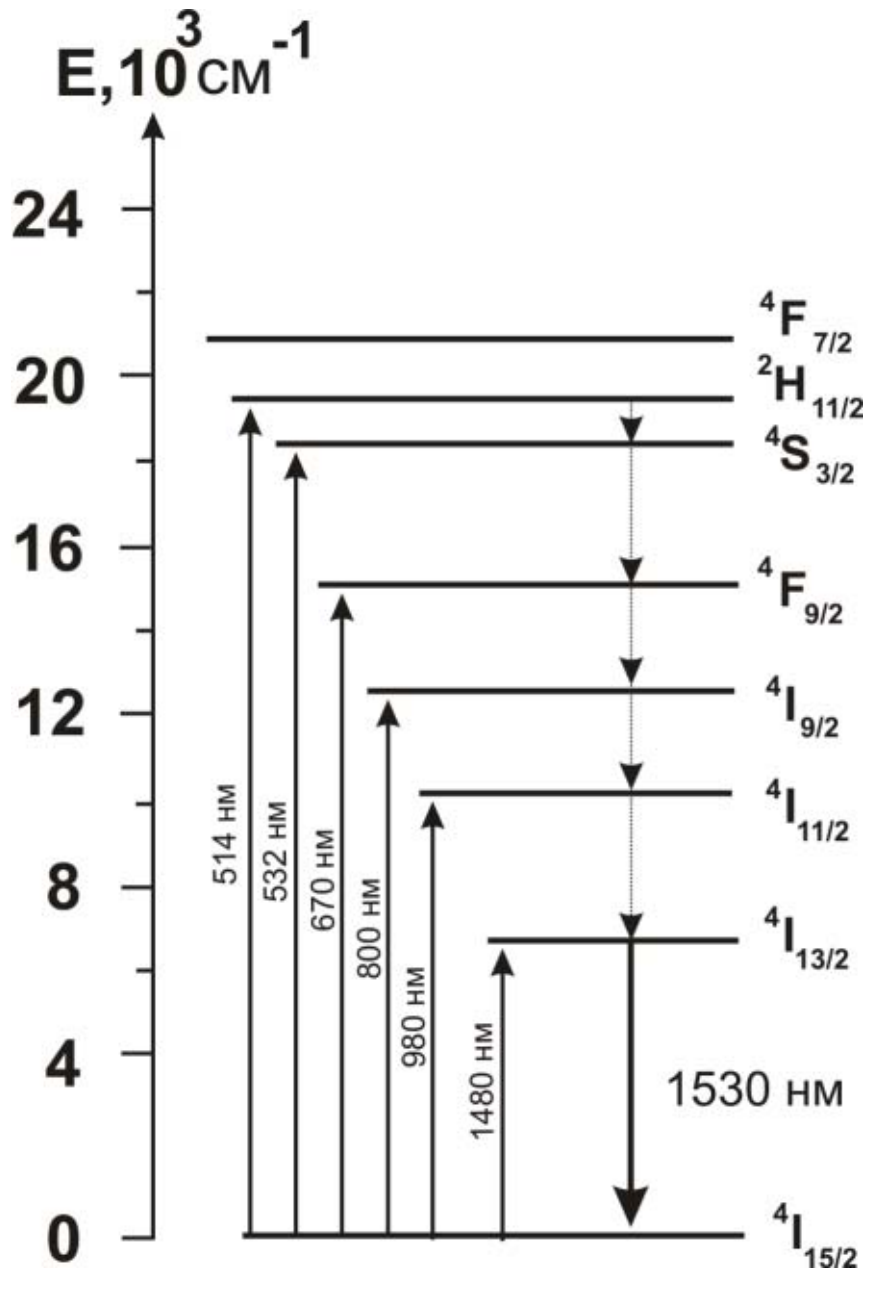

а

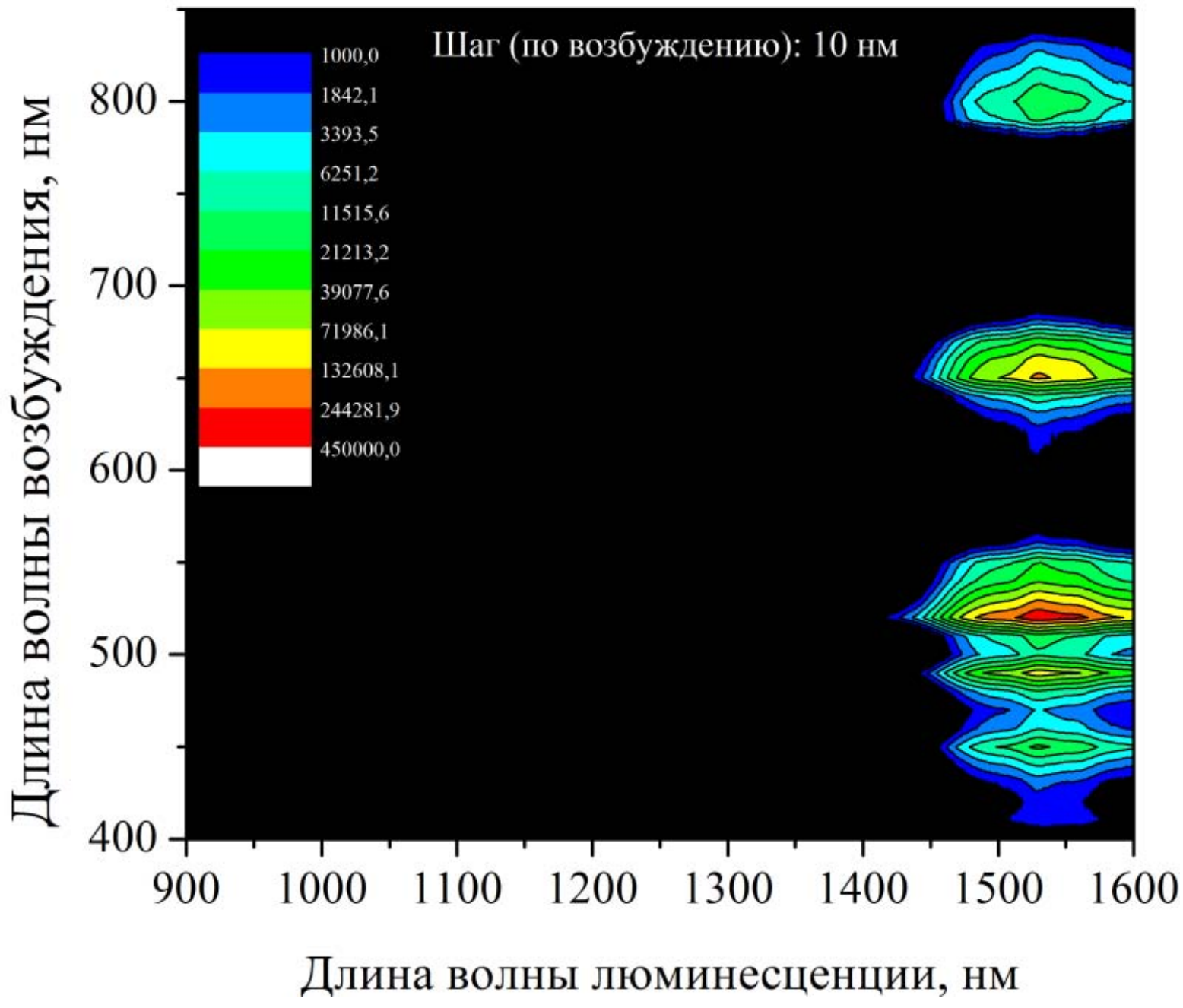

б

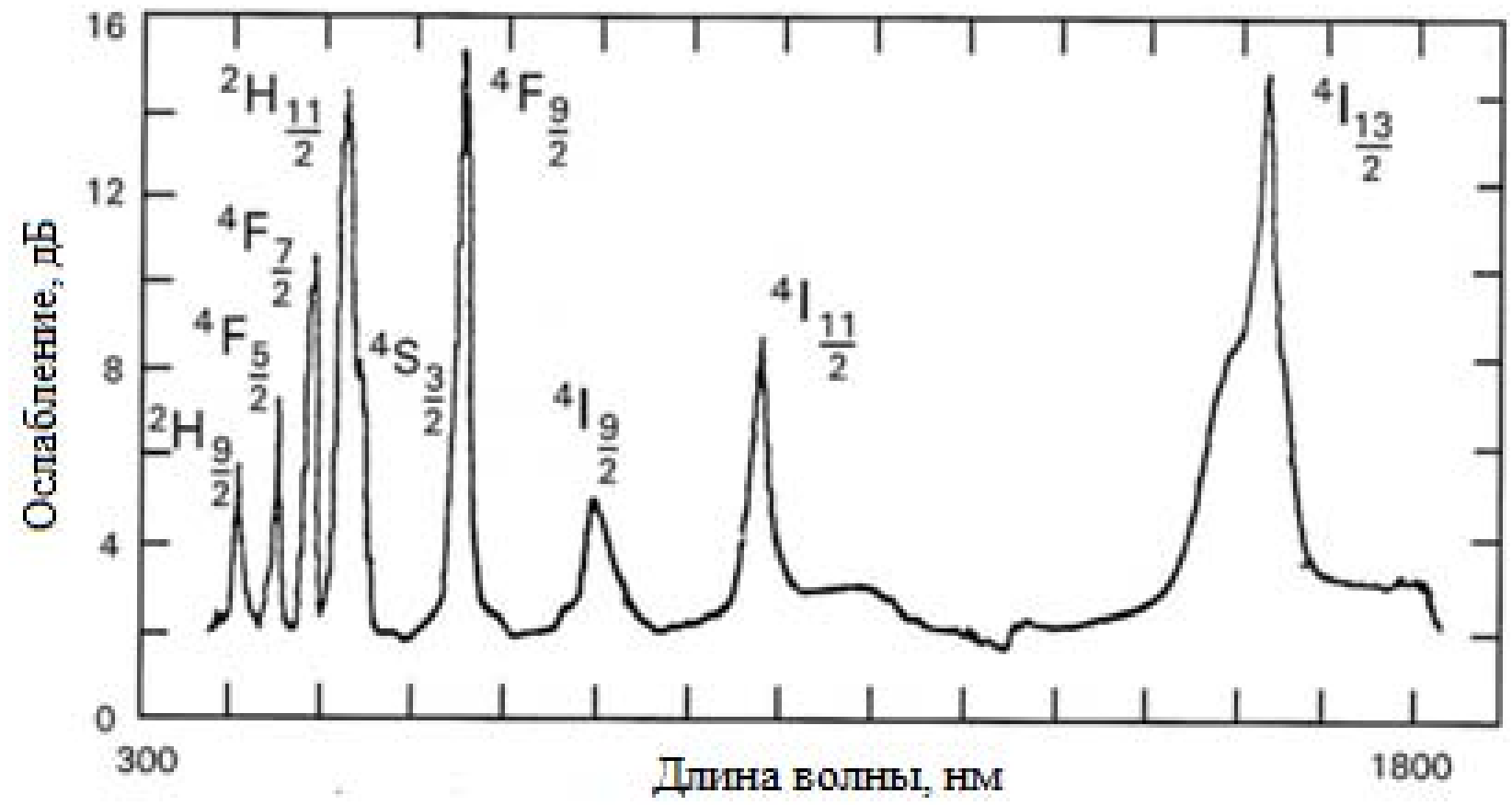

в

Рис. 2. 3 а) Схема энергетических уровней иона $Er^{3+}$ (линиями со стрелками вверх указываются основные полосы возбуждения; сплошной линией со стрелкой вниз – основной излучательный переход эрбия; пунктирными линиями со стрелками вниз – безызлучательные переходы)[3]; б) трехмерный спектр возбуждения-эмиссии люминесценции для стекла, легированного ионами $Er^{3+}$, полученный на нашей установке; в) типичный спектр поглощения $Er^{3+}$ [124].

В качестве "пробной" задачи нами был получен трехмерный спектр возбуждения-эмиссии люминесценции стекла, легированного эрбием, который представлен на Рис. 2. 3 б. Схема энергетических уровней иона $Er^{3+}$ в стекле хорошо известна (см. Рис. 2. 3 а). Для того, чтобы определить их положение, в принципе достаточно измерить, например, спектр поглощения материала (Рис. 2. 3 в), или похожий на него спектр возбуждения ИК люминесценции на переходе $^4I_{13/2} \rightarrow {}^4I_{15/2}$. Полученный в результате наших измерений трехмерный спектр возбуждения-эмиссии люминесценции стекла с $Er^{3+}$ (Рис. 2. 3 б) показывает положение максимумов люминесценции на одном и том же переходе $^4I_{13/2} \rightarrow {}^4I_{15/2}$ при возбуждении на различные энергетические уровни $^4I_{9/2}$, $^4F_{9/2}$, $^4S_{3/2}$, $^2H_{11/2}$, $^4F_{7/2}$.

Характерная ИК люминесценция эрбия в области 1530 нм появлялась при возбуждении на длинах волн: ≈450, ≈490, ≈530, ≈670 и ≈800 нм, попадающих в полосы поглощения ионов эрбия (Рис. 2. 3 в). Положение максимумов люминесценции по длинам волн возбуждения сразу указывает на положение известных энергетических уровней иона $Er^{3+}$ (Рис. 2. 3 а). Таким образом, используя трехмерные спектры люминесценции, мы имеем наглядное представление о наличии различных полос люминесценции и совокупности полос возбуждения люминесценции на необходимой длине волны. Построение спектров значительно упрощает процедуру анализа спектрально-люминесцентных свойств активной среды.

## 2.4 Временные зависимости затухания люминесценции: схема измерений

При измерении времен жизни люминесценции в качестве источника возбуждения использовался параметрический генератор излучения, работающий в области 410 – 2500 нм. Длительность импульса накачки составляла 10 нс, частота повторения импульсов – 50 Гц, энергия импульса – более 1 мДж. Измерения временных зависимостей затухания ИК люминесценции проводились с помощью InGaAs фотодетектора со временем отклика 10 нс. Для устранения рассеянного излучения накачки использовались фильтры, представляющие собой пластины кристаллического кремния различной толщины. В видимом спектральном диапазоне в качестве приемника использовался фотоэлектронный умножитель (со временем отклика 50 нс), который устанавливался после монохроматора МДР-2 (решетка 1200 шт/мм). Использование монохроматора позво-ляло проводить сканирование по длине волны люминесценции. Все измерения были выполнены при комнатной температуре.

## 2.5 Измерение спектров антистоксовой люминесценции при ступенчатом двухквантовом возбуждении

Способность акустооптического фильтра вырезать из непрерывного спектра излучения одновременно 2 длины волны позволила реализовать схему ступенчатого возбуждения. Для регистрации люминесценции при ступенчатой двухквантовой накачке использовалась схема подобная той, что была представлена на Рис. 2. 2.

Отличие от приведенной схемы в том, что возбуждение образца осуществлялось одновременно на двух длинах волн – $\lambda_{ex1}$ и $\lambda_{ex2}$, причем длина волны одного излучения ($\lambda_{ex1}$) оставалась неизменной, а другого ($\lambda_{ex2}$) – сканировалась с шагом 10 нм. Диапазон изменения длин волны возбуждения $\lambda_{ex2}$ составлял 1100-2000 нм; диапазон регистрации люминесценции – 400-1000 нм. Средняя мощность излучения на каждой длине волны возбуждения составляла 200-700 мкВт. Суммарная мощность возбуждающего излучения при использовании источника суперконтинуума не превосходила 1.5 мВт. В некоторых экспериментах в качестве источников излучения также использовались диодные и волоконные лазеры.

Для возбуждения антистоксовой люминесценции в наших экспериментах использовалось явление поглощения из возбужденного состояния, схематично представленное на Рис. 2. 4. Кванты возбуждающего излучения $h\nu_1$ и $h\nu_2$ вызывают последовательно переходы $A_{12}$ и $A_{23}$, после чего становится возможным наблюдение антистоксовой люминесценции за счет излучательного перехода $L_{31}$. Длина волны одного возбуждающего излучения $\lambda_{ex1}$ устанавливалась равной разности основного и первого возбужденного уровней ВАЦ. При этом схема регистрации излучения люминесценции настраивалась на длину волны перехода со второго возбужденного уровня в основное состояние. Таким обра-зом, снимались спектры возбуждения люминесценции со второго возбужденного состояния по отношению к длине волны $\lambda_{ex2}$.

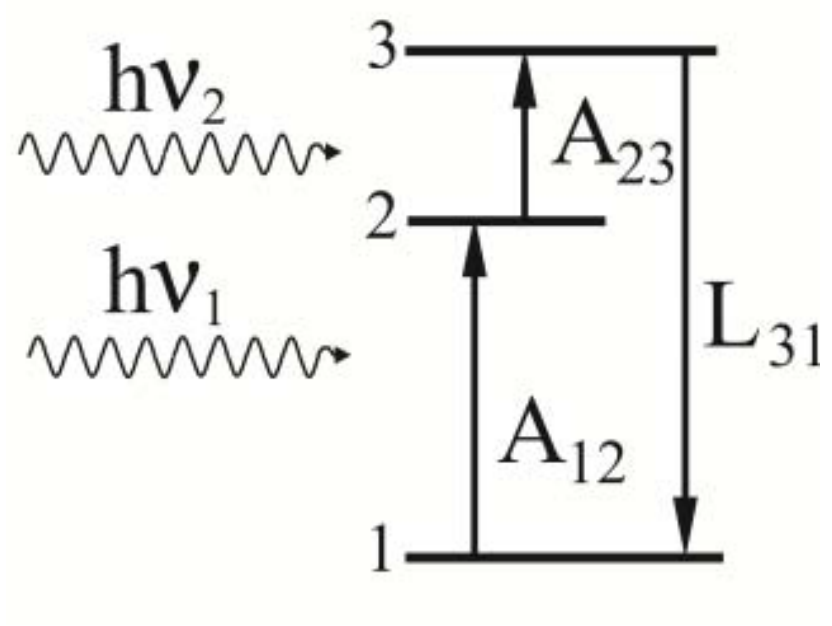

Рис. 2. 4 Схематическое изображение процесса последовательного двухквантового возбуждения. 1, 2 и 3 –любые уровни энергии ВАЦ.

При обработке полученных результатов проводилась нормировка спектров люминесценции на функцию пропускания каналов волоконного делителя и спектральную чувствительность спектрометра. Все результаты, кроме тех, где это оговорено особо, получены при комнатной температуре.

# ГЛАВА III. ОПТИЧЕСКИЕ СВОЙСТВА СВЕТОВОДОВ С СЕРДЦЕВИНОЙ ИЗ v-$SIO_2$, ЛЕГИРОВАННОГО ВИСМУТОМ [125, 126, 127, 128, 129, 130].

В данной главе будут приведены экспериментальные результаты, касающиеся исследования оптических свойств световодов из v-$SiO_2$, легированного висмутом, будет приведена структура энергетических уровней ВАЦ в $SiO_2$ световоде. Также будет показано, что на основном переходе ВАЦ в $SiO_2$ световоде возможно получить лазерную генерацию. Кроме того, в данной главе будет рассмотрено влияние легирования оксидов алюминия и фосфора на оптические свойства ВАЦ.

## 3.1 Абсорбционные свойства световода из кварцевого стекла, легированного висмутом, в видимой и ближней ИК области спектра

Спектр поглощения световода из кварцевого стекла, легированного висмутом, приведен на Рис. 3. 1. Для такого типа световода характерным является наличие в спектрах поглощения пьедестала, на котором расположены полосы поглощения ВАЦ. Видно, что уровень поглощения в пьедестале возрастает в коротковолновую сторону. Подобная же картина наблюдается при введении добавок оксида алюминия или фосфора.

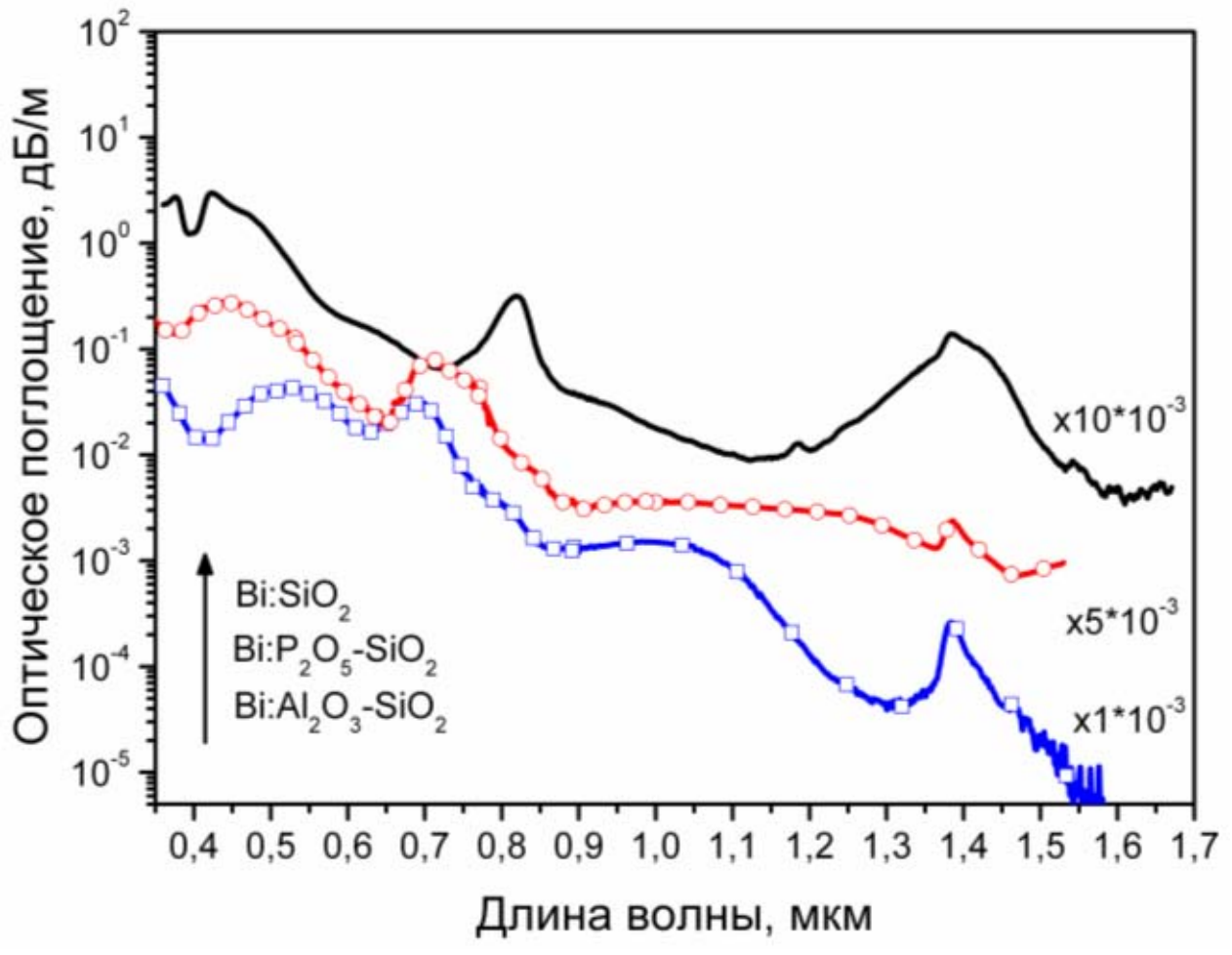

Рис. 3. 1 Спектры оптического поглощения SBi, PSBi и ASBi световодов. Во избежание сильного взаимного перекрытия спектров поглощения и наглядности соответствующие значения поглощения были умножены на коэффициенты, приведенные справа от каждой линии.

Уровень поглощения в пьедестале зависит от концентрации висмута. В случае алюмосиликатных световодов увеличение концентрации висмута приводит к интенсивному росту поглощения в пьедестале по сравнению с полосами поглощения ВАЦ, что отрицательно сказывается на КПД лазеров [131].

Для ВАЦ формирующихся в SBi световоде типичными являются полосы на 1410 нм, 820 нм, плечи на 600 и 480 нм, а также достаточно узкие пики на 420 и 380 нм, между которыми наблюдается резкий провал [120]. В области 1385 нм наблюдается характерный пик поглощения гидроксильных групп, содержащихся в сетке стекла. Подобный спектр поглощения наблюдался в световоде аналогичного состава, изготовленного и по другой технологии (FCVD) [132]. При введении оксида алюминия или фосфора в сердцевину световода из кварцевого стекла с висмутом происходит изменение спектров поглощения. В этом случае полосы поглощения в спектрах ASBi и PSBi световодов становятся гораздо шире, чем в случае SBi световода. Характерные полосы для алюмосиликатных световодов располагаются на 500, 700, 800 и 1000 нм; для фосфоросиликатных – 450, 730, плечо на 820 нм и комплексная полоса в области 1000-1400 нм. В спектре поглощения ASBi световода в около 1400 нм, помимо полосы OH-групп, также наблюдается полоса ВАЦ. Появление этой полосы, также как и полосы на 800 нм, обусловлено наличием кремния в стекле.

## 3.2 Спектрально-люминесцентные свойства ВАЦ в v-$SiO_2$ при T=77 K и T=300 K

Подробное изучение ВАЦ в столь широкой спектральной области (220-1600 нм) проводилось впервые. Экспериментальные результаты люминесцентных свойств ВАЦ были представлены в виде трехмерных спектров. Метод измерения и построения таких графиков приведен в Главе 2. Данные спектры показывают распределение интенсивности люминесценции в зависимости от длины волны возбуждения и длины волны люминесценции ($I_{lum}(\lambda_{ex}, \lambda_{em})$) для различных образцов.

Трехмерные спектры возбуждения-эмиссии люминесценции для ВАЦ в v-$SiO_2$ стекле при комнатной температуре и температуре жидкого азота представлены на Рис. 3. 2. При анализе таких спектров следует обратить внимание на следующие особенности. Во-первых, на приведенных зависимостях отношение максимального и минимального значений интенсивностей люминесценции, отображенных на каждом графике, выбрано ~100. Это ограничивает объем информации, отображаемой на каждом спектре, а именно это касается областей с низкими значениями интенсивности люминесценции. Но без такого ограничения спектры люминесценции получаются практически не читаемыми. Поэтому следует учесть, что на приведенных 3-мерных спектрах отображены только основные пики наблюдаемой люминесценции. Во-вторых, на всех приведенных графиках $I_{lum}$ ($\lambda_{ex}$, $\lambda_{em}$) наблюдается линия, проходящая по диагонали $\lambda_{ex} = \lambda_{em}$ через весь график, которая соответствует рассеянному излучению возбуждения. На спектрах присутствует также излучение второго порядка дифракции рассеянного излучения накачки в виде участка линии $\lambda_{em}=2\lambda_{ex}$. В-третьих, как уже указывалось выше, положение каждого пика характеризуется двумя параметрами: $\lambda_{ex}$, $\lambda_{em}$. В дальнейшем для обозначения максимума люминесценции будут указываться его буквенное обозначение в таблице (или на рисунке) и соответствующие ему длины волн возбуждения и эмиссии, например А ($\lambda_{ex}^{max}$, $\lambda_{em}^{max}$).

Используемые обозначения наблюдаемых на Рис. 3. 2 пиков люминесценции $I_{lum}(\lambda_{ex}, \lambda_{em})$, и соответствующие максимумам этих пиков длины волн возбуждения и эмиссии для SBi образца представлены в Табл. 3. 1.

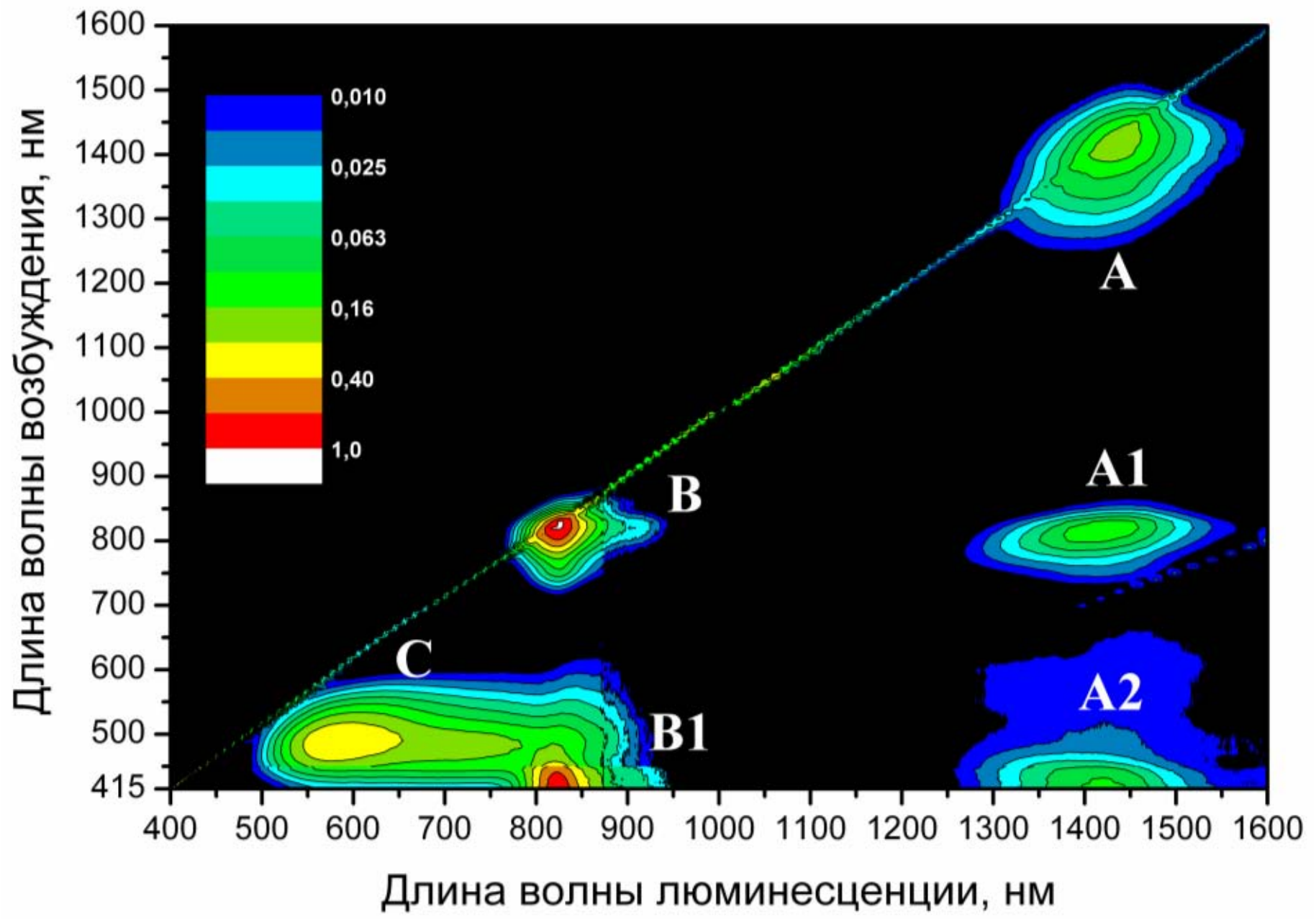

а

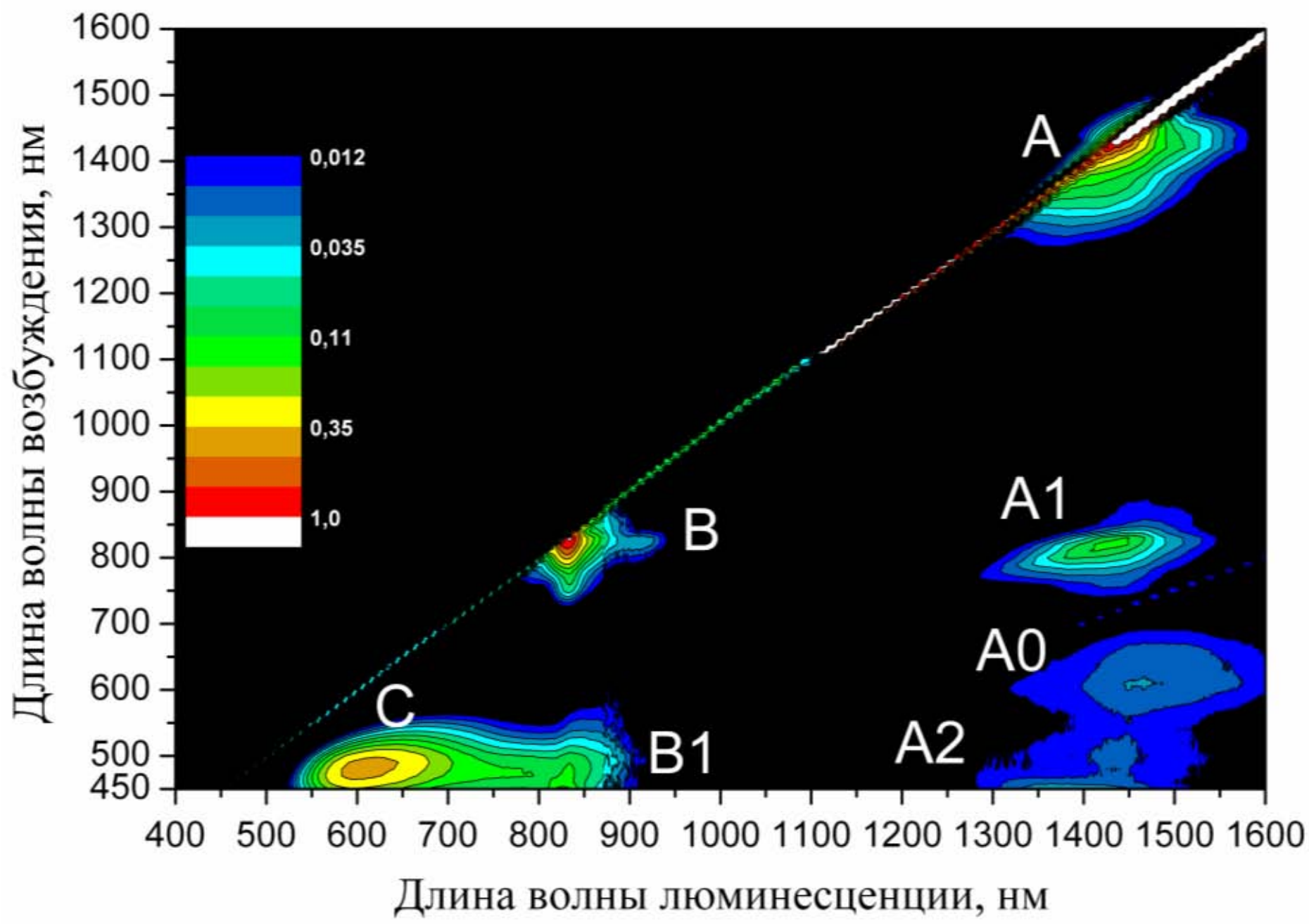

б

Рис. 3. 2 Трехмерные спектры возбуждения-эмиссии люминесценции SBi световода при комнатной температуре (а) и при температуре жидкого азота (б). Цветом указана интенсивность люминесценции в относительных единицах.

Табл. 3. 1. Основные максимумы люминесценции SBi световода: обозначения, длины волн возбуждения и эмиссии при T=300 K и 77 K

| Обозначение пика | $\lambda_{ex}^{max}$, нм | | $\lambda_{em}^{max}$, нм | |
|---|---|---|---|---|
| | 300 K | 77 K | 300 K | 77 K |
| A | 1415 | 1425 | 1430 | 1435 |
| A1 | 823 | 821 | 1430 | 1430 |
| A2 | 420 | <450 | 1430 | ≤1430 |
| A0 | — | 620 | — | 1480 |
| B | 823 | 823 | 827 | 833 |
| B1 | 420 | <450 | 827 | 830 |
| C | 470 | 480 | 603 | 605 |

На спектре возбуждения-эмиссии люминесценции, принадлежащем SBi образцу, наблюдаются 6 основных максимумов люминесценции в видимой и ИК-области спектра при T = 300 K: A, A1, A2, B1, B2 и C(Рис. 3. 2 а). Пики A(1415 нм, 1430 нм) и B(823 нм, 827 нм) отличаются малым стоксовым сдвигом, который значительно меньше их ширины. Они практически лежат на диагонали $\lambda_{ex} = \lambda_{em}$. Пики же A1, A2, B1 и C, напротив, характеризуются значительным стоксовым сдвигом между длиной волны возбуждения люминесценции и наблюдаемой длиной волны люминесценции. Все упоминавшиеся выше пики серий «A» и «B» дают люминесценцию в полосах с максимумами на длинах волн 1430 нм и 827 нм и имеют попарно одинаковые длины волн возбуждения люминесценции (B и A1 – 823 нм, B1 и A2 – 420 нм или несколько короче).

Основные максимумы зависимости $I_{lum}$ ($\lambda_{ex}$, $\lambda_{em}$) для SBi световода при комнатной температуре наблюдаются и при T=77 K (Рис. 3. 2 б). При снижении температуры значительно уменьшается антистоксова часть пиков люминесценции A и B, происходит сужение полос возбуждения люминесценции, а ширина полос эмиссии люминесценции остается практически неизменной. В этом случае можно более отчетливо наблюдать слабые максимумы B′(820 нм, 910 нм) и B″(760 нм, 830 нм), расположенные вблизи пика B(820 нм, 830 нм). Один из них (B′) расположен в области более длинных волн эмиссии люминесценции, чем основной максимум (B), но с одинаковыми длинами волн возбуждения, а другой максимум (B″) – в области более коротких длин волн возбуждения с одинаковой длиной волны эмиссии люминесценции (по отношению к B). Менее отчетливо максимумы B′ и B″ наблюдаются и при T=300 K. При этом максимум B имеет вид сглаженного креста. При 77 K появляется новый сравнительно слабый широкий пик ИК люминесценции с максимумом A0(620 нм; 1480 нм), который практически не наблюдался при комнатной температуре. Пик A2 становится значительно слабее и, по-видимому, смещается по длинам волн возбуждения в коротковолновую сторону.

Расширение исследуемого диапазона оптических свойств ВАЦ в УФ область позволило получить дополнительные данные, а именно, положение новых максимумов ИК и видимой люминесценции.

Известно, что возбуждение в УФ области может сопровождаться возникновением люминесценции, обусловленной наличием собственных дефектов сетки стекла. Для устранения неоднозначности в определении пиков, относящихся к ВАЦ или к собственным дефектам сетки стекла, были измерены спектры возбуждения-эмиссии люминесценции для одинаковых по составу образцов, один из которых содержал висмут, а другой – нет. На Рис. 3. 3 приведен такой спектр, полученный для среза заготовки нелегированного кварцевого стекла. В этом случае наблюдался единственный пик люминесценции T2 ($\lambda_{ex}$=252 нм, $\lambda_{em}$=279 нм). Появление этой люминесценции обусловлено синглет-синглетным переходом кварцевого кислородно-дефицитного центра (ККДЦ) [133, 134]. Слабая интенсивность люминесценции триплет-синглетного перехода ККДЦ (люминесценция на длине волны около 460 нм) вследствие того, что сила осциллятора триплет-синглетного перехода в ККДЦ на 7 порядков меньше, чем синглет-синглетного перехода, не позволяла наблюдать ее в наших экспериментах.

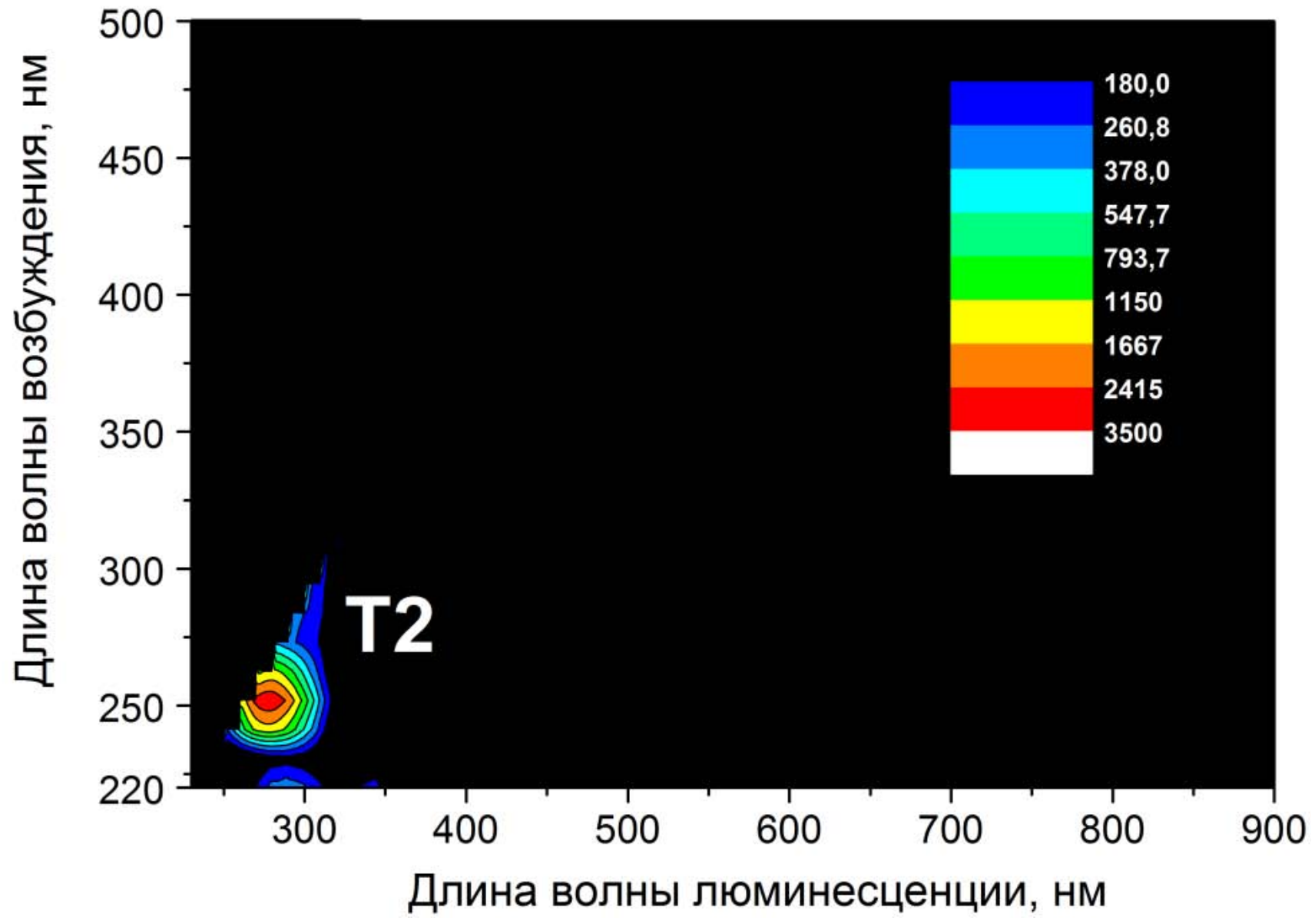

Рис. 3. 3 Трехмерный спектр возбуждения-эмиссии люминесценции чистого кварцевого стекла без висмута.

В отличие от чистого кварцевого стекла, трехмерный спектр возбуждения-эмиссии люминесценции кварцевого стекла, содержащего висмут, состоит из большего числа максимумов. На Рис. 3. 4 приведен спектр возбуждения-эмиссии люминесценции в области 240-1600 нм легированного висмутом образца SBi. Фактически полученный спектр сочетает в себе спектрально-люминесцентные характеристики заготовки и вытянутых из них волоконных световодов. Вообще говоря, люминесцентные свойства заготовок и световодов могут отличаться, поскольку волоконные световоды, по сравнению с волоконными заготовками, проходят дополнительно ряд технологических операций, включающих несколько циклов нагревания до ~ 2000 С с последующим охлаждением. Для образцов, исследованных в настоящей работе, были проведены серия измерений по выявлению возможных расхождений люминесцентных свойств. В результате было установлено, что спектры люминесценции заготовок волоконных световодов практически не отличаются от спектров люминесценции одномодовых волоконных световодов, полученных из этих заготовок, для длин волн возбуждения в диапазоне от 450 до 900 нм. Данное обстоятельство позволило предположить, что спектры люминесценции для указанных заготовок и световодов не отличаются (или отличаются незначительно) во всем исследованном диапазоне длин волн от 250 до 2000 нм. Поэтому в дальнейшем мы представляем результаты измерений, выполненных в настоящей работе как единую зависимость интенсивности люминесценции ($I_{lum}$) от длины волны возбуждения ($\lambda_{ex}$) и длины волны эмиссии ($\lambda_{em}$) люминесценции $I_{lum}(\lambda_{ex}, \lambda_{em})$.

Из Рис. 3. 4 видно, что при возбуждении на длинах волн короче 400 нм возникают попарно расположенные (на одной горизонтальной линии – т.е. соответствующие одной длине волны возбуждения) пики ИК люминесценции B2(376 нм; 827 нм) и A3(375 нм; 1417 нм); B3(≈240 нм; 827 нм) и A4(≈240 нм; ≈1417 нм).

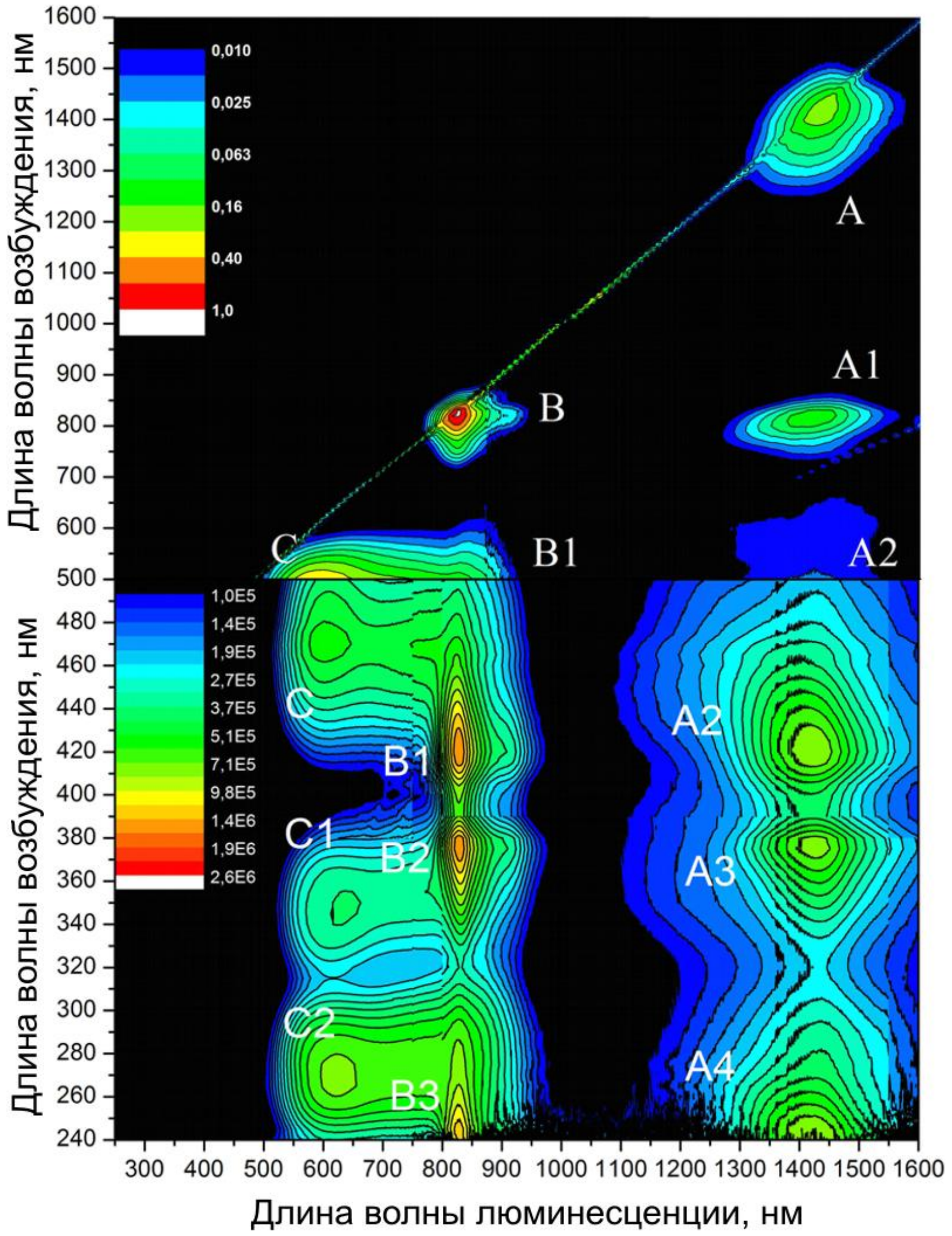

Рис. 3. 4 Трехмерные спектры возбуждения-эмиссии люминесценции SBi образца. Цветом указана интенсивность люминесценции в относительных единицах.

Основные параметры новых максимумов приведены в Табл. 3. 2. Некоторые расхождения в положении пиков красной и ИК люминесценции для разных диапазонов может быть частично обусловлены использованием различного спектрального оборудования.

Табл. 3. 2 Основные максимумы люминесценции ВАЦ SBi образца: обозначения, длины волн возбуждения и эмиссии при Т=300 K и 77 K

| Обозначение пика | $\lambda_{ex}^{max}$, нм | | $\lambda_{em}^{max}$, нм | |
|---|---|---|---|---|
| | 300 K | 77 K | 300 K | 77 K |
| A | 1415 | 1425 | 1430 | 1435 |
| A1 | 823 | 821 | 1430 | 1430 |
| A2 | 420 | <450 | 1430 | ≤1430 |
| A3 | 375 | | 1417 | |
| A4 | 240 | | 1417 | |
| A0 | — | 620 | — | 1480 |
| B | 823 | 823 | 827 | 833 |
| B1 | 420 | <450 | 827 | 830 |
| B2 | 376 | | 827 | |
| B3 | 240 | | 827 | |
| C | 470 | 480 | 603 | 605 |
| C1 | 348 | | 635 | |
| C2 | 270 | | 623 | |

Сравнивая спектры возбуждения люминесценции и оптических потерь (Рис. 3. 5), можно констатировать, что полосы возбуждения основных пиков люминесценции совпадают с полосами поглощения. А совпадение спектров возбуждения люминесценции на 830 нм и 1410 нм доказывает их принадлежность единому ВАЦ.

Характерные полосы люминесценции и поглощения в области 830, 1410 нм в стеклах и световодах подобного состава неоднократно наблюдались в работах различных авторов [119, 120, 135, 136], проводивших свои исследования параллельно с выполнением данной работы.

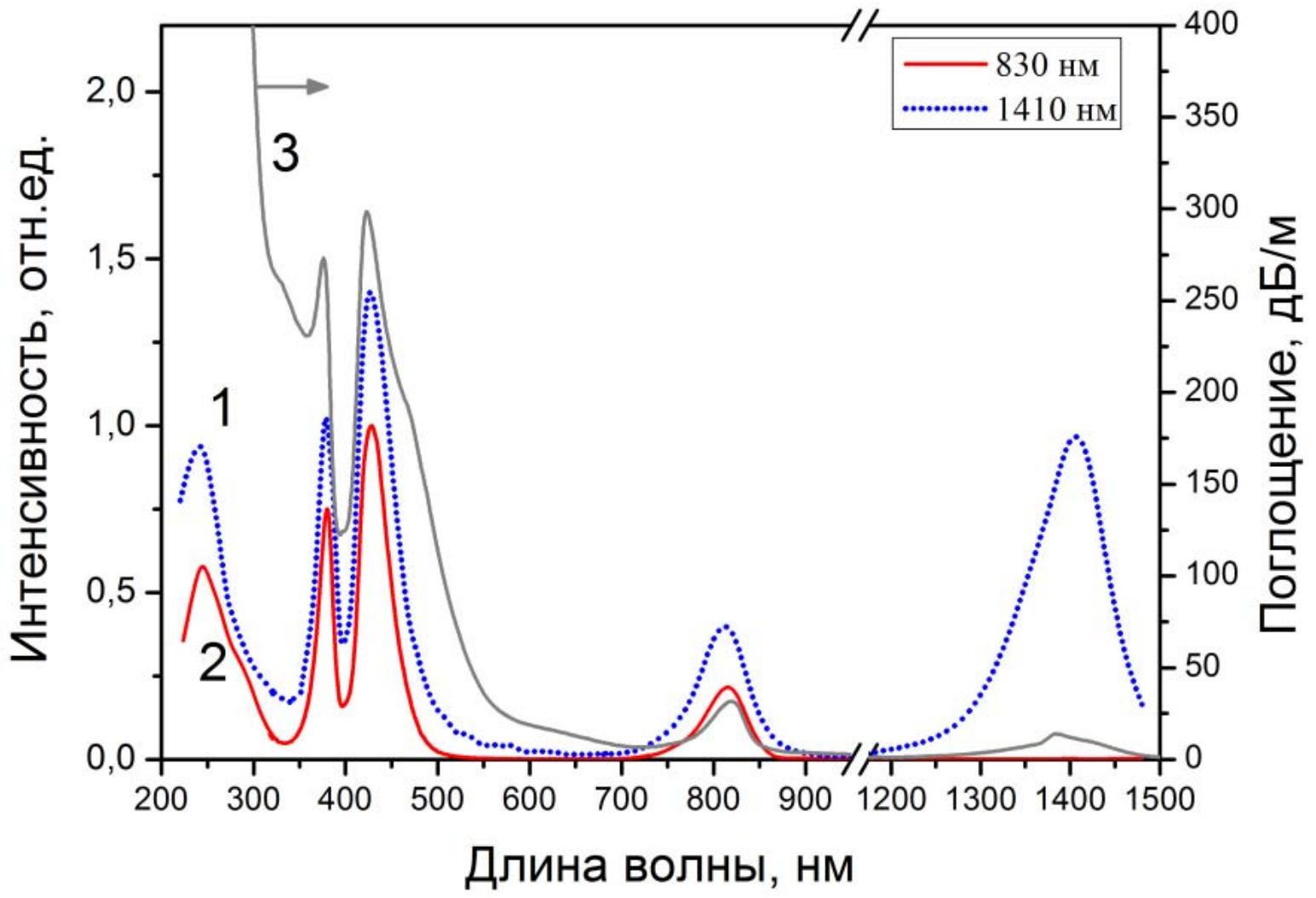

Рис. 3. 5 Спектры возбуждения люминесценции на 1410 и 830 нм (1 и 2) и спектр поглощения (3) SBi световода.

Отдельно следует остановиться на совместно появляющейся с ИК люминесценцией, красной люминесценции (3 пика красной люминесценции C(470 нм; 603 нм), C1(348 нм; 635 нм), C2(270 нм; 623 нм)). Серия пиков C значительно отличается от серий пиков A и B тем, что с ними по длине волны возбуждения не совпадают никакие пики в ИК области спектра. Это указывает на отсутствие связи между источником этой люминесценции и ИК висмутовыми активными центрами в SBi образце.

Ранее уже высказывалось предположение о принадлежности пика красной люминесценции в SBi световоде ионам $Bi^{2+}$ [120]. Дополнительные аргументы дает сравнение полученных нами спектров возбуждения люминесценции SBi световода на длине волны 600 нм со спектрами возбуждения люминесценции ионов $Bi^{2+}$ в кристаллах [28]. Спектр возбуждения красной люминесценции SBi световода приведен на Рис. 3. 6.

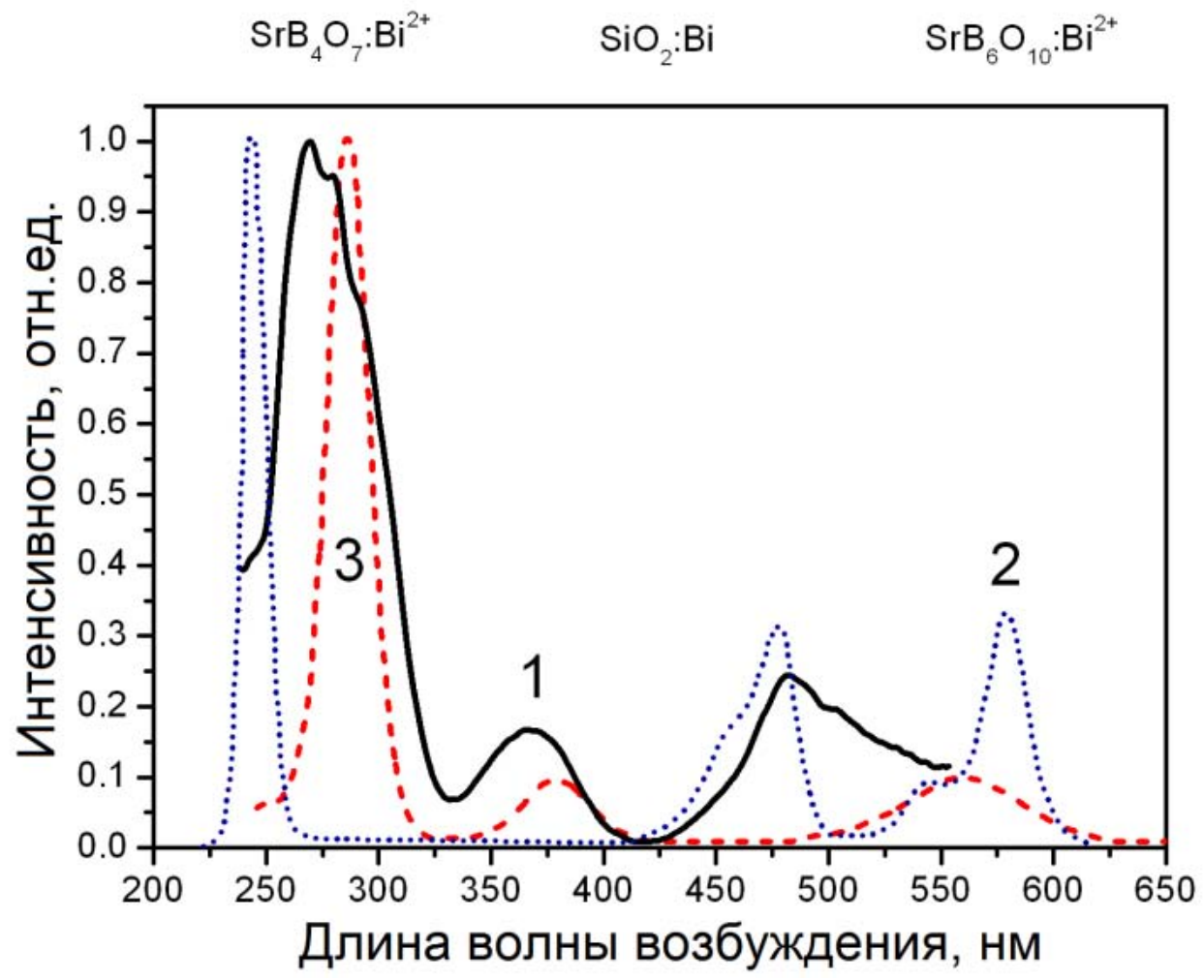

Рис. 3. 6 Спектры возбуждения красной люминесценции: SBi световод (1), в кристаллах $SrB_6O_{10}:Bi^{2+}$ (2) и $SrB_4O_7:Bi^{2+}$ (3) [28].

На том же графике представлены спектры возбуждения люминесценции ионов $Bi^{2+}$ в кристаллах $SrB_6O_{10}:Bi^{2+}$ и $SrB_4O_7:Bi^{2+}$. Максимум спектра люминесценции в кристалле $SrB_6O_{10}:Bi^{2+}$ приходится на 660 нм, в кристалле $SrB_4O_7:Bi^{2+}$ на 588 нм. Спектры возбуждения ионов $Bi^{2+}$ в кристаллах в диапазоне 225-600 нм состоят из трех пиков, которые соответствуют переходам в ионе $Bi^{2+}$ (см. Рис. 1. 8): $^2P_{1/2}\rightarrow{}^2S_{1/2}$, $^2P_{1/2}\rightarrow{}^2P_{3/2}(2)$, $^2P_{1/2}\rightarrow{}^2P_{3/2}$ (1) (в порядке убывания энергии перехода). Люминесценция наблюдается на переходе $^2P_{3/2}$ (1)$\rightarrow{}^2P_{1/2}$.

Максимум красной люминесценции в SBi образце находится на длине волны 590 нм. Спектр возбуждения этой люминесценции имеет 3 пика. Причем следует отметить, что различия в положении пиков люминесценции и возбуждения для двух кристаллов существенно больше, чем различия в положении этих максимумов между спектрами кристалла $SrB_6O_{10}:Bi^{2+}$ и SBi световода. Наблюдаемое качественное и в существенной мере количественное подобие спектров люминесценции и возбуждения является подтверждением принадлежности пика люминесценции C(480 нм, 590 нм) ионам $Bi^{2+}$ в сетке кварцевого стекла. И, следовательно, ионы $Bi^{2+}$ не связаны, по-видимому, с ИК висмутовыми активными центрами.

## 3.3 Времена жизни люминесценции и схема энергетических уровней ВАЦ в v - $SiO_2$

Не вызывает сомнений, что появление этих висмутовых центров обусловлено наличием кремния в стекле. Поэтому далее будет встречаться обозначение для ВАЦ в $SiO_2$ стекле – висмутовые активные центры, ассоциированные с кремнием или ВАЦ-Si. Анализируя полученные трехмерные спектры возбуждения-эмиссии люминесценции для ВАЦ в $SiO_2$ стекле (Рис. 3. 4), нами были определены энергии основных энергетических уровней и излучательных переходов между ними. В результате была построена схема энергетических уровней ВАЦ-Si (Рис. 3. 7).

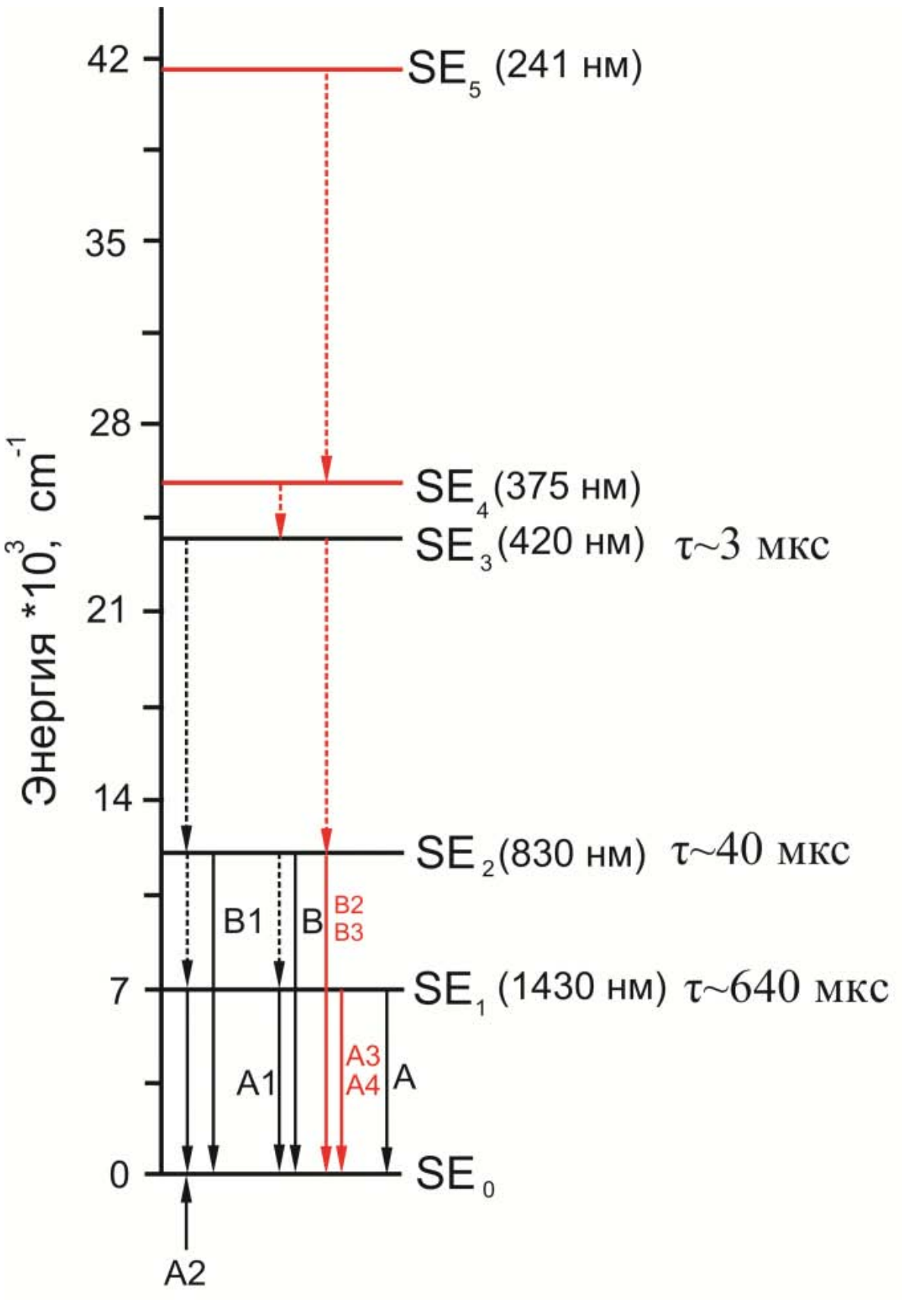

Рис. 3. 7 Схема энергетических уровней ВАЦ-Si. На схеме сплошными линиями со стрелками, направленными вниз, обозначены наблюдаемые (излучательные) переходы при T=300 К. Пунктиром обозначены переходы, люминесценция которых не наблюдалась. Справа указаны измеренные времена жизни первых трех возбужденных уровней.

В данном случае уровни изображены в виде линий, поскольку, как выше отмечалось, в этом случае электрон-фононным взаимодействием можно пре-небречь (из-за малого стоксового сдвига люминесценции). Основные оптические переходы (сплошные линии), которые наблюдаются в эксперименте, также приведены на Рис. 3. 7. Пунктирными линиями обозначены переходы, которые не наблюдались в эксперименте, потому что либо длина волны (энергия) кванта выходила за пределы диапазона измерения (например, переход с уровня $SE_2 \rightarrow SE_1$), либо эти переходы были безызлучательными.

Совместно появляющаяся с ИК люминесценцией, красная люминесценция (3 пика красной люминесценции C(470 нм; 603 нм), C1(348 нм; 635 нм), C2(270 нм; 623 нм)) в отличии от групп пиков А и В, имеет спектр возбуждения, который не соответствует построенной схеме уровней энергии ВАЦ-Si. Данный факт указывает, что происхождение ИК и красной люминесценции имеют различную природу. Как было показано (Рис. 3. 6), спектры возбуждения красной люминесценции (время жизни ≈ 3-5 мкс) близки к спектрам возбуждения ионов двухвалентного висмута в кристаллах [28, 49, 137], что позволяет сделать вывод о том, что пики люминесценции C-C2 на Рис. 3. 4 соответствуют люминесценции ионов $Bi^{2+}$.

Времена жизни люминесценции определялись по экспериментально измеренным временным зависимостям затухания люминесценции (Рис. 3. 8). Возбуждение осуществлялось импульсами длительностью 10 нс. Длина волны возбуждения выбиралась исходя из значений длин волн поглощения, соответствующих переходам $SE_0 \rightarrow SE_1$; $SE_0 \rightarrow SE_2$; $SE_0 \rightarrow SE_3$. На Рис. 3. 8 а приведены полученные временные зависимости затухания люминесценции на 1430 нм (а) и на 830 нм (б). Видно, что времена жизни люминесценции около ≈1400 нм составляло 620-640 мкс и слабо зависело от длины волны возбуждения. Следует отметить для пиков A1, A2 (излучательный переход $SE_2 \rightarrow SE_0$) после выключения импульса накачки наблюдалось возрастание (вместо затухания люминесценции) сигнала со временем ≈40 мкс (Рис. 3. 8 а). Такое поведение может быть объяснено наличием вышерасположенного уровня (более короткоживущего, чем основной), который после прекращения импульса накачки начинает рассе-ляться со временем ≈ 40 мкс. В данном случае таким уровнем является уровень $SE_2$, время жизни которого, как и ожидалось, составляет 40 мкс (Рис. 3. 8 б)

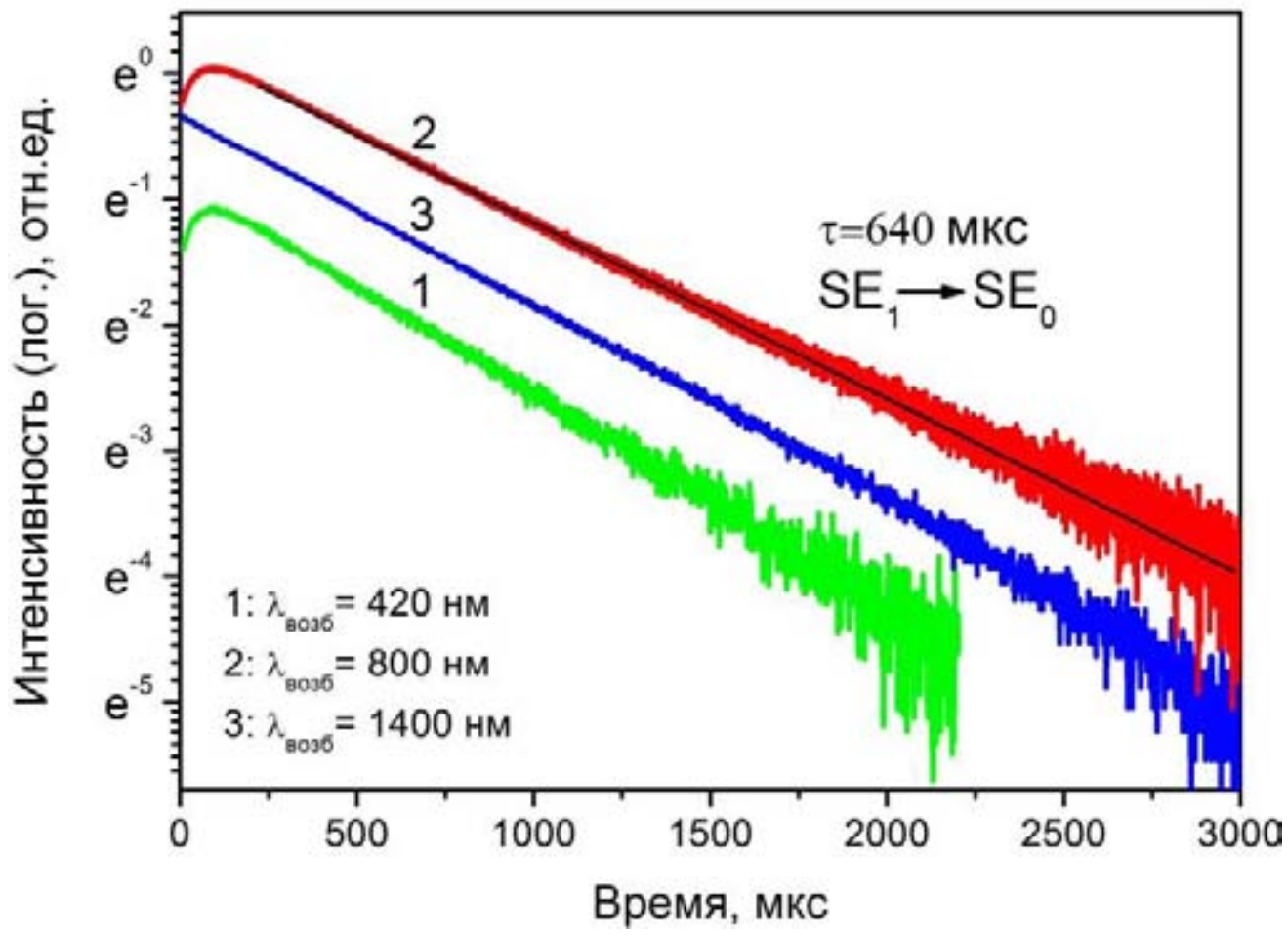

а

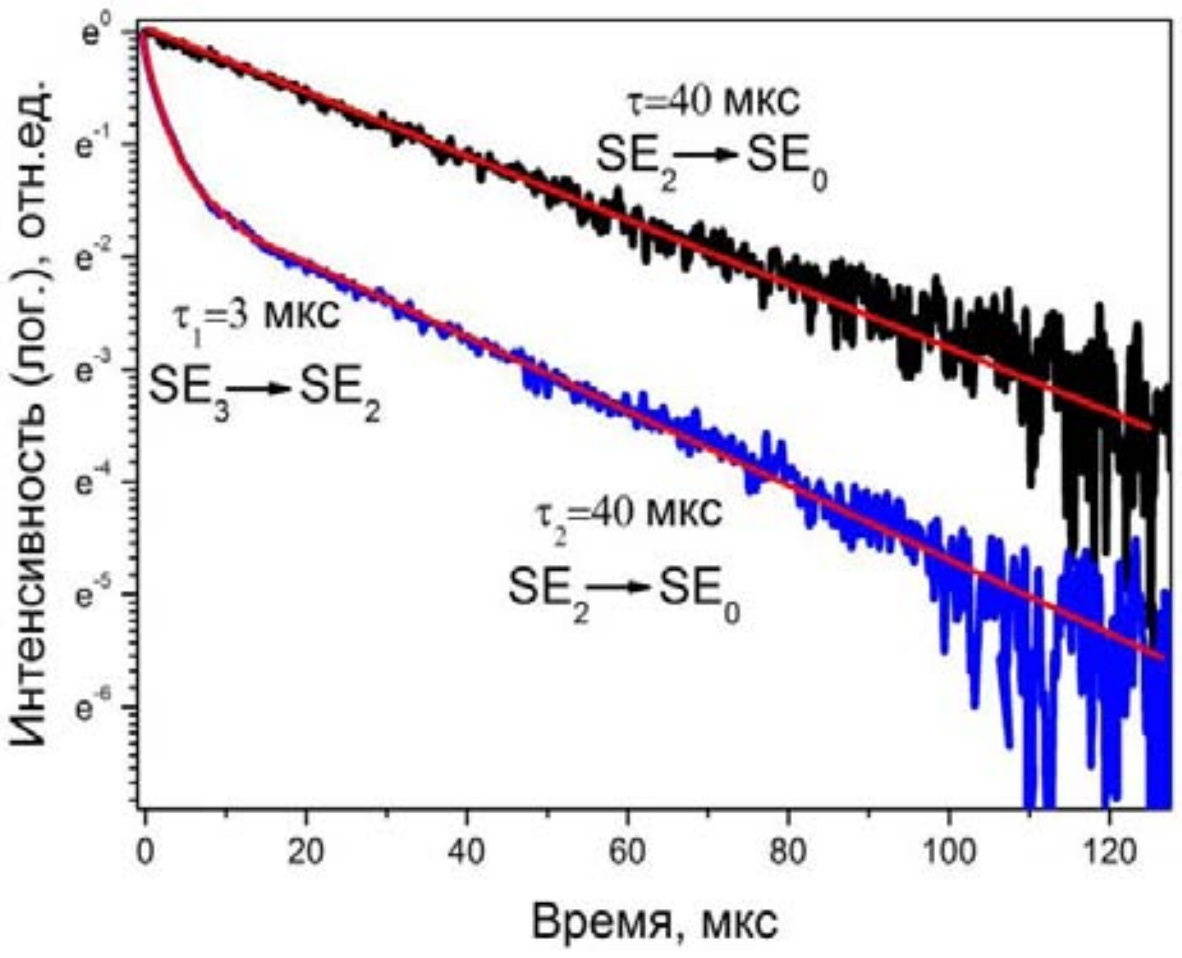

б

Рис. 3. 8 Временные зависимости затухания люминесценции на длине волны $\lambda_{em}$ : а) 1430 нм; б) 830 нм. На графиках приведены времена жизни люминесценции и соответствующие оптические переходы, характерные для ВАЦ-Si.

В отличие от люминесценции на ≈1400 нм, кривые затухания люминесценции на 830 нм существенно различались при изменении длины волны возбуждения. Если при прямом возбуждении в уровень $SE_2$ наблюдается люминесценция на 830 нм со временем 40 мкс, то при смещении длины волны возбуждения на 420 нм (следующий уровень $SE_3$), затухание люминесценции происходит с двумя временами жизни (3 и 40 мкс). Появление короткого времени связано с наличием дополнительного излучательного перехода с одинаковой энергией. И действительно, расстояние между уровнями $SE_3$ и $SE_2$ является близким по энергии с расстоянием между уровнями $SE_2$ и $SE_0$. Значения времен жизни люминесценции основных пиков, наблюдаемых на спектре возбуждения-эмиссии люминесценции приведены в Табл. 3. 3.

Следовательно, люминесценция пиков B1, B2 и B3 состоит из двух полос с временами 3-5 и 40-50 мкс. Таким образом, обнаруженное короткое время жизни принадлежит энергетическому уровню $SE_3$. Полученные экспериментальные результаты (люминесценция, возбуждение и времена жизни) полностью укладываются в единую схему энергетических уровней для ВАЦ-Si и являются дополнительным аргументом в пользу правильности построенной схе-мы энергетических уровней. Отметим, что позже похожие результаты, но в меньшем объеме, были получены другими авторами, исследовавшими пористые пленки из кварцевого стекла, легированного висмутом [138].

Табл. 3. 3 Измеренные времена жизни для основных пиков люминесценции ВАЦ-Si.

| Обозначение пика | $\lambda_{ex}^{max}$, нм | $\lambda_{em}^{max}$, нм | $\tau$, мкс |
|---|---|---|---|
| A | 1415 | 1430 | 630 |
| A1 | 823 | 1430 | 640* |
| A2 | 420 | 1430 | 640* |
| A3 | 375 | 1426 | 630* |
| A4 | 240 | 1426 | 620* |
| B | 823 | 827 | 40 |
| B1 | 420 | 827 | 3 и 40 |
| B2 | 376 | 830 | 10 и 50 |
| B3 | 240 | 830 | 5 и 45 |
| C | 480 | 590 | 3 |
| C1 | 350 | 620 | 3 |
| C2 | 270 | 630 | 3.5 |

* обозначены кривые, на первоначальном участке которых наблюдается нарастание интенсивности сигнала.

## 3.4 Оптическое усиление и лазерная генерация на ВАЦ, ассоциированных с кремнием (ВАЦ-Si)

При анализе схемы энергетических уровней ВАЦ-Si (Рис. 3. 7) с точки зрения возможности получения лазерной генерации в первую очередь обращает на себя внимание переход $SE_1 \rightarrow SE_0$. Большое время жизни верхнего уровня (~ 1 мс) позволяет получать значительную его заселенность при сравнительно низких интенсивностях излучения накачки. Правда нижний уровень этого перехода – $SE_0$ – является основным, с бесконечным временем жизни, но эта трудность при большом уширении (по сравнению с кристаллами) энергетических уровней в стекле обходится использованием квазичетырехуровневой схемы генерации, аналогично той, что и в широко распространенных иттербиевых лазерах.

Поэтому в настоящей работе была предпринята успешная попытка получить лазерную генерацию на переходе $SE_1 \rightarrow SE_0$. В результате впервые была показана возможность получения непрерывной лазерной генерации в волоконном световоде с сердцевиной из стеклообразного $SiO_2$, легированного висмутом, без других легирующих добавок.

Висмутовый волоконный лазер (ВВЛ) был собран по стандартной линейной схеме (см. Рис. 3. 9). Резонатор лазера состоял из активного SBi световода длиной 8 м и волоконных брэгговских решеток (БР) на длину волны $\lambda_{Bi}$=1460 нм, приваренных к концам световода. Одна из решеток (БР100) имела коэффициент отражения, близкий к 100%, а другая (БР30), являющаяся выходным зеркалом резонатора – 30%.

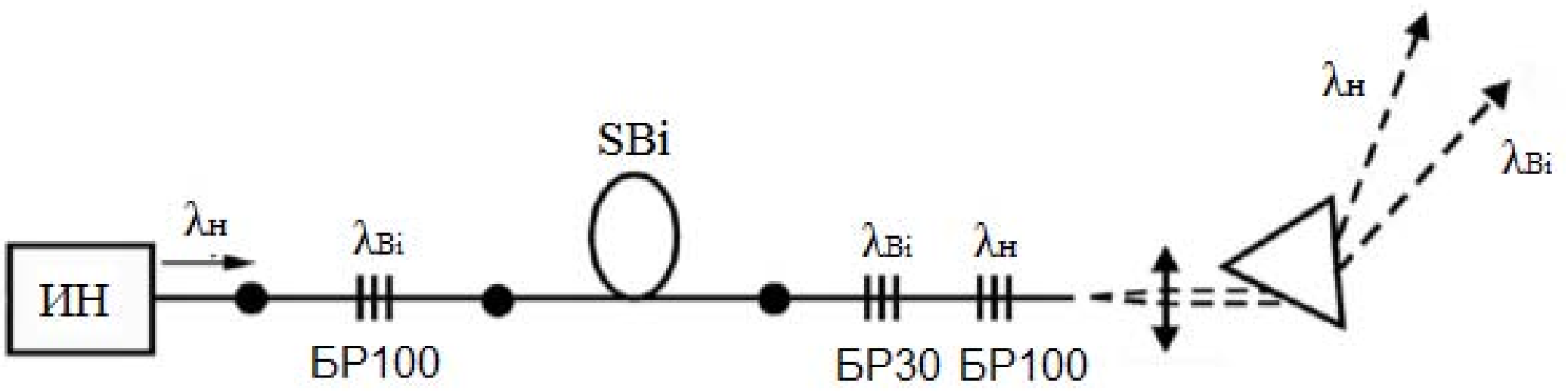

Рис. 3. 9 Схема висмутового волоконного лазера, работающего в непрерывном режиме. ИН – источник накачки; БР – брэгговская решетка (справа указан коэффициент отражения). Точками обозначены места сварки световодов.

В качестве источника накачки (ИН) служил висмутовый волоконный лазер с $\lambda_н$ = 1340 нм на фосфоросиликатном световоде, легированном висмутом, который возбуждался рамановским волоконным лазером с длиной волны 1230 нм. Излучение ИН вводилось через БР100 с $\lambda_{Bi}$ в сердцевину активного световода. Дополнительная БР100 с $\lambda_н$, имеющая высокий (близким к 100%) коэффициент отражения на длине волны накачки, размещалась на выходе ВВЛ для возвращения непоглощенной доли излучения накачки обратно в резонатор. Спектральная ширина всех БР была ~0.5 нм.

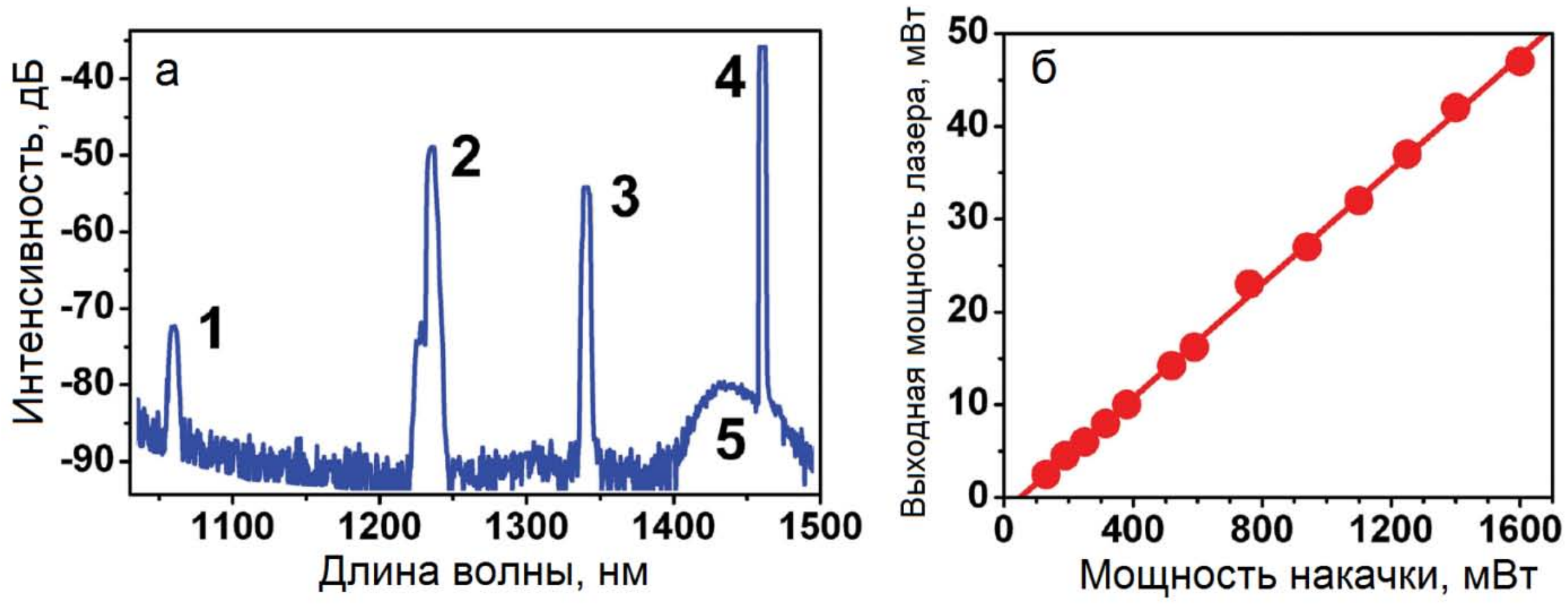

Рис. 3. 10 а) Спектр излучения на выходе ВВЛ. 1 – непоглощенная накачка рамановского лазера (Yb волоконный лазер с λ=1058 нм); 2 – непоглощенное излучение рамановского лазера; 3 – излучение висмутового лазера на фосфоросиликатном световоде; 4 – излучение ВВЛ на активном SBi световоде; 5 – люминесценция висмутового SBi световода; б) зависимость мощности выходного излучения ВВЛ на длине волны 1460 нм от мощности поглощенного излучения накачки на длине волны 1340 нм.

Спектр излучения на выходе ВВЛ представлен на Рис. 3. 10 а. Помимо линии генерации лазера (4) и люминесценции (5), в спектре выходного излучения можно наблюдать линии накачки (1-3). Зависимость выходной мощности ВВЛ от поглощенной мощности накачки показана на Рис. 3. 10 б. Линия генерации ВВЛ выделялась из спектра выходного излучения с помощью стандартной дисперсионной призмы. Дифференциальный кпд ВВЛ оказался равным ≈3 % по отношению к поглощенной мощности накачки. Порог лазерной генерации составил 50 мВт, а максимальная выходная мощность на длине волны 1460 нм – 47 мВт.

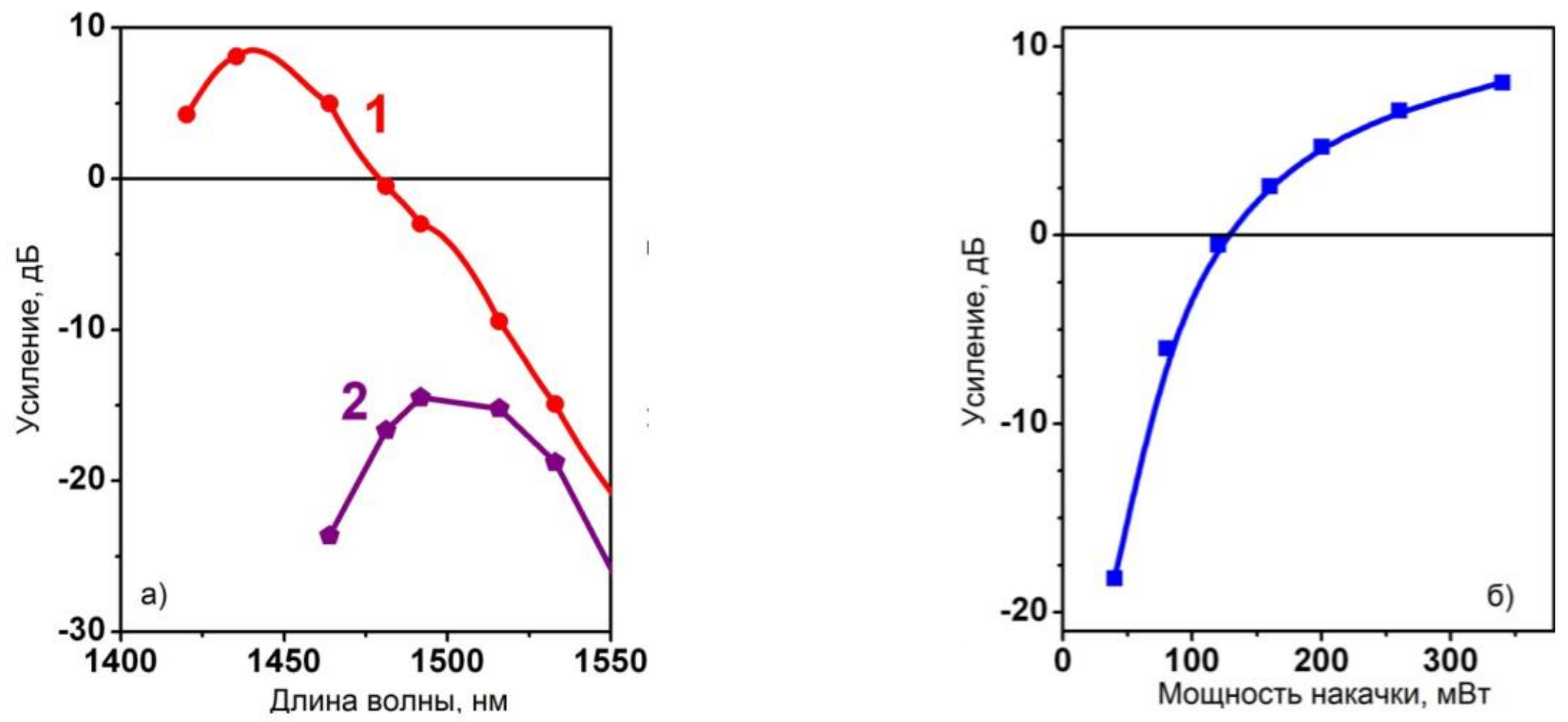

Рис. 3. 11 а) Спектр усиления SBi световода при мощности накачки 340 мВт (кривая 1) и 35 мВт (кривая 2); б) зависимость коэффициента усиления световода на $\lambda_{макс}$=1440 нм от мощности накачки.

Результаты выполненных измерений оптического усиления в SBi световоде длиной 15.2 м при накачке непрерывным волоконным рамановским лазером на длине волны 1320 нм приведены на Рис. 3. 11 а. Измерения проводились по стандартной схеме, подробное описание схемы измерений можно найти в [139]. При небольших уровнях накачки поглощение в световоде превышает усиление (кривая 2). При мощности накачки 120 мВт усиление в световоде компенсирует потери, а при 340 мВт усиление приближается к насыщению (см. Рис. 3. 11 б) и составляет в максимуме около 8 дБ (кривая 1 на Рис. 3. 11 а). При этом положительный коэффициент усиления наблюдался в диапазоне длин волн от 1410 до 1470 нм. Пик усиления приходился на длину волны $\lambda_{макс}$=1440 нм. Максимальное усиление на единицу длины достигало 0.52 дБ/м. Ширина полосы усиления по уровню – 3 дБ составила примерно 40 нм. Усилительные и генерационные характеристики SBi световодов ограничиваются сравнительно высоким уровнем ненасыщающегося поглощения. Кроме того, большое поглощение в области 1.35-1.42 мкм обусловленное значительным содержанием OH-групп, что так же снижает эффективность лазера.

Что касается возможности получения генерации на других переходах ВАЦ-Si, то обращает на себя внимание сравнительно высокое время жизни уровня $SE_2$ (~ 40 мкс). По-видимому, используя его как верхний лазерный уро-вень, можно получить генерацию по квазичетырехуровневой схеме на переходе $SE_2 \rightarrow SE_0$, $\lambda \approx 830$ нм и на переходе $SE_2 \rightarrow SE_1$, $\lambda \approx 2000$ нм при условии наличия необходимого источника накачки. К сожалению, для этого необходим одномодовый лазер с выходной мощностью более 1 Вт в диапазоне длин волн ~ 780 – 800 нм. Использование же объемного Ti-сапфирового лазера для накачки волоконного неперспективно.

### 3.5 Люминесцентные свойства ВАЦ в алюмо- и фосфоросиликатных световодах, легированных висмутом

По алюмосиликатным световодам с висмутом были проведены многочисленные экспериментальные исследования по изучению их оптических свойств в видимой и ИК области спектра (см., например, [99] и ссылки в ней). Для комплексных систем стекол (стекло (v-$SiO_2$) + активатор (Bi) + легирующая примесь (Al или P)) использование дискретного набора двумерных спектров люминесценции не дает возможности идентифицировать структуру уровней ВАЦ. В данной работе была предпринята попытка получить подробные данные о люминесцентных свойствах ВАЦ в алюмо- (ВАЦ-Al) и фосфоросиликатных (ВАЦ-P) световодах с помощью построения трехмерных спектров возбуждения-эмиссии люминесценции.

Введение дополнительных примесей в виде $P_2O_5$ или $Al_2O_3$ на уровне единиц молярных процентов приводит к значительному изменению спектров люминесценции по сравнению с SBi образцом (Рис. 3. 12 и Рис. 3. 14). Особенность висмутовых люминесцентных центров, формирующихся в алюмосиликатных и фосфоросиликатных стеклах, заключается в том, что для них характерным является наличие гораздо более широких полос люминесценции, чем для ВАЦ-Si. Значительные стоксовы сдвиги частоты большинства линий свидетельствуют о существенном влиянии в данном случае электрон-фононного взаимодействия. ВАЦ-Al и ВАЦ-P отличаются от ВАЦ-Si также значительно более сильной зависимостью спектра люминесценции от длины волны возбуждения для некоторых полос люминесценции, о чем свидетельствует значитель-ный наклон этих полос люминесценции по отношению к оси длин волн эмиссии (подобное поведение полос люминесценции регистрировалось ранее для некоторых стекол и световодов, легированных висмутом [102, 140]).

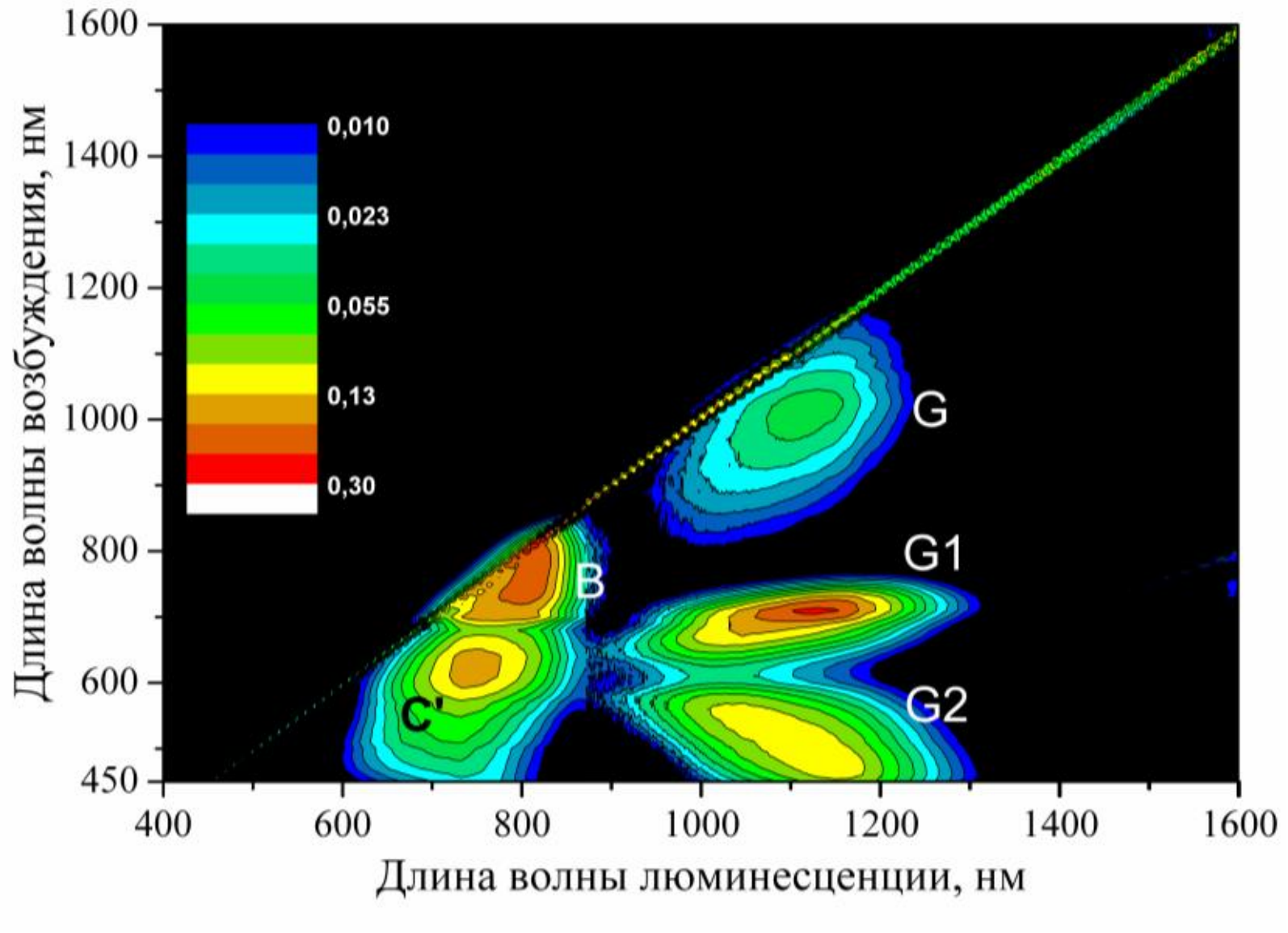

а

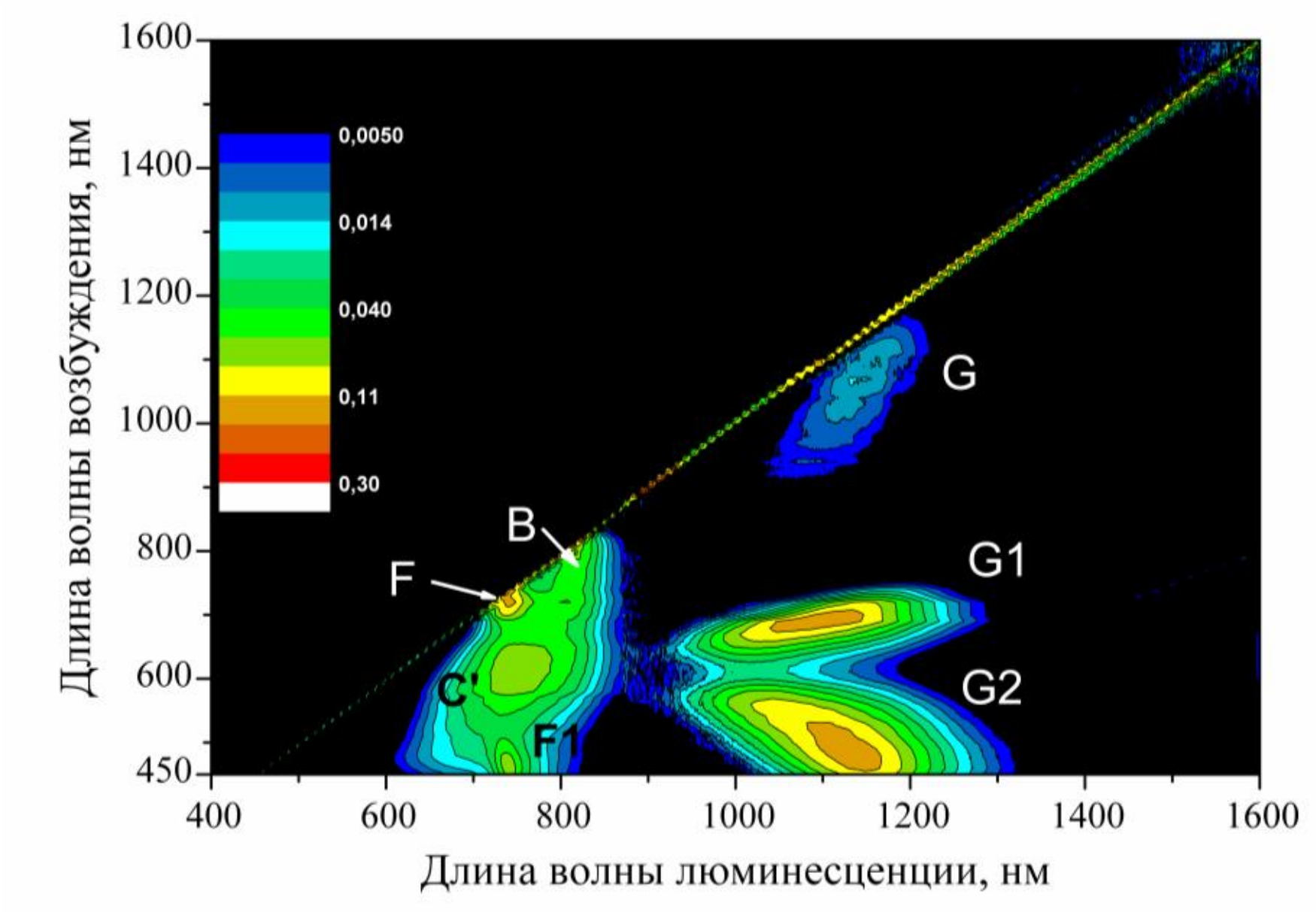

б

Рис. 3. 12 Трехмерные спектры возбуждения-эмиссии люминесценции для ASBi световода при комнатной температуре (а) и при температуре жидкого азота (б)

На Рис. 3. 12 представлен трехмерный спектр возбуждения-эмиссии люминесценции для ASBi (450 нм $\leq \lambda_{ex} \leq$ 1600 нм) при комнатной температуре (а) и температуре жидкого азота (б). При комнатной температуре полученный спектр состоит из 5 максимумов, три из которых расположены в области 1 мкм и два – в области 700-800 нм. Положения указанных максимумов люминесценции приведены в Табл. 3. 4. При снижении температуры образца наблюдалось сужение полос люминесценции в области 700-800 нм, что позволило определить некоторые дополнительные полосы (F, F1). Однако полученных данных оказалось недостаточно для построения (даже в упрощенном виде) системы уровней ВАЦ-Al.

Табл. 3. 4 Обозначения, длины волн возбуждения и люминесценции основных максимумов люминесценции ВАЦ ASBi световода.

| Обозначение пика | $\lambda_{ex}^{max}$, нм | | $\lambda_{em}^{max}$, нм | |
|---|---|---|---|---|
| | 300 К | 77 К | 300 К | 77 К |
| G | 1020 | 1080 | 1120 | 1150 |
| G1 | 705 | 688 | 1130 | 1100 |
| G2 | 510 | 490 | 1100 | 1130 |
| B | 780 | 780 | 815 | 820 |
| F | — | 720 | — | 740 |
| F1 | — | 460 | — | 740 |
| C′ | 620 | 610 | 745 | 745 |

Для получения дополнительной информации о люминесцентных свойства ВАЦ-Al были проведены измерения спектров люминесценции при возбуждении в УФ области (240 нм $\leq \lambda_{ex} \leq$ 500 нм). Трехмерный спектр возбуждения-эмиссии люминесценции ASBi световода 240 нм $\leq \lambda_{ex} \leq$ 1600 нм приведен на Рис. 3. 13. Значения $\lambda_{ex}$ и $\lambda_{em}$, соответствующие максимумам люминесценции, наблюдаемых нами в ASBi образце при возбуждении в области длин волн 240 – 500 нм, приведены в Табл. 3. 5.

При возбуждении в УФ области наблюдается полоса С'', которая по длине волны люминесценции (она близка к $\lambda_{em}$ полосы С') может быть отнесена к люминесценции двухвалентного висмута в алюмосиликатной матрице. Кроме

Табл. 3. 5 Основные пики люминесценции, наблюдаемые в ASBi образце (Рис. 3. 13) при 240 нм ≤ $\lambda_{ex}$ ≤ 500 нм.

| Обозначение пика | $\lambda_{ex}$, нм | $\lambda_{em}$, нм |
|---|---|---|
| G2 | 510 | 1100 |
| C'' | 297 | 780 |

того, смещение по $\lambda_{ex}$ в коротковолновую область позволило полностью прописать пик G2 (по сравнению с Рис. 3. 12 а). Обнаружено, что при такой же пороговой чувствительности схемы регистрации, как и в случае с SBi (Рис. 3. 4), в образцах ASBi не регистрируется люминесценция при $\lambda_{em}$ > 1 мкм при возбуждении УФ излучением.

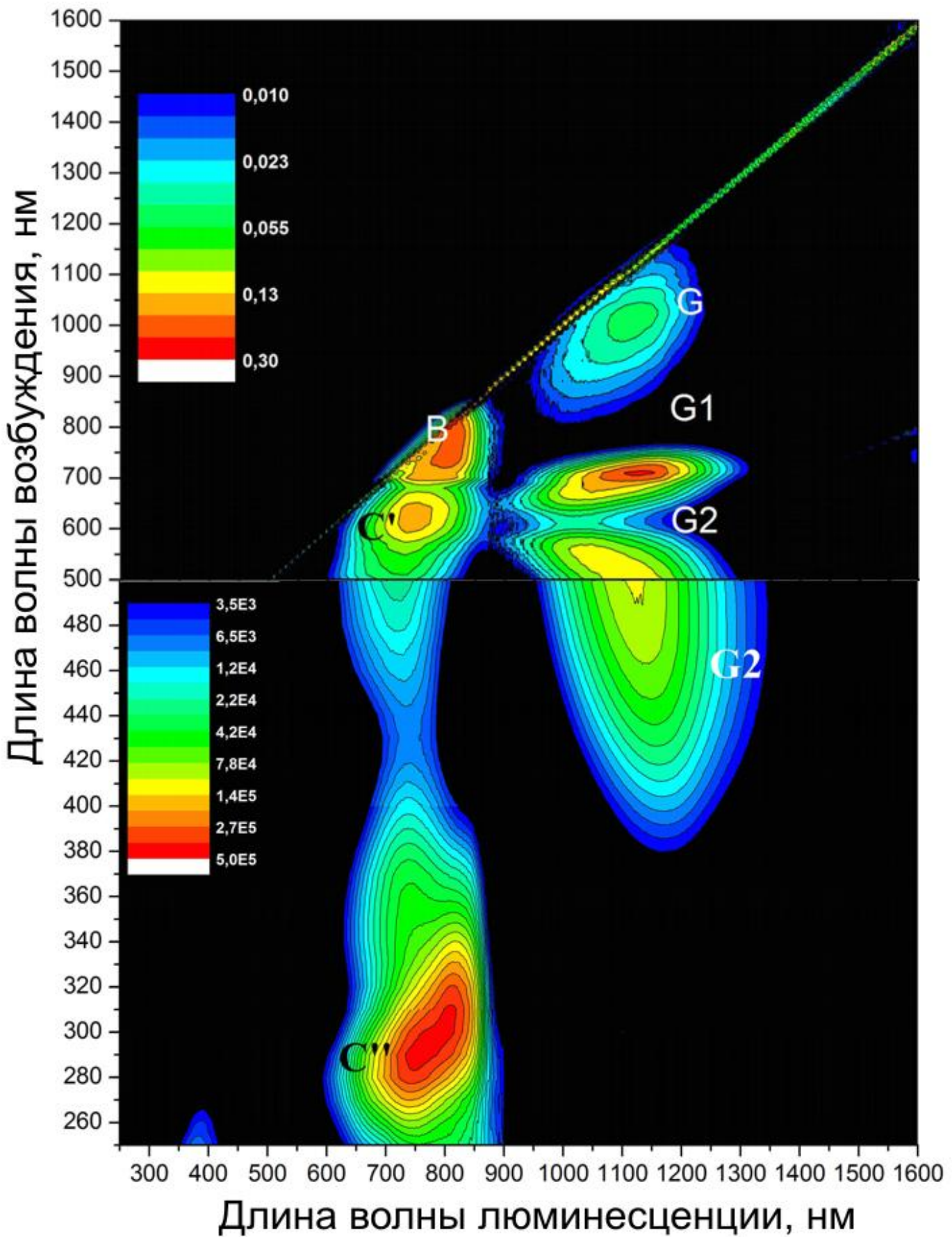

Рис. 3. 13 Трехмерный спектр возбуждения-эмиссии люминесценции ASBi образца при возбуждении в области длин волн от 250 до 1600 нм.

Полученные результаты достаточно подробно отражают люминесцентные свойства ASBi образца, но не позволяют построить схему энергетических уровней такого типа образцов. Таким образом, показано, что построение трехмерных графиков возбуждения-эмиссии люминесценции не является достаточным условием для определения схемы уровней ВАЦ в световодах с сердцевиной из многокомпонентного стекла, в частности, алюмосиликатного стекла. В этом случае требуется проведения дополнительных исследований.

Более детально остановимся на результатах, полученных для фосфоросиликатного (PSBi) образца. На Рис. 3. 14 приведены трехмерные спектры возбуждения-эмиссии люминесценции PSBi световода. Значения $\lambda_{ex}$ и $\lambda_{em}$, соответствующие максимумам пиков люминесценции, наблюдаемых нами в PSBi образце приведены в Табл. 3. 6.

Табл. 3. 6. Обозначения, длины волн возбуждения и люминесценции основных максимумов люминесценции PSBi световода при T=300 и 77 К (Рис. 3. 14).

| Обозначение пика | $\lambda_{ex}^{max}$, нм | | $\lambda_{em}^{max}$, нм | |
|---|---|---|---|---|
| | 300 К | 77 К | 300 К | 77 К |
| I | 1065 | 1056 | 1195 | 1180 |
| I1 | 750 | 753 | 1250 | 1260 |
| I2 | ≈460 | ≈460 | ≈1245 | ≈1240 |
| I3 | (1266-1427) | (1241-1409) | (1283-1429) | (1282-1425) |
| B | 813 | 814 | 824 | 825 |
| B1 | 450 | 450 | 825 | 825 |
| F | 772 | 776 | 782 | 787 |
| F1 | — | 450 | — | 787 |
| C″ | 523 | 520 | 760 | 772 |

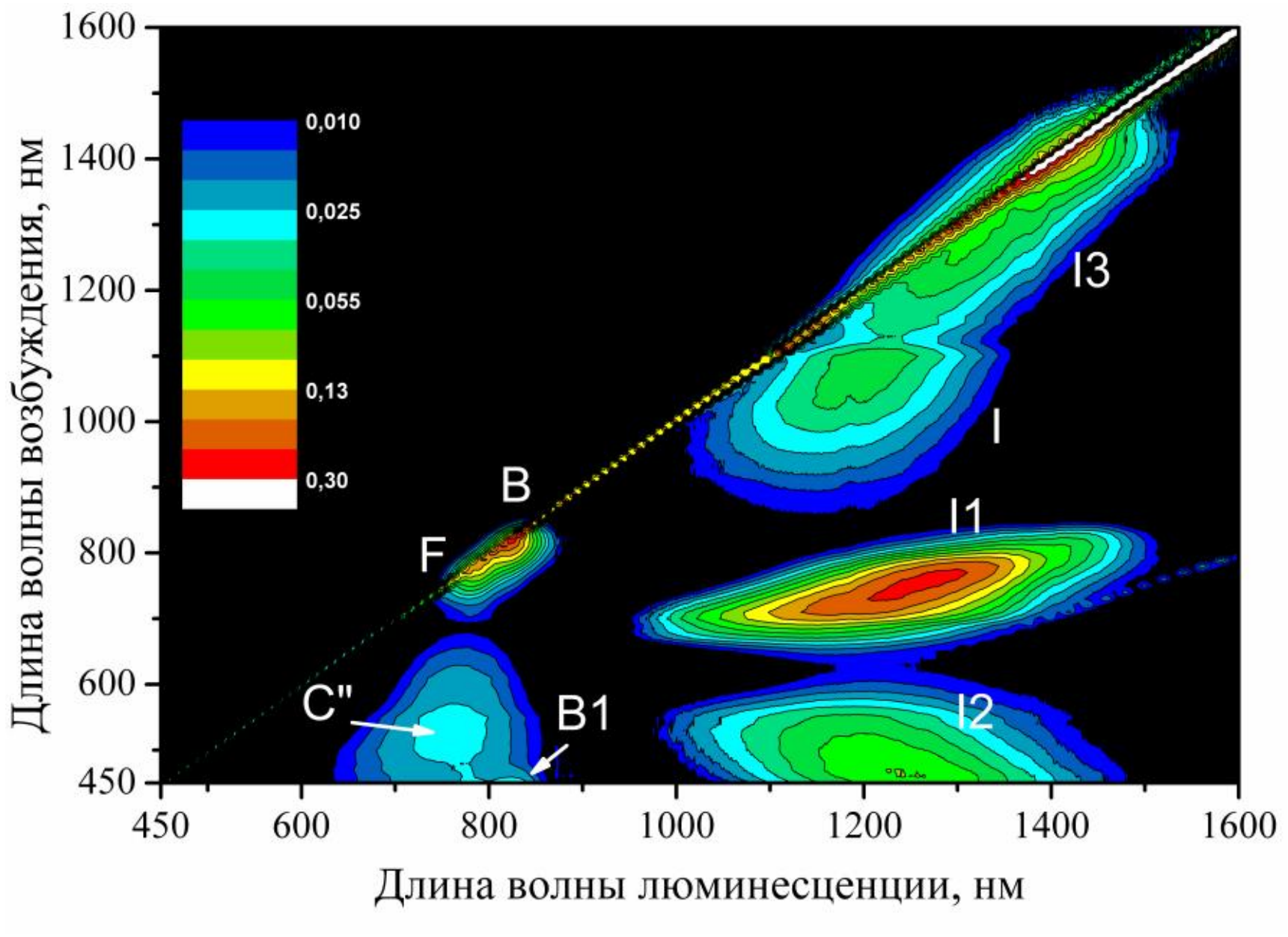

а

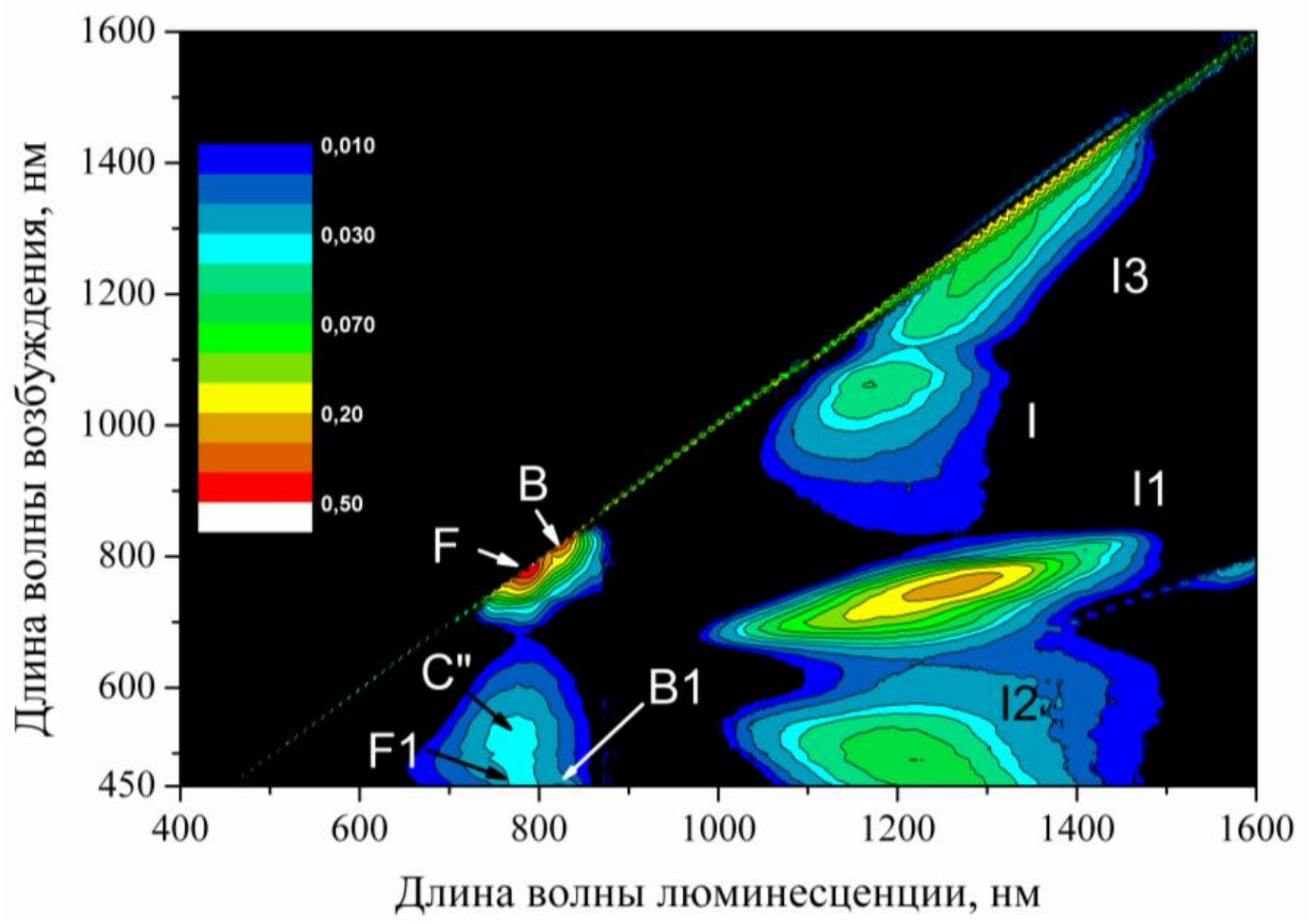

б

Рис. 3. 14 Трехмерные спектры возбуждения-эмиссии люминесценции PSBi световода при комнатной температуре (а) и при температуре жидкого азота (б)

Для большинства максимумов люминесценции, наблюдаемых на трехмерном спектре PSBi световода, характерным является несимметричность контуров распределения интенсивности люминесценции. Данное обстоятельство указывает на сложную структуру (несколько близко расположенных взаимно перекрывающихся максимумов люминесценции) этих максимумов (в частности, для I – I3). Данная картина сохраняется и для результатов измерений, по-лученных в низкотемпературных условиях (см. Рис. 3. 14 б), что не позволяет определить положение энергетических уровней ВАЦ-P в PSBi световоде в отличие от SBi световода. Для SBi образца наблюдается система пиков люминесценции, расположенных в виде трапеции (A-A4 и B-B3) и попарно имеющих одну длину волны возбуждения (кроме «непарного» пика A). В спектре люминесценции PSBi такая система не наблюдается.

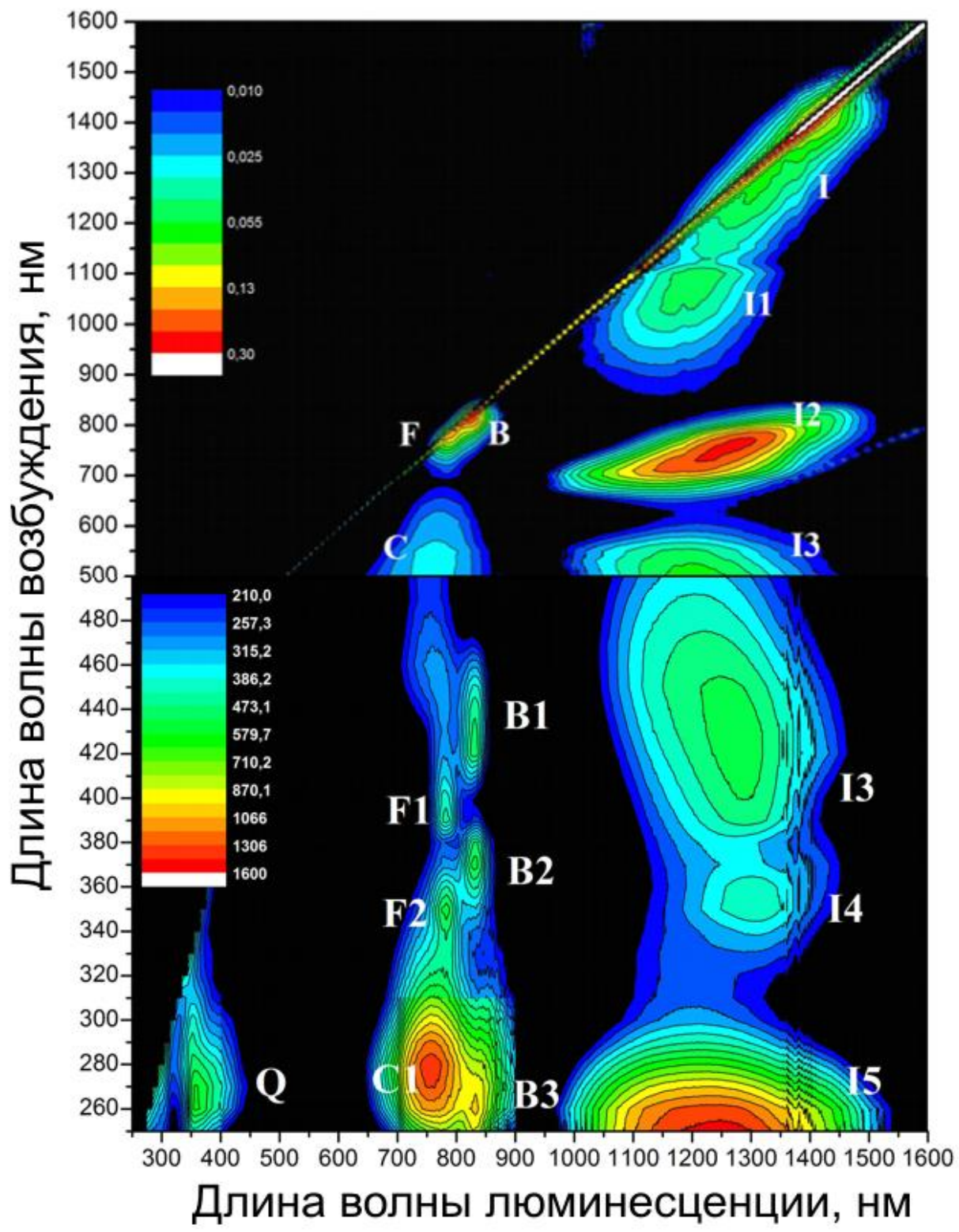

Рис. 3. 15 Трехмерный спектр возбуждения-эмиссии люминесценции PSBi образца при возбуждении в области длин волн от 250 до 1600 нм.

Как и в случае с ASBi образцом, исследовались люминесцентные свойства PSBi образца при возбуждении УФ излучением с длинами волн 240 нм $\leq \lambda_{ex} \leq$ 500 нм. Измеренный спектр возбуждения-эмиссии люминесценции PSBi приведен на Рис. 3. 15. С учетом полученных данных ИК люминесценция в области около 1250 - 1300 нм наблюдалась при возбуждении на длинах волн, начиная с ≈ 240 нм (серия пиков люминесценции I – I5). Помимо максимумов серии I, на графике присутствуют пики серии B-B3, которые по их положению могут быть отнесены к ВАЦ-Si. Люминесценция на 760 нм (пик C) соответствует двухвалентному висмуту. Пик C1($\lambda_{em}$=757 нм), лежащий на од-ной вертикали с С ($\lambda_{em}$=757 нм), по-видимому, также показывает люминесценцию двухвалентного висмута при возбуждении в УФ области. Наблюдается также люминесценция в УФ области (пик Q на $\lambda_{em}$=360 нм), положение которой по $\lambda_{ex}$ и $\lambda_{em}$, соответствует люминесценции на переходе $^3P_1 \rightarrow {}^1S_0$ иона $Bi^{3+}$, наблюдавшейся ранее в стеклах и кристаллах (см. Глава I данной работы). Наконец, наблюдается серия пиков люминесценции F-F2, которая по своему положению подобна серии B-B3 (возможно, что пик F3 – аналог B3 не наблюдается из-за присутствия яркого пика C1).

Значения $\lambda_{ex}$ и $\lambda_{em}$, соответствующие максимумам пиков люминесценции, наблюдаемых нами в PSBi образце, при возбуждении в области длин волн 240 – 500 нм, приведены в Табл. 3. 7.

Табл. 3. 7 Основные пики люминесценции, наблюдаемые в PSBi образце (Рис. 3. 15) при 240 нм $\leq \lambda_{ex} \leq$ 450 нм.

| Обозначение пика люминесценции | $\lambda_{ex}$, нм | $\lambda_{em}$, нм |
|---|---|---|
| I3 | 427 | 1277 |
| I4 | 353 | 1300 |
| I5 | ≈250 | ≈1250 |
| F1 | 390 | 783 |
| F2 | 350 | 783 |
| C1 | 278 | 757 |
| Q | 265 | 360 |

Если предположить, что структура ВАЦ-P в какой-то степени подобна структуре ВАЦ-Si, то, основываясь на положении пиков F-F2 по длинам волн возбуждения мы можем получить оценку энергий 2-ого, 3-ого и 4-ого возбужденных состояний ВАЦ-P. Положение 1-ого возбужденного состояния ВАЦ-P можно оценить как по положению полосы I (Рис. 3. 15), так и по спектральному диапазону длин волн генерации, полученному на ВАЦ-P [141].

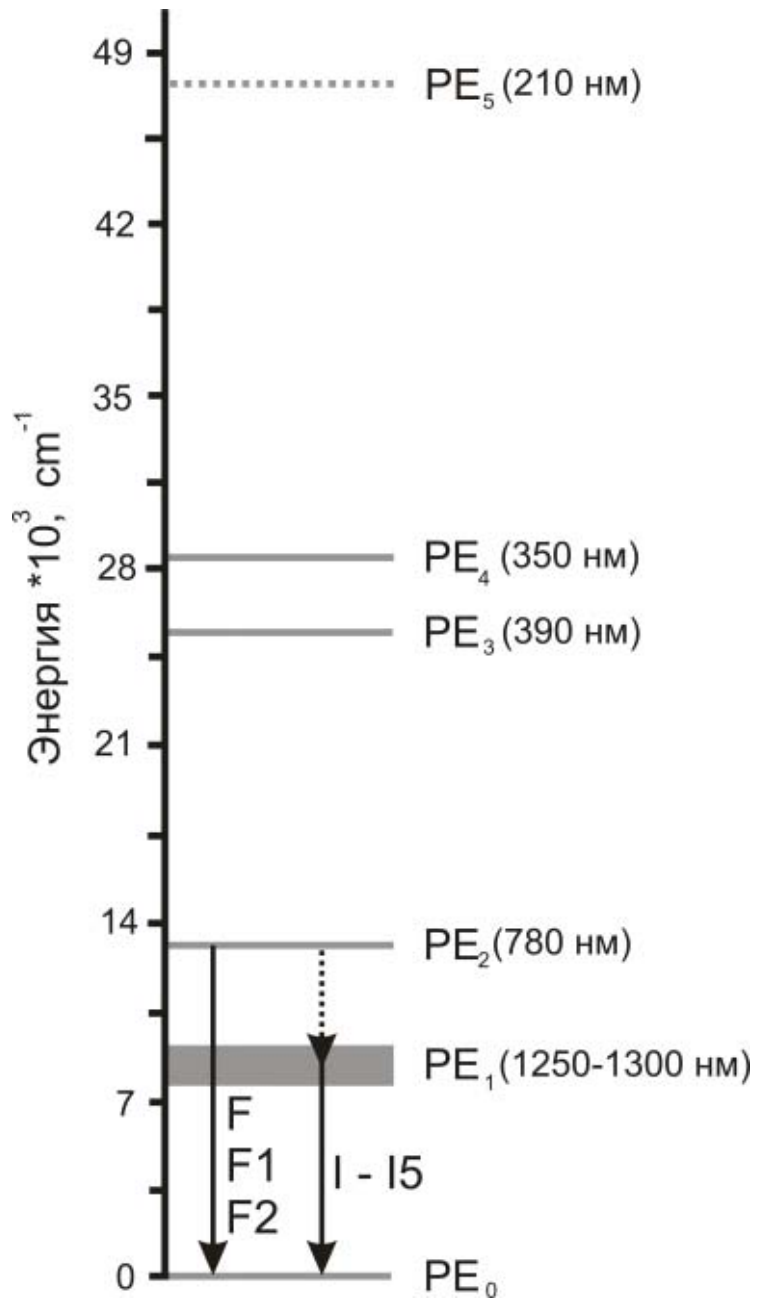

Рис. 3. 16 Схема энергетических уровней ВАЦ-Р. Сплошной линией со стрелкой обозначены излучательные переходы, наблюдаемые в данной работе, пунктирной – ненаблюдаемые. Уровень $PE_5$ изображен пунктирной линией, поскольку его достоверное положение в эксперименте не было установлено.

Информацию о положении 5-ого энергетического уровня ВАЦ-Р можно получить по длине волны возбуждения ИК полосы I5. В результате мы получаем схему уровней ВАЦ-Р, приведенную на Рис. 3. 16. Следует отметить, что в этой схеме, как и в схеме уровней ВАЦ-Si, выполняется соотношение $2PE_2 \approx PE_3$. Но, разумеется, данная схема в настоящее время может рассматриваться только как некоторое первое приближение для описания уровней ВАЦ-Р, на что указывает сложный спектр возбуждения ИК излучения в области 1 – 1,5 мкм.

# ГЛАВА IV. ОПТИЧЕСКИЕ СВОЙСТВА СВЕТОВОДОВ С СЕРДЦЕВИНОЙ ИЗ v-$GEO_2$, ЛЕГИРОВАННОГО ВИСМУТОМ [125, 126, 128, 129, 130, 142]

## 4.1 Абсорбционные свойства световода из германатного стекла, легированного висмутом

Структура германатного стекла, также как и кварцевого, преимущественно состоит из тетраэдров. Поэтому при легировании германатного стекла висмутом не исключено, что свойства такого стекла будут близки к свойствам стекла из чистого $SiO_2$ с висмутом. Таким образом, были исследованы волоконные световоды с сердцевиной из $GeO_2$, легированного висмутом, и оболочкой из кварцевого стекла. Из-за большого различия коэффициентов теплового расширения кварцевого и германатного стекол было практически невозможно получить пригодный для оптических измерений срез заготовки германатного световода – его сердцевина всегда растрескивалась при механической обработке среза. Поэтому мы исследовали только германатные световоды с висмутом. Концентрация висмута в них была также менее 0.02 ат.%.

На Рис. 4. 1 представлен спектр поглощения такого световода с сердцевиной из v-$GeO_2$, легированного висмутом. Для сравнения на том же графике приведены спектры поглощения для различных типов световодов. Полученный спектр поглощения GBi в отличие от SBi световода не имеет отдельных четких полос, и, как и все предыдущие рассмотренные типы висмутовых световодов, не позволяют сделать какие-либо определенные выводы об энергетических уровнях активных центров.

Однако можно выделить некоторые характерные для GBi образца области, в которых отсутствовало поглощение в других типах световодов: 950 нм и ≈1600-1650 нм. В видимом диапазоне наблюдается сложная полоса с максимумом около 450 нм и “провал” (минимум поглощения) около 425 нм (в SBi световоде аналогичный, по-видимому, минимум наблюдается на ≈400 нм).

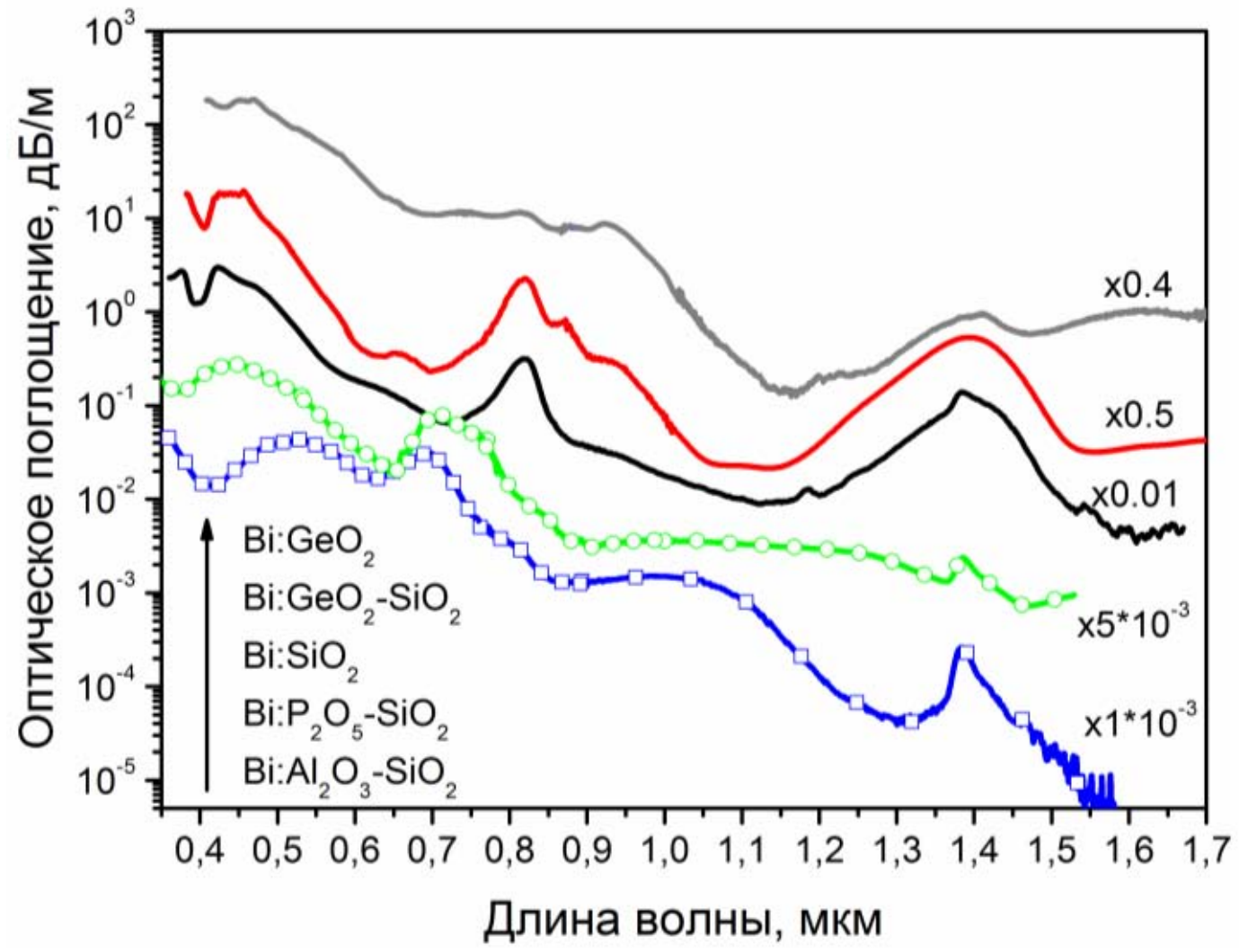

Рис. 4. 1 Спектры оптического поглощения ASBi, PSBi, SBi, GSBi и GBi световодов. Из-за сильного перекрытия исходных спектров поглощения, для наглядного представления соответствующие значения поглощения были умножены на коэффициенты, приведенные справа от каждой линии.

Аналогичные особенности наблюдались для образца с невысоким содержанием германия GSBi – германосиликатного световода ($5GeO_2$ – $\approx 95SiO_2$), легированного висмутом. Следует отметить, что спектр поглощения данного световода практически полностью повторял спектр поглощения SBi световода. Это свидетельствует о том, что несмотря на введение заметного количества оксида германия, в сердцевине световода происходит в основном формирование активных центров, ассоциированных с кремнием (ВАЦ-Si).

## 4.2 Люминесцентные свойства ВАЦ-Ge в УФ, видимом и ближнем ИК спектральных диапазонах

По сравнению с SBi световодом, GBi световод имеет более сложную картину люминесценции (см. Рис. 4. 2). Данное обстоятельство является несколько неожиданным, поскольку составы сердцевины этих световодов подобны и отличаются только заменой атомов кремния на атомы германия. В сердцевины этих световодов не вводились никакие дополнительные легирующие добавки. Но в спектре на Рис. 4. 2 а наблюдаются пики A, B, и B1, которые по длинам волн очень близки к одноименным пикам на Рис. 3. 2 а, хотя и

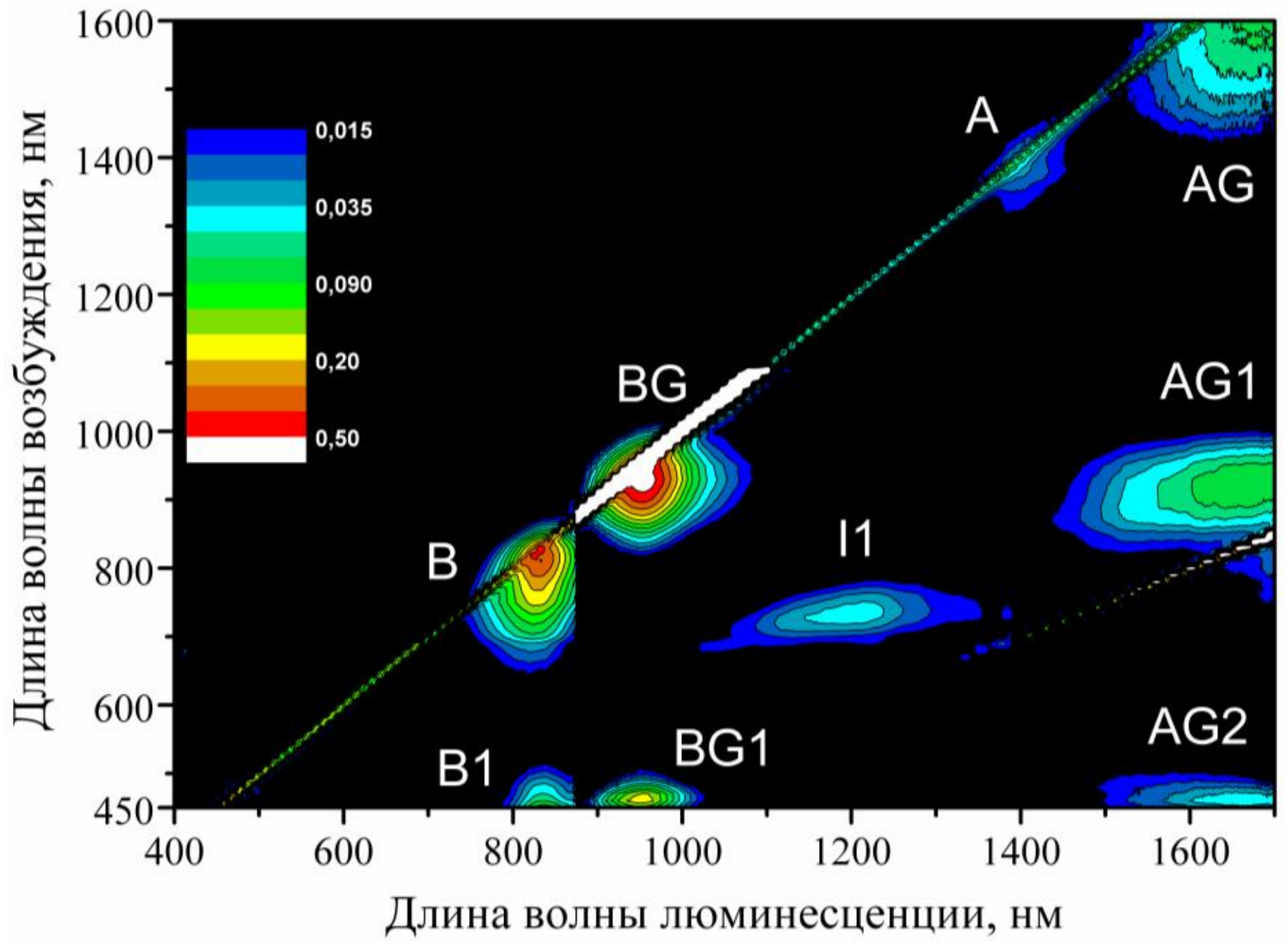

а

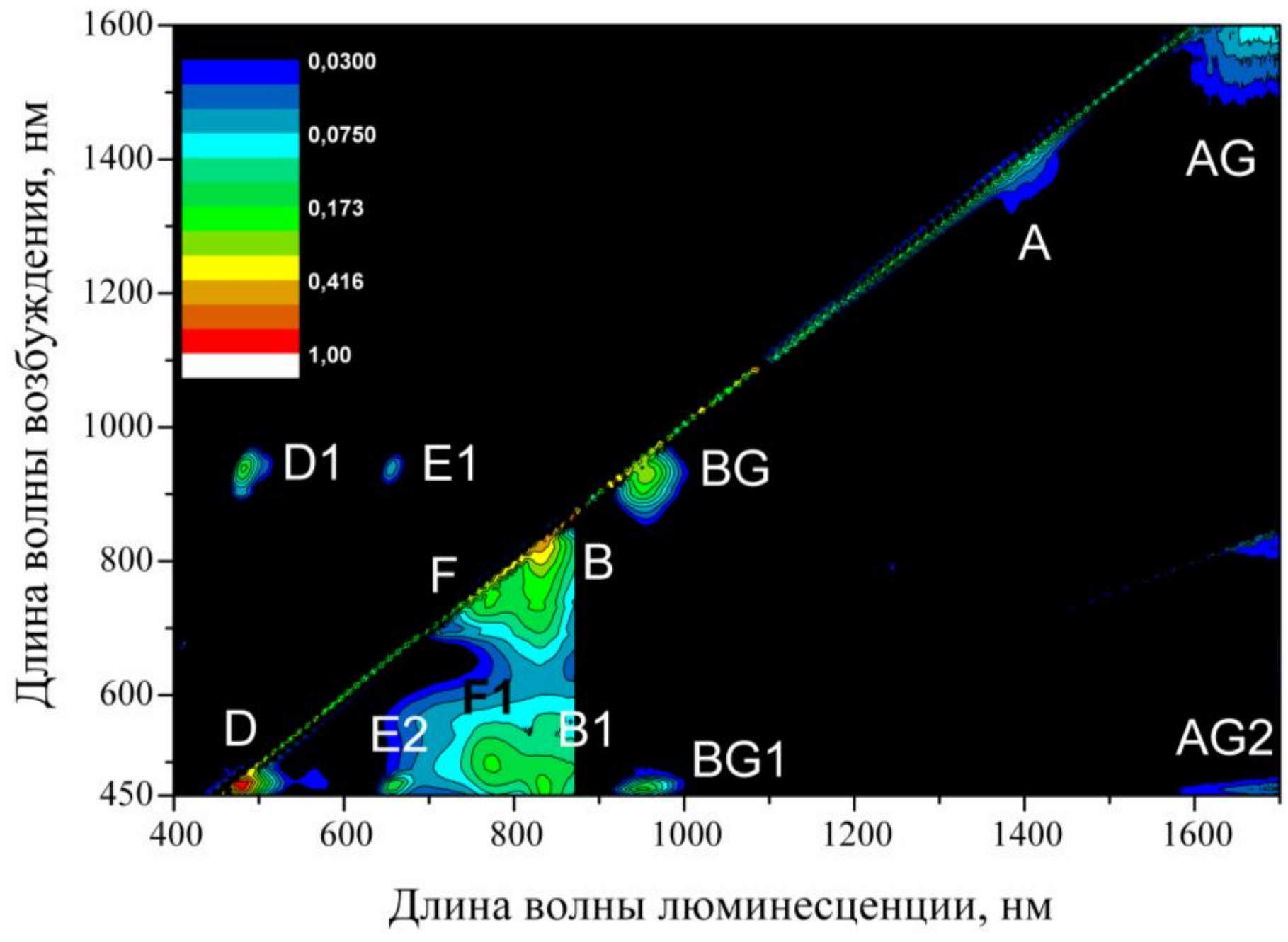

б

Рис. 4. 2 Трехмерные спектры возбуждения-эмиссии люминесценции GBi световода при комнатной температуре (а) и при температуре жидкого азота (б) сравнительно менее яркие. Учитывая, что сердцевина GBi световода граничит со стеклом отражающей оболочки, состоящей в основном из $SiO_2$, и 1) поле моды волоконного световода проникает в эту оболочку, 2) атомы висмута могут диффундировать в области около границы сердцевины, состоящие в основном из $SiO_2$, то вполне вероятно, что данные пики (пики люминесценции A, B, B1) обусловлены присутствием в GBi световоде ВАЦ, ассоциированных с кремнием. Следует также отметить, что положение максимума люминесценции I1 для GBi стекла и PBSi стекла подобны (см. Рис. 3. 15). Вероятно, что появление этого максимума связано именно с присутствием фосфора, который вводился в дополнительный слой в процессе изготовления (см. Глава 2).

Табл. 4. 1 Основные максимумы люминесценции ВАЦ GBi световода: обозначения, длины волн возбуждения и эмиссии при T=300 K и 77 K

| Обозначение пика | $\lambda_{ex}^{max}$, нм | | $\lambda_{em}^{max}$, нм | |
|---|---|---|---|---|
| | 300 K | 77 K | 300 K | 77 K |
| A | 1390 | 1390 | 1410 | 1435 |
| B | 815 | 820 | 830 | 835 |
| B1 | 450 | 450 | 830 | 835 |
| AG | >1600 | >1600 | ≈1670 | ≈1670 |
| AG1 | 925 | — | 1670 | — |
| AG2 | 463 | 463 | 1670 | 1670 |
| BG | 925 | 925 | 955 | 955 |
| BG1 | 463 | 460 | 955 | 955 |
| D | — | 463 | — | 480 |
| D1 | — | 940 | — | 482 |
| E1 | — | 940 | — | 655 |
| E2 | — | 463 | — | 655 |
| F | — | 750 | — | 775 |
| F1 | — | 500 | — | 775 |
| I1 | 740 | — | 1190 | — |

Для GBi световода характерно появление пиков AG, AG1, AG2, BG и BG1 (Табл. 4. 1), которые расположены аналогично пикам A, A1, A2, B и B1 на спектре SBi световода (Рис. 3. 2 а), но с небольшим смещением в длинноволно-вую область, как по длинам волн возбуждения, так и по длинам волн эмиссии. Исходя из подобия структуры стекол из v-$SiO_2$ и v-$GeO_2$, эти пики можно отнести к ВАЦ, ассоциированных с германием (ВАЦ-Ge).

При измерении люминесценции при T=77 K в спектрах GBi световода появляется дополнительно ряд узких максимумов, расположенных в видимом спектральном диапазоне (максимумы D, D1, E1, E2, F1, F2, см. Рис. 4. 2 б). Визуально можно наблюдать яркое синее свечение GBi световода при вводе в сердцевину излучения накачки на длине волны 925 нм мощностью около 1 мВт. В действительности, при возбуждении на длинах волн около 925 нм наблюдается две полосы антистоксовой люминесценции: синяя D1(940, 482) и более слабая красная E1 (940, 655). Такая же пара полос люминесценции возникает при возбуждении на длине волны 460 нм (пики D и E2). Рядом с пиком люминесценции B при T=300 K появляется новый пик F, также с малой величиной стоксова сдвига. Кроме того, люминесценция с максимумом на той же длине волны, что и пик F, возникает при возбуждении на 500 нм. Пики F, F1 также, по-видимому, относятся к ВАЦ-P (Рис. 3. 15).

Возникновение остальных новых линий люминесценции GBi световодов при низкотемпературных измерениях по сравнению с измерениями при комнатной температуре может быть объяснено как результат уменьшения вероятности безызлучательной релаксации возбужденных энергетических уровней ВАЦ с уменьшением температуры (снижение влияния температурного тушения). По результатам проведенных экспериментов была построена схема ВАЦ-Ge, состоящая из 3 энергетических уровней (Рис. 4. 3). На диаграмме приведены излучательные переходы, в результате которых наблюдалась стоксовая и антистоксовая люминесценция.

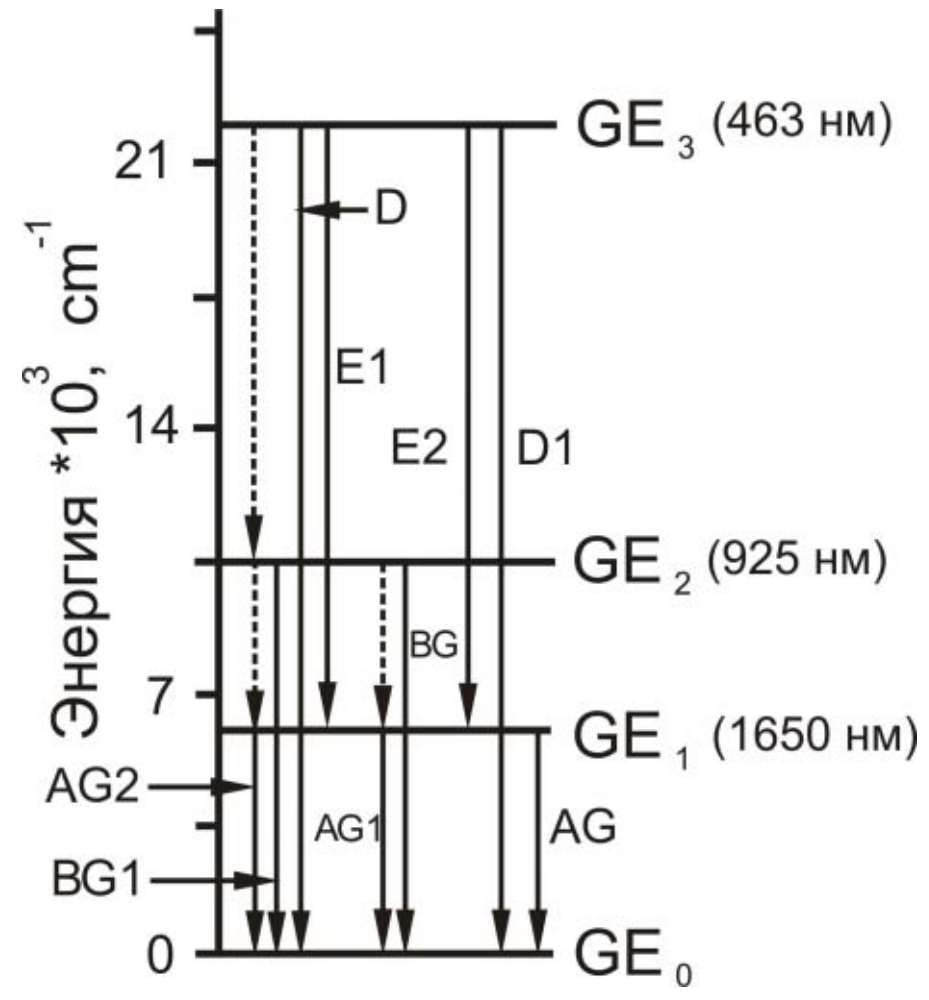

Рис. 4. 3 Схема низколежащих энергетических уровней ВАЦ-Ge. На схемах сплошными линиями со стрелками, направленными вниз, обозначены наблюдаемые излучательные переходы при T=300 и 77 К. Пунктиром обозначены переходы, люминесценция которых не наблюдалась.

Данные по люминесценции GBi световодов при T=77 К находятся в полном соответствии со схемой энергетических уровней ВАЦ-Ge, подобной схеме энергетических уровней ВАЦ-Si. В частности, серия максимумов D появляется в результате перехода $GE_3 \rightarrow GE_0$ в ВАЦ-Ge, а серия максимумов E – $GE_3 \rightarrow GE_1$ (Рис. 4. 3).

Таким образом, в GBi световоде обнаружено присутствие одновременно как ВАЦ-Ge, так и ВАЦ-Si, аналогично тому, как в алюмосиликатном волоконном световоде, легированном висмутом, были обнаружены как висмутовые активные центры, ассоциированные с кремнием, так и ассоциированные с алюминием [143]. Ранее в Главе 3 уже было показано сосуществование нескольких типов активных центров, на примере PSBi стекла, в котором присутствуют два типа ВАЦ (ВАЦ-Si и ВАЦ-P).

Известно, что возбуждение УФ излучением германосиликатного стекла без висмута часто приводит к появлению интенсивной синей люминесценции, что является следствием присутствия в стекле германиевых кислороднодефицитных центров (ГКДЦ). Синее свечение таких стекол легко наблюдается невооруженным глазом. Спектр люминесценции $I_{lum}(\lambda_{em}, \lambda_{ex})$ образца, по составу повторяющего GBi из Табл. 2. 1, но без висмута, представлен на Рис. 4. 4 а.

Люминесценция ГКДЦ характеризуется тремя интенсивными (примерно на порядок более яркими, чем люминесценция ККДЦ на Рис. 3. 3) максимумами T (333 нм, 384 нм), T1 (242 нм, 382 нм), T2 (255 нм, 286 нм). Структура и оптические свойства ГКДЦ изучены достаточно подробно во многих работах, например [133, 144]. Известно, что пики люминесценции T и T1 обусловлены триплет-синглетными переходами, а T2 – синглет-синглетным переходом ГКДЦ. В наших экспериментах интенсивность люминесценции в пиках T2 и T1 была сравнима и примерно в три раза превосходила интенсивность люминесценции в пике T.

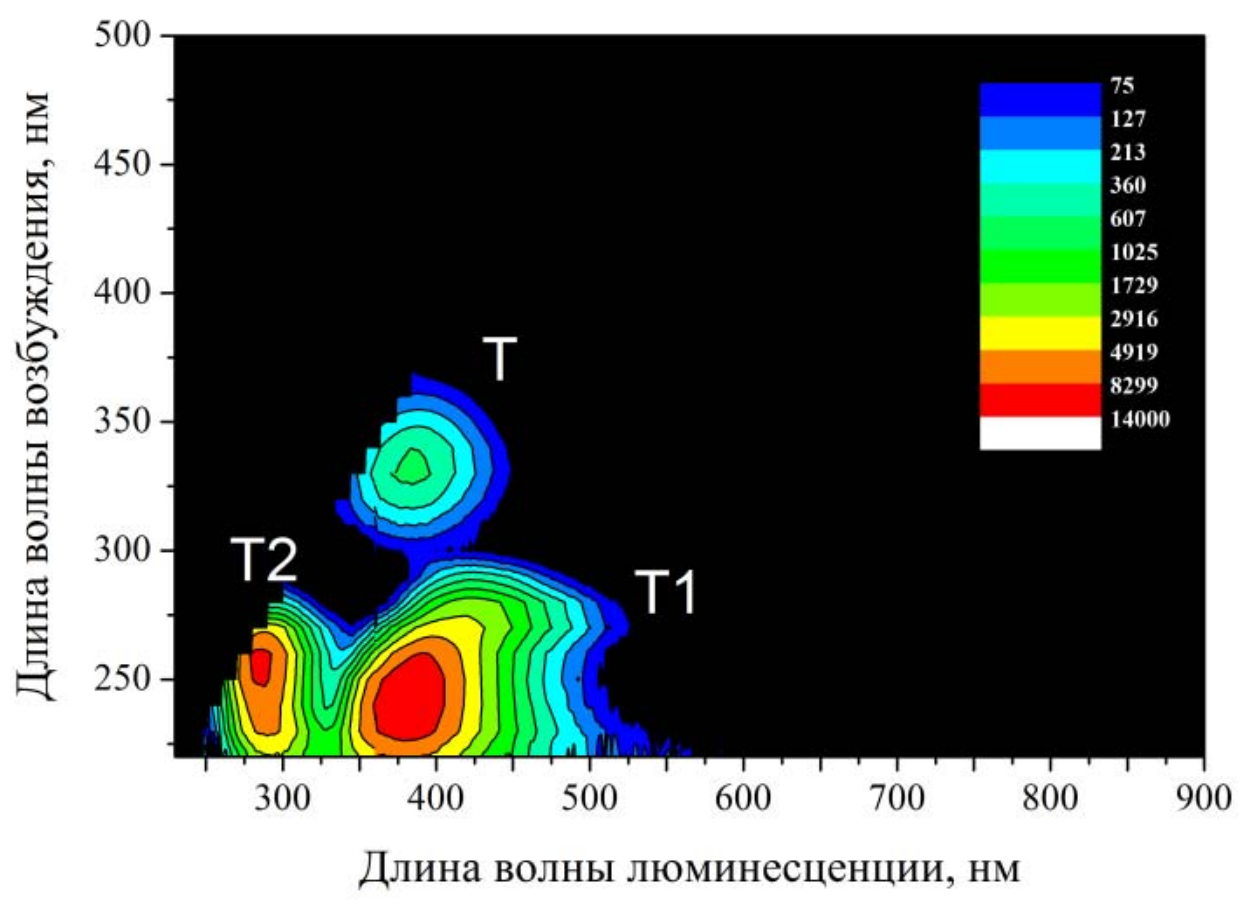

а

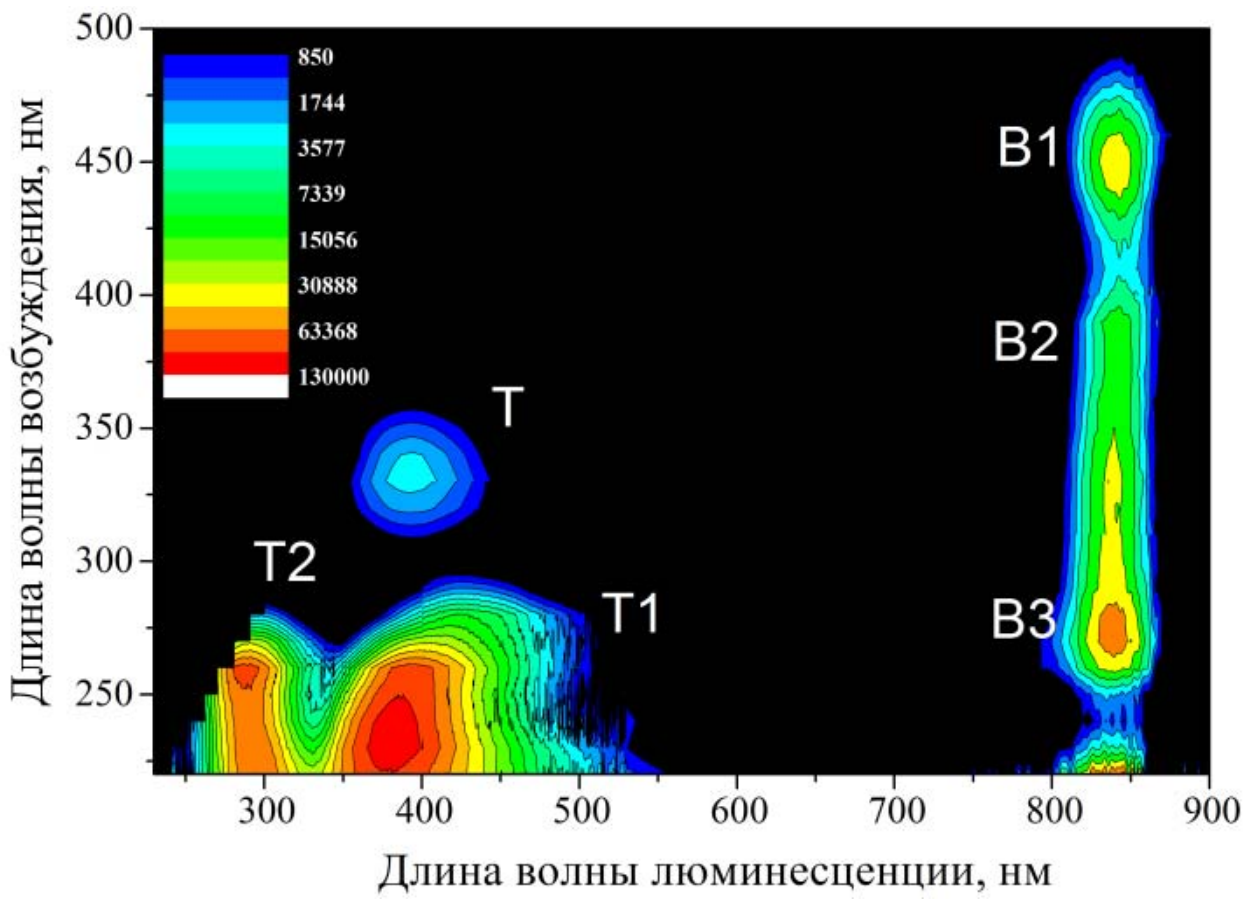

б

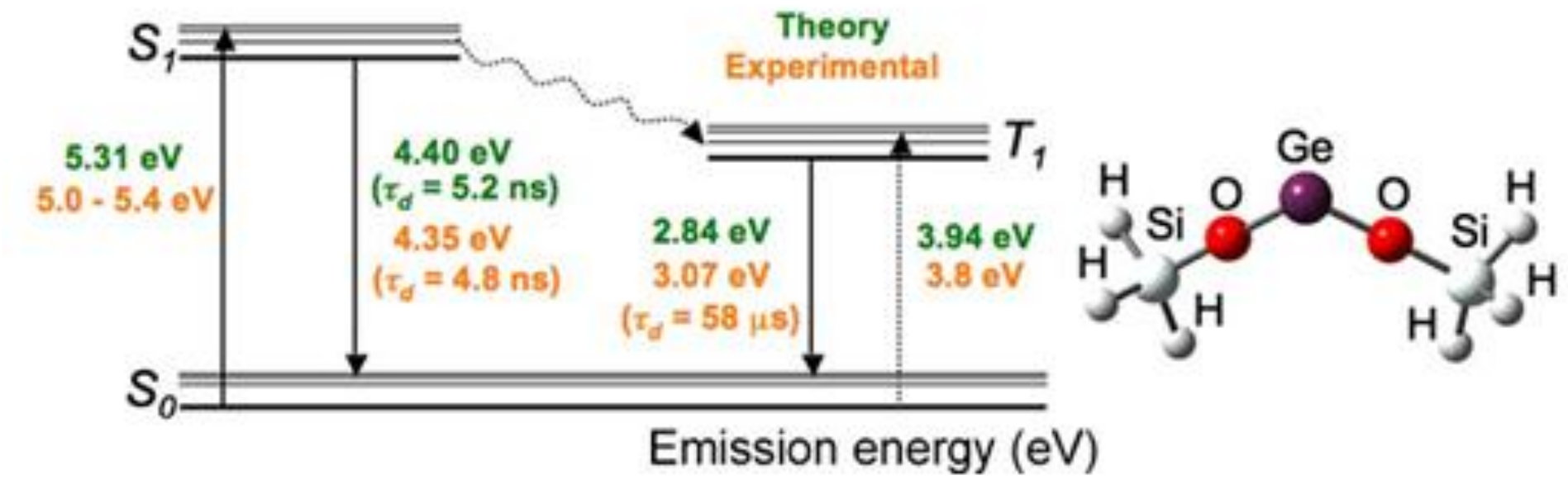

в

Рис. 4. 4 Трехмерные спектры возбуждения-эмиссии: а) германосиликатного стекла без висмута; б) GBi стекла; в) модель структуры германиевого кислородно-дефицитного центра (ГКДЦ) и схема энергетических уровней с основными переходами [145].

Выше было показано сосуществование ВАЦ-Si и ВАЦ-Ge в GBi образце, что позволяло проводить сравнение положений основных максимумов люминесценции указанных ВАЦ. Однако из-за малого содержания ВАЦ-Si в герма-натном стекле не удавалось наблюдать некоторые характерные максимумы люминесценции ВАЦ-Si, которые хорошо регистрируются в германосиликатном (GSBi) образце. Поэтому для сравнения положений основных максимумов люминесценции ВАЦ-Si и ВАЦ-Ge исследовались люминесцентные свойства GSBi образца.

Спектр люминесценции $I_{lum}(\lambda_{em}, \lambda_{ex})$ легированного висмутом GSBi стекла представлен на Рис. 4. 5. Он содержит гораздо большее число полос люминесценции, чем SBi (Рис. 3. 4).

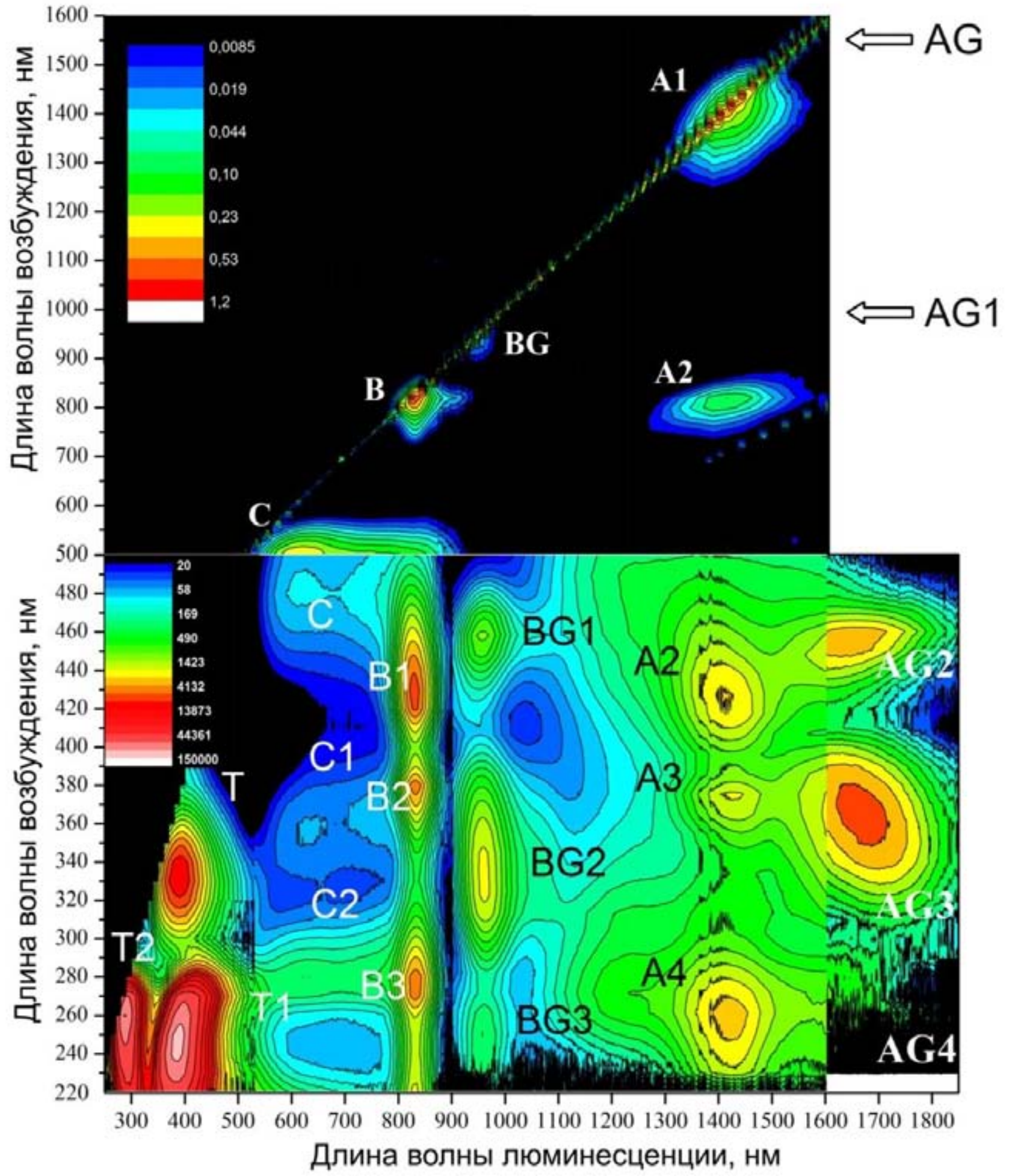

Рис. 4. 5 Трехмерный спектр возбуждения-эмиссии люминесценции GSBi образца. Стрелками указаны положения 2 максимумов (AG и AG1), которые не наблюдались в эксперименте из-за низкой интенсивности люминесценции.

1) Здесь присутствуют все пики люминесценции ВАЦ-Si (серии A и B). Это естественно, поскольку сердцевина заготовки GSBi, и полученного из нее световода, содержит большое количество $SiO_2$.

2) Выше было показано, что в световоде с германатной (состоящей в основном из $GeO_2$) сердцевиной наблюдается система пиков люминесценции висмутовых активных центров, ассоциированных с германием (ВАЦ-Ge), подобная системе пиков ВАЦ-Si. Но только некоторые из них наблюдаются на Рис. 4. 5 – BG, BG1, AG2. По-видимому, из-за сравнительно небольшой концентрации $GeO_2$ в нашем световоде интенсивность пиков AG, AG1 (их положение обозначено на Рис. 4. 5 по данным Рис. 4. 2 а) была недостаточной для их регистрации на фоне более яркой люминесценции. Но на Рис. 4. 5 мы видим и не наблюдавшиеся ранее пики люминесценции BG2, BG3, AG3 и AG4, которые, судя по положению полос эмиссии люминесценции, также принадлежат ВАЦ-Ge. Следует отметить, что наличие мощной линии поглощения ГКДЦ в GSBi световоде [144] (и их синяя люминесценция) может приводить к смещению наблюдаемого положения пиков люминесценции по оси $\lambda_{ex}$. В частности, наблюдается смещение $\lambda_{ex}$ максимума B3 для световода GSBi на 30 нм в длинноволновую сторону по сравнению с аналогичным пиком B3 в чисто кварцевом световоде SBi (с ≈240 нм для SBi до ≈270 нм для GSBi).

3) На спектре люминесценции GSBi световода наблюдается серия пиков красной люминесценции C (C, C1 и C2), как и для SBi световода, которые относятся к люминесценции двухвалентного висмута. Только в случае GSBi их относительная интенсивность существенно ниже. Поэтому можно предположить, что при увеличении концентрации оксида германия концентрация двухвалентного висмута снижается. Тем более, что в световоде с чисто германатной сердцевиной красная люминесценция не регистрировалась вообще (Рис. 4. 4 б). На основании данных Рис. 4. 5 можно примерно определить положение еще двух энергетических уровней центра ВАЦ-Ge, но с существенно меньшей точностью из-за влияния полос ГКДЦ. Полученная схема уровней ВАЦ-Ge приведена на Рис. 4. 6 (пунктиром обозначены уровни, положение которых определено с меньшей по сравнению с другими точностью из-за положения ГКДЦ). Если обычная точность определения положения уровня в наших экспериментах, как уже указывалось, определяется фактически шагом по $\lambda_{ex}$ получения спектров люминесценции (10 нм), то для уровней $GE_4$ и $GE_5$ (Рис. 4. 6) ошибка может быть в несколько раз больше.

4) На Рис. 4. 5 также наблюдаются три пика (серия T), соответствующих люминесценции ГКДЦ (как на Рис. 4. 4 б).

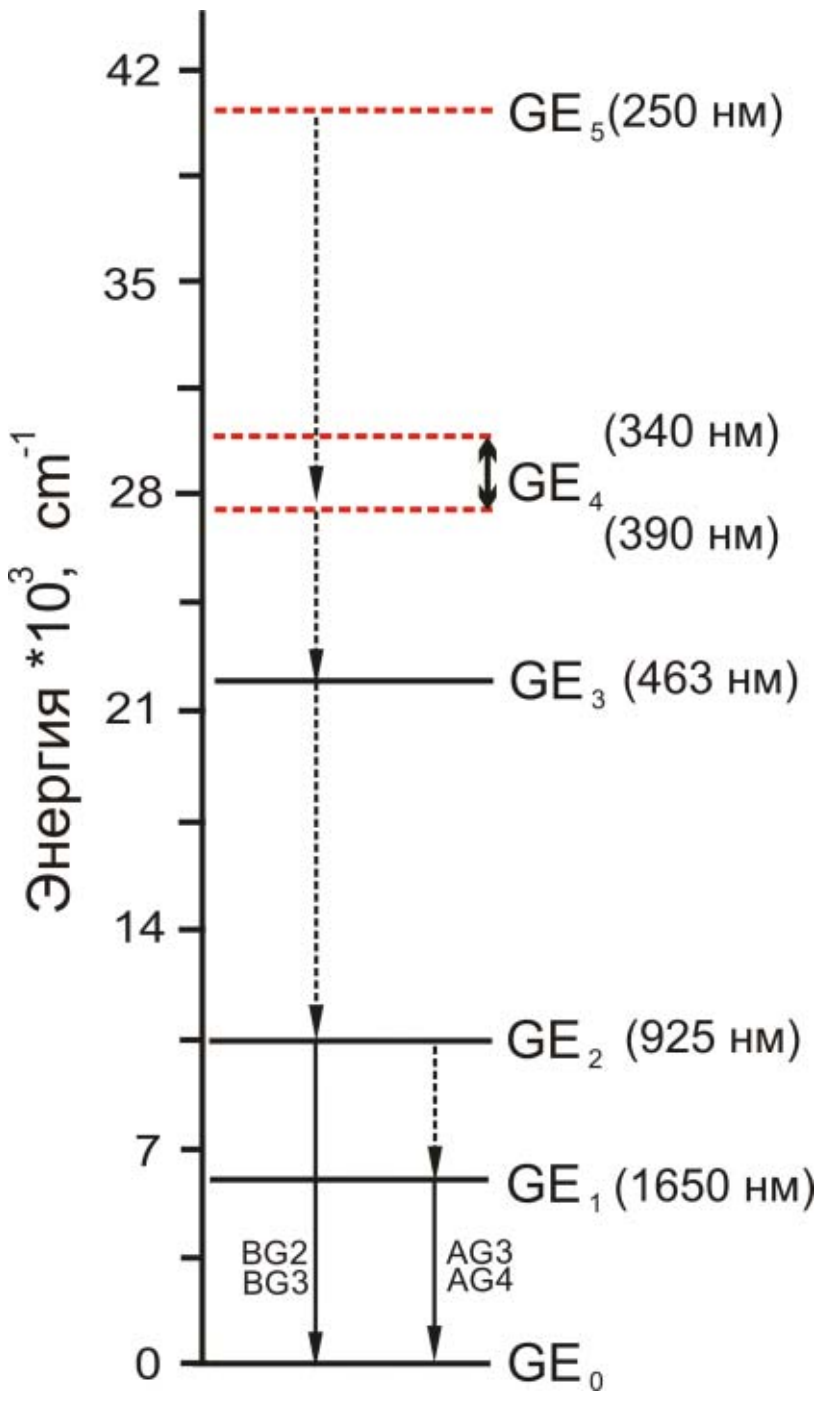

Рис. 4. 6 Схема энергетических уровней ВАЦ-Ge. Сплошные линии со стрелкой указывают оптические переходы, в результате которых наблюдается люминесценция при возбуждении УФ излучением, штриховые линии со стрелкой – ненаблюдаемые переходы. Штриховыми линиями обозначены уровни, положение которых определено неоднозначно.

Значения $\lambda_{ex}$ и $\lambda_{em}$, соответствующие максимумам пиков люминесценции, наблюдаемых нами в GSBi образце, при возбуждении в области длин волн 240 – 500 нм приведены в Табл. 4. 2. Положение пиков люминесценции образца при $\lambda_{ex} \geq 450$ нм было приведено в Табл. 4. 1.

Табл. 4. 2 Основные пики люминесценции, характерные для GSBi световода (Рис. 4. 5) при 240 нм ≤ $\lambda_{ex}$ ≤ 500 нм.

| Обозначение пика люминесценции | $\lambda_{ex}$, нм | $\lambda_{em}$, нм |
|---|---|---|
| AG2 | 455 | 1646 |
| AG3 | 365 | 1665 |
| BG1 | 460 | 956 |
| BG2 | 333 | 956 |
| BG3 | ≈247 | 956 |
| T | 333 | 387 |
| T1 | 244 | 387 |
| T2 | 260 | 290 |

Интересно отметить, что в положении уровней как ВАЦ-Si, так и ВАЦ-Ge выполняется соотношение $GE_2$-$GE_0$ ≈ $GE_3$-$GE_2$, то есть третий возбужденный уровень расположен приблизительно вдвое выше, чем второй [146].

Таким образом, получены данные о люминесцентных свойствах ВАЦ, ассоциированных с германием. Определены положения новых энергетических уровней таких центров. Из построенной схемы уровней для ВАЦ-Ge на основании экспериментально полученных данных следует, что каждый энергетический уровень ВАЦ-Ge расположен ниже соответствующего уровня, принадлежащего ВАЦ-Si. В УФ области данное правило выполняется не для всего набора новых уровней. Один из уровней (исходя из положения пиков BG2 и AG3) выше, чем аналогичный уровень ВАЦ-Si. Вероятно, что искажение энергии возбуждения непосредственно связано с влиянием со стороны ГКДЦ.

## 4.3 Схема энергетических уровней, принадлежащих ВАЦ в световоде с сердцевиной из v-$GeO_2$

Полученные результаты в разделе 4. 2, по аналогии с SBi световодом, позволяют построить схему энергетических уровней для ВАЦ в GBi световоде.

Данная схема уровней с учетом новых (расположенных в УФ области) приведена на Рис. 4. 6. Новые энергетические уровни на графике обозначены штриховыми линиями, поскольку их расположение может быть искажено по вышеуказанным причинам. Кроме того, положение уровня $GE_4$ может отличаться в разных образцах. На данной диаграмме приводятся только переходы, которые могут быть в такой системе при возбуждении УФ излучением на $GE_4$ и $GE_5$.

Для ВАЦ-Ge были измерены временные зависимости затухания люминесценции (переходы $GE_2 \rightarrow GE_0$ и $GE_1 \rightarrow GE_0$). Оказалось, что кривые затухания люминесценции для германатного стекла не могут быть описаны одной экспонентой. В частности для полосы 1650 нм присуще 2 компоненты с временами 30 и 500 мкс, а для 950 нм – несколько компонент с временами от единиц мкс до 70 мкс.

Предполагается, что имеющиеся различия, во-первых, могут быть связаны с наличием дополнительных каналов возникновения люминесценции между возбужденными уровнями. В частности, $\lambda_{em}$=950 нм может появляться в результате перехода $GE_2 \rightarrow GE_0$, а также $GE_3 \rightarrow GE_2$. Кроме того, к наблюдаемым различиям могут быть причастны процессы поглощения квантов энергии накачки ВАЦ в возбужденном состоянии, которые возможны для данного типа световодов. Второй причиной может быть то, что структура германатного стекла не повторяет полностью структуру кварцевого стекла, а имеет некоторые особенности, в частности, координационное число германия может быть как 4, так и 6. Результатом этого может быть частичное изменение оптических свойств ВАЦ-Ge.

Важно отметить, что построенная схема энергетических уровней принадлежит ВАЦ-Ge, на которых уже ранее была получена лазерная генерация на переходе $GE_1 \rightarrow GE_0$ с накачкой на уровень $GE_2$ [118]. Лазерная генерация была получена на длине волны 1550 нм.

# ГЛАВА V. УФ ПОГЛОЩЕНИЕ И ЛЮМИНЕСЦЕНЦИЯ В ВИСМУТОВЫХ СВЕТОВОДАХ ПРИ УФ И ПОСЛЕДОВАТЕЛЬНОМ ДВУХКВАНТОВОМ ИК ВОЗБУЖДЕНИИ [128, 130, 146, 147]

## 5.1 УФ поглощение в световодах, легированных висмутом.

Известно, что стекла, легированные висмутом, имеют полосу поглощения в УФ области на длине волны в окрестности 250 нм, причем точное положение полосы зависит от состава стекла [75]. Эта полоса объясняется поглощением излучения ионами $Bi^{3+}$ на переходе ${}^1S_0 \rightarrow {}^3P_1$ [148]. В данной работе были измерены положения максимумов соответствующих УФ полос поглощения в стеклах различного состава, а именно, SBi, ASBi, PSBi. На вставке Рис. 5. 1 приведены спектры поглощения заготовок в области 190 – 350 нм, измеренные со спектральным разрешением 1 нм. Поглощение в SBi стекле составляло порядка $10^4$ дБ/м. Стекло без висмута не имело интенсивных полос (~$10^4$ дБ/м) в УФ области (Рис. 5. 1 (вставка)). Введение оксида германия в чисто кварцевое стекло (даже без висмута) приводило к поглощению на длинах волн около 240 нм, по уровню которое значительно превосходило поглощение в SBi образце. Появление столь высокого уровня поглощения объясняется присутствием германиевых кислородно-дефицитных центров (ГКДЦ) [144]. Для иллюстрации этого факта на Рис. 5. 1 приведен спектр поглощения германосиликатной заготовки (без висмута) с содержанием $GeO_2$ всего 0.8 мол.%. В этом случае наблюдался пик поглощения ГКДЦ на длине волны 242 нм и амплитудой ~$10^5$ дБ/м. Очевидно, что в GBi и GSBi заготовках поглощение ГКДЦ препятствует наблюдению полос поглощения, обусловленных присутствием висмута.

Спектры поглощения в УФ области алюмосиликатного и фосфоросиликатного стекол с висмутом мало отличались от того, что наблюдалось для чисто кварцевого стекла, содержащего висмут. В обоих случаях возникала полоса поглощения в области 220 – 240 нм, обусловленная присутствием висмута. Спектральное положение максимума и ширина полосы поглощения в области 230 нм зависели от состава стекла: SBi – 227 нм (FWHM – 24,5 нм); ASBi – 238 нм (34 нм) и PSBi – 234 нм (41 нм). Следует отметить, что данная полоса распола-галась на склоне другой более интенсивной полосы поглощения, максимум который лежал в области длин волн короче 190 нм.

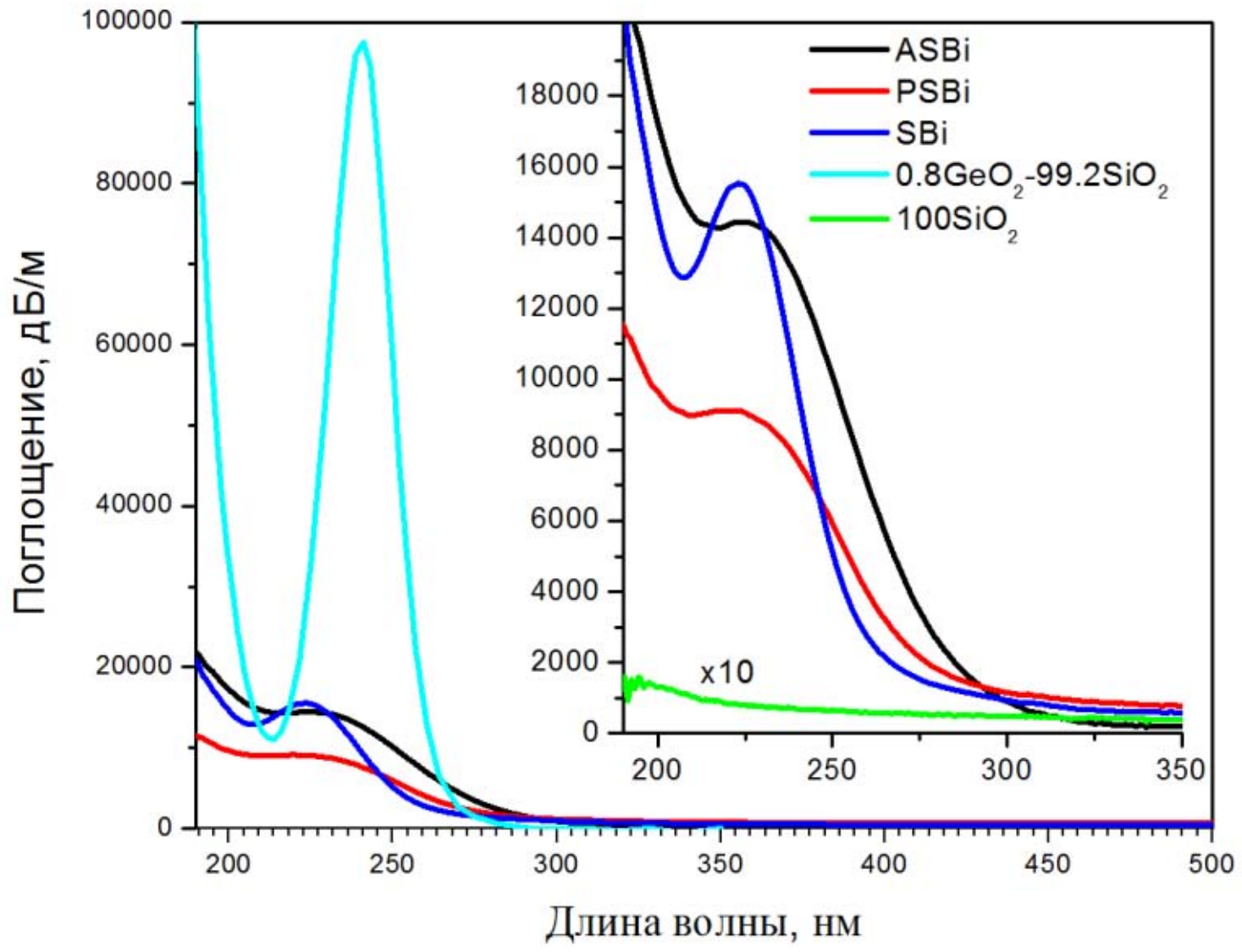

Рис. 5. 1 Спектры поглощения в УФ области отрезков заготовок волоконных световодов. На вставке детально показаны полосы поглощения ВАЦ.

Представляло интерес проведение сравнительного анализа спектров поглощения образцов в УФ области и спектров возбуждения ИК люминесценции. Трехмерные спектры возбуждения-эмиссии люминесценции (Рис. 3. 4, Рис. 3. 13, Рис. 3. 15) не содержат такой информации, поскольку диапазон длин волн возбуждения был ограничен 240 нм. Из-за низкой интенсивности источника излучения возбуждения в диапазоне 200 – 240 нм возникала необходимость значительного увеличения времени накопления сигнала в каждой точке. Поэтому измерения спектров возбуждения ИК люминесценции в спектральном диапазоне 200-240 нм проводились со временем накопления сигнала, которое на ≈ 2 порядка превосходило время накопления для результатов, представленных на трехмерных графиках в Главах III и IV. Результаты измерений для образцов SBi, PSBi и ASBi представлены на Рис. 5. 2. Для сравнения здесь же показаны спектры поглощения этих образцов. Следует отметить, что для SBi образца спектры возбуждения люминесценции на 1410 нм и 830 нм мало отличаются от спектра поглощения (Рис. 5. 2 а). В случае PSBi образца (Рис. 5. 2 б) спектр возбуждения имеет более сложную структуру, чем спектр поглощения, но в среднем по-добен ему. Интенсивность полосы возбуждения в области 200 – 250 нм в SBi и PSBi образцах существенно превышает интенсивность остальных полос, наблюдаемых в приведенных на Рис. 5. 2 а и б спектрах (375 и 420 нм для SBi и 440 нм для PSBi). В случае же ASBi образца полоса возбуждения люминесценции в УФ области существенно ниже полос на 500 и 700 нм. Следовательно, можно сделать вывод, что интенсивные полосы возбуждения ИК люминесценции в УФ области в значительной степени совпадают с полосами поглощения ионов $Bi^{3+}$ для образцов SBi и PSBi. В случае алюмосиликатного образца такое соотношение не имеет места.

Приведенные выше факты не позволяют исключать участие ионов $Bi^{3+}$в формировании ВАЦ по меньшей мере в кварцевом и фосфоросиликатном стекле.

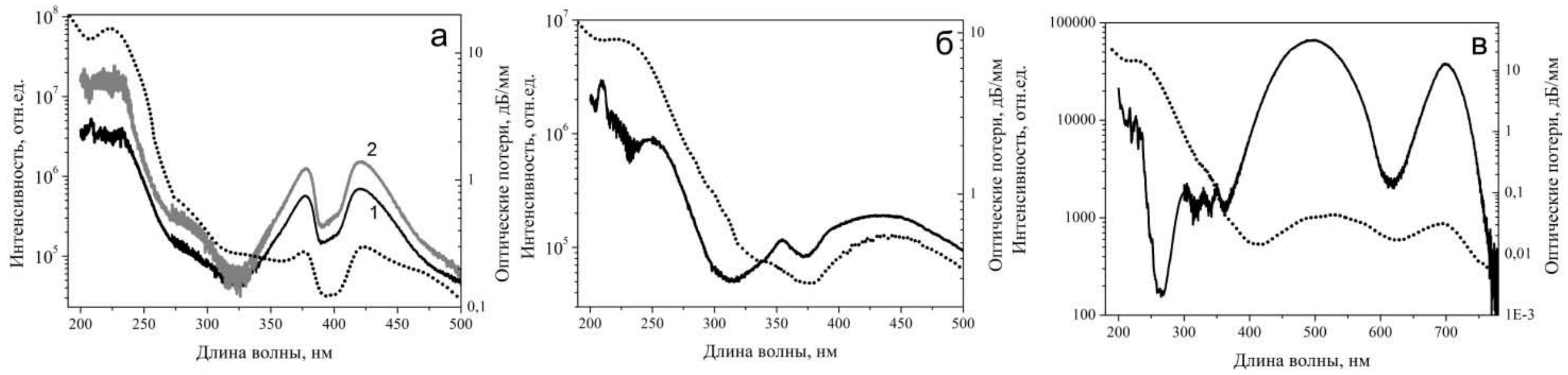

Рис. 5. 2 Спектры возбуждения ИК люминесценции (сплошная линия) в SBi(а, 1-$\lambda_{em}$=1410 нм, 2-$\lambda_{em}$=830 нм), PSBi(б, $\lambda_{em}$=1250 нм) и ASBi(в, $\lambda_{em}$=1150 нм). На каждом графике представлены спектры оптических потерь для тех же образцов (пунктирная линия).

Отметим, что возможное участие $Bi^{3+}$ в формировании ВАЦ уже рассматривалось ранее в [85] (где $Bi^{3+}$ рассматривался в виде составляющей димера $Bi_2^{5+}$) и в [149] (в виде составляющей димера $Bi^{3+}$+$Bi^{2+}$, разделенного анионной вакансией).

## 5.2 Антистоксовая люминесценция ВАЦ в v-$GeO_2$ и v-$SiO_2$ световодах при ступенчатом двухквантовом возбуждении

Результаты по измерениям спектров возбуждения-эмиссии люминесценции ВАЦ в SBi и GBi световодах (см. Главы III и IV) позволили определить положения энергетических уровней и энергии переходов между ними (схемы энергетических уровней). Для обоих типов световодов ВАЦ обладают подобными по структуре схемами энергетических уровней. Различие заключается только в том, что положение уровней для GBi световода расположены несколько ниже. Поэтому в дальнейшем при изложении материала в этом разделе будут использоваться одинаковые обозначения ($E_0$, …, $E_3$) для уровней ВАЦ в обоих световодах.

В настоящем разделе представлены экспериментальные данные, позволяющие независимо проверить схемы энергетических уровней ВАЦ-Si и ВАЦ-Ge (Рис. 5. 3). Суть выполненных экспериментов заключается в регистрации антистоксовой люминесценции при совпадении энергий возбуждающего излучения с расстояниями между соответствующими энергетическими уровнями (схема эксперимента подробно описана в Главе II).

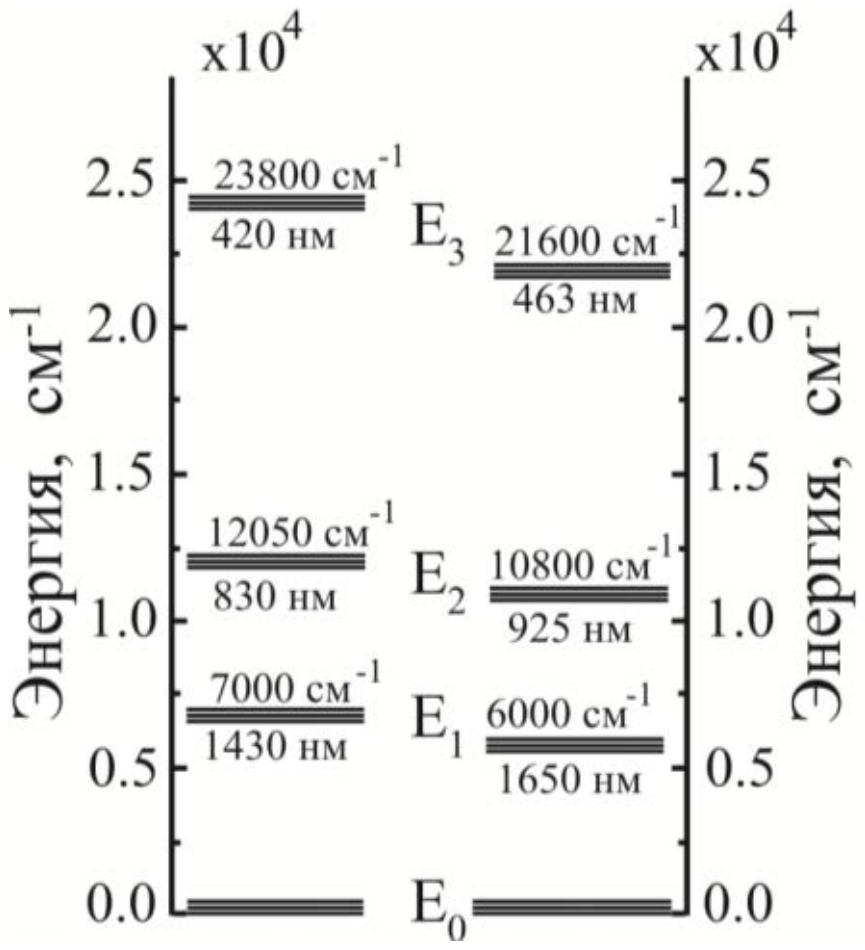

Рис. 5. 3 Схемы энергетических уровней ВАЦ-Si (слева) и ВАЦ-Ge (справа). Около каждой линии, обозначающей энергетический уровень указана длина волны и энергия излучения, соответствующие переходам с данного уровня в основное состояние.

Для этого в эксперименте использовалось излучение с двумя различными длинами волн $\lambda_{ex1}$ и $\lambda_{ex2}$. Длина волны первого возбуждающего излучения ($\lambda_{ex1}$), для SBi световода выбиралась равной ≈1350 и ≈1900 нм, а для GBi – ≈1650 и ≈2000 нм, соответственно. Длина волны второго возбуждающего излучения ($\lambda_{ex2}$) изменялась в широком диапазоне. При этом схема регистрации излучения люминесценции настраивалась на длину волны перехода со второго возбужденного уровня в основное состояние (для SBi - это 830 нм, а для GBi – 950 нм). Таким образом, снимались спектры возбуждения люминесценции со второго возбужденного состояния по отношению к длине волны $\lambda_{ex2}$.

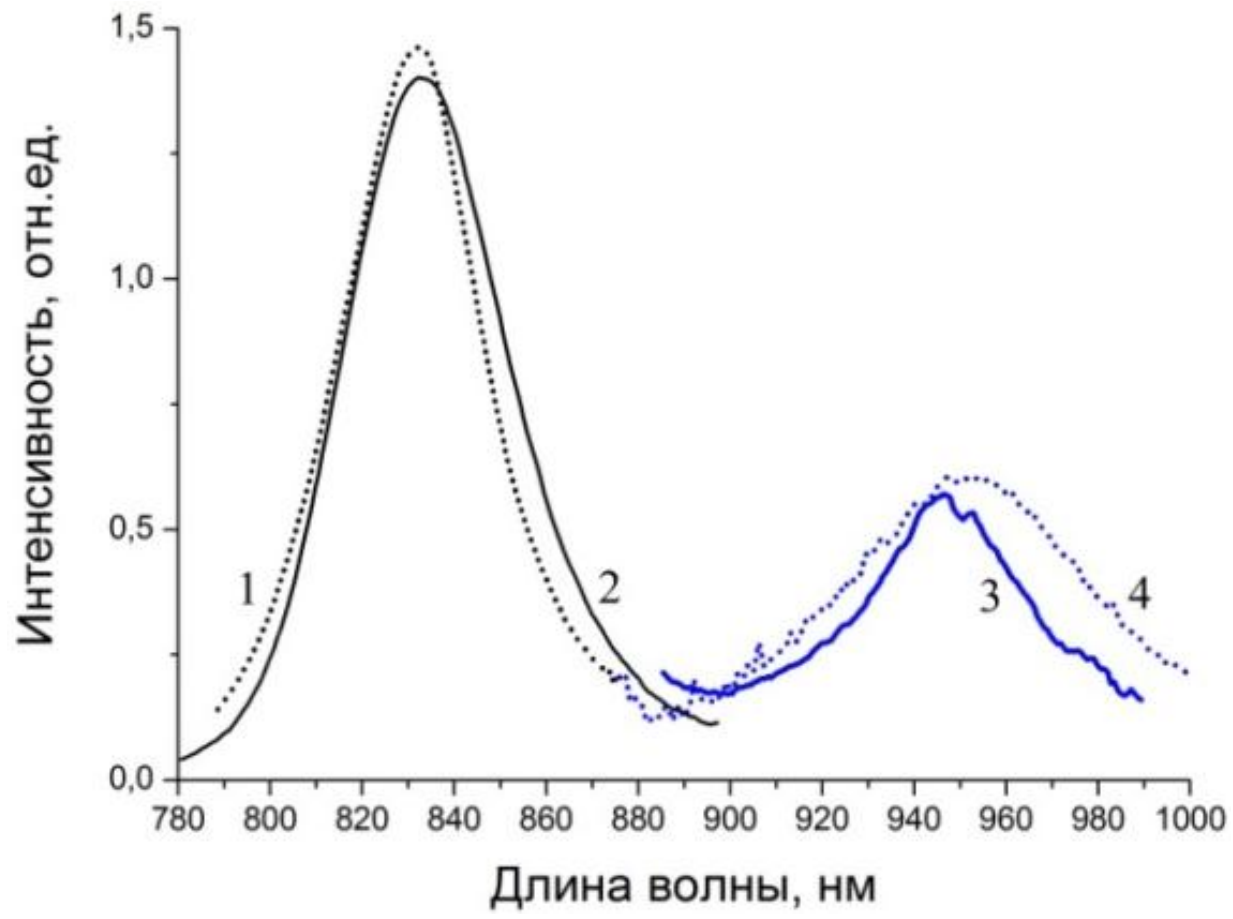

Рис. 5. 4 Спектры люминесценции SBi и GBi световодов при ступенчатом двух-квантовом возбуждении (кривые 2 и 3), и при одноквантовом возбуждении (кривые 1 и 4).

Во всех реализованных схемах эксперимента при поглощении из возбужденного состояния нами наблюдалась антистоксовая люминесценция. На Рис. 5. 4 приведены спектры люминесценции в SBi и GBi световодах при ступенчатом возбуждении их на длинах волн $\lambda_{ex1}$= 1900 нм и $\lambda_{ex2}$=1400 нм для SBi и $\lambda_{ex1}$= 2000 нм и $\lambda_{ex2}$=1650 нм для GBi световодов. В этих экспериментах наблюдались полосы антистоксовой люминесценции с максимумами ($\lambda_{em}$) на 830 нм (SBi) и 945 нм (GBi). При этом соблюдается соотношение для энергий квантов излучения $h\nu(\lambda_{ex1}) + h\nu(\lambda_{ex2}) \approx h\nu(\lambda_{em})$. Соотношение выдерживается с точностью до ширин полос возбуждения и люминесценции, и до величины стоксова сдвига между длинами волн одноквантового возбуждения и эмисии люминесценции, которые для рассматриваемых световодов сравнительно невелики, как показано выше. Для сравнения на Рис. 5. 4 приведены спектры полос люминесценции тех же световодов, но при одноквантовом возбуждении на 780 нм (SBi) и 850 нм (GBi). Подобие спектров люминесценции при одно- и двухквантовом возбуждении показывает, что в обоих случаях появление полос люминесценции связано с одним и тем же активным центром.

Были измерены спектры возбуждения антистоксовой люминесценции в зависимости от одной из длин волн возбуждения. На Рис. 5. 5 показаны зависимости изменения интенсивности люминесценции на 830 нм для SBi и 950 нм – для GBi от длины волны возбуждения $\lambda_{ex2}$ в случае ступенчатого возбуждения. Как и следовало ожидать, эти зависимости носят резонансный характер и соответствуют переходам между уровнями $E_0 \rightarrow E_1$ и $E_1 \rightarrow E_2$.

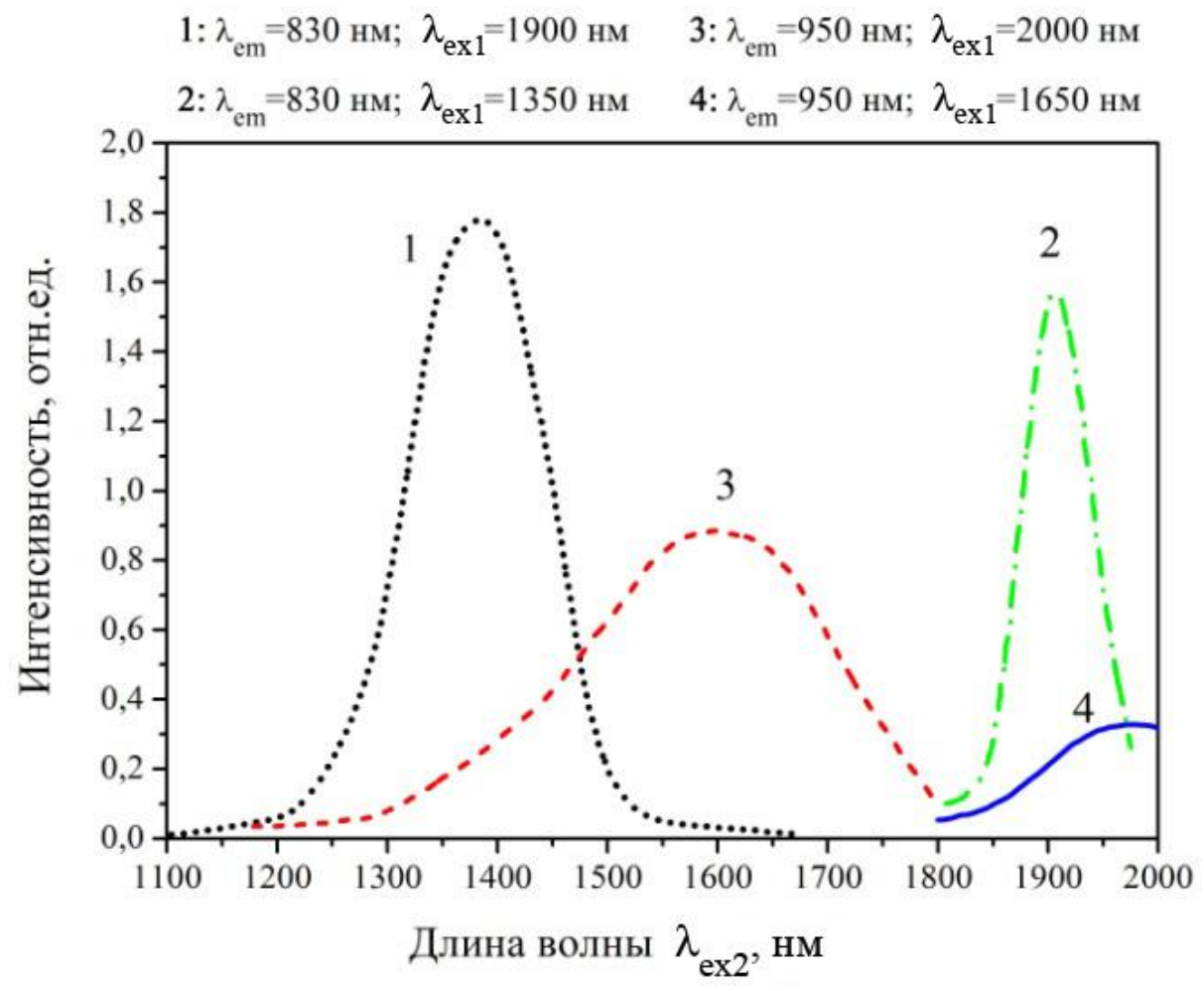

Рис. 5. 5 Зависимости интенсивностей люминесценции на длине волны $\lambda_{em}$=830 нм (SBi) и 950 нм (GBi) от длины волны возбуждения $\lambda_{ex2}$ при постоянном значении длины волны возбуждения $\lambda_{ex1}$.

Если, для SBi образца, в качестве излучения с неизменным значением энергии ($\lambda_{ex1}$) фотонов выбрать излучение с длиной волны 1900 нм, то максимальное значение интенсивности люминесценции на 830 нм достигалось на длине волны $\lambda_{ex2}$=1380 нм (Рис. 5. 5, кривая 1). В другом эксперименте максимум интенсивности люминесценции на 830 нм был расположен на длине волны $\lambda_{ex2}$=1900 нм при возбуждении излучением с $\lambda_{ex1}$ = 1350 нм (Рис. 5. 5, кривая 2). Кривые 3 и 4, приведенные на Рис. 5. 5, относятся к световоду из германатного стекла с висмутом. Видно, что в отличие от SBi резонансные максимумы для GBi более широкие и сдвинуты в длинноволновую область – около 1600 нм и 2000 нм. Положение максимумов соответствует переходам внутри одного активного центра со схемой уровней, представленой на Рис. 5. 3. Отметим, что переход между уровнями $E_1$ и $E_2$ в явном виде ранее не наблюдался.

Система уровней рассматриваемых ВАЦ содержит еще один, выше расположенный энергетический уровень $E_3$ (Рис. 5. 3). Положение этого уровня таково, что в случае германатного стекла, переход ВАЦ на данный уровень осуществляется при поглощении кванта с энергией около 21700 $см^{-1}$ (460 нм), а в случае кварцевого стекла - 23800 $см^{-1}$ (420 нм). При одноквантовом возбуждении люминесценция с уровня $E_3$ наблюдалась только в GBi и только при T=77 K [125]. При комнатной температуре такая люминесценция не наблюдалась. Спектр люминесценции в этом случае при T=77 K состоит из узких полос в видимой области: голубой (480 нм) и красной (650 нм). Появление полос голубой и красной люминесценции обусловлено оптическими переходами ВАЦ $E_3 \rightarrow E_0$ и $E_3 \rightarrow E_1$.

Особенностью расположения энергетических уровней ВАЦ в кварцевом и германатном стеклах является то, что второй $E_2$ и третий $E_3$ возбужденные уровни расположены таким образом, что энергии переходов $E_0 \rightarrow E_2$ и $E_2 \rightarrow E_3$ приблизительно равны (см. Рис. 5. 3). В этом случае для ступенчатого возбуждения на уровень $E_3$ достаточно излучения с одной длиной волны: $\lambda_{ex1}= \lambda_{ex2} \approx 830$ нм для SBi и 945 нм для GBi образца. В качестве примера на Рис. 5. 6 представлен спектр антистоксовой люминесценции легированного висмутом германатного световода при возбуждении на 925 нм, полученного при температуре 77К. Этот спектр представлен нами ранее в Главе IV в виде трехмерного спектра эмиссии-возбуждения люминесценции (Рис. 4. 2), а на Рис. 5. 6 приведено его сечение при длине волны возбуждения 950 нм.

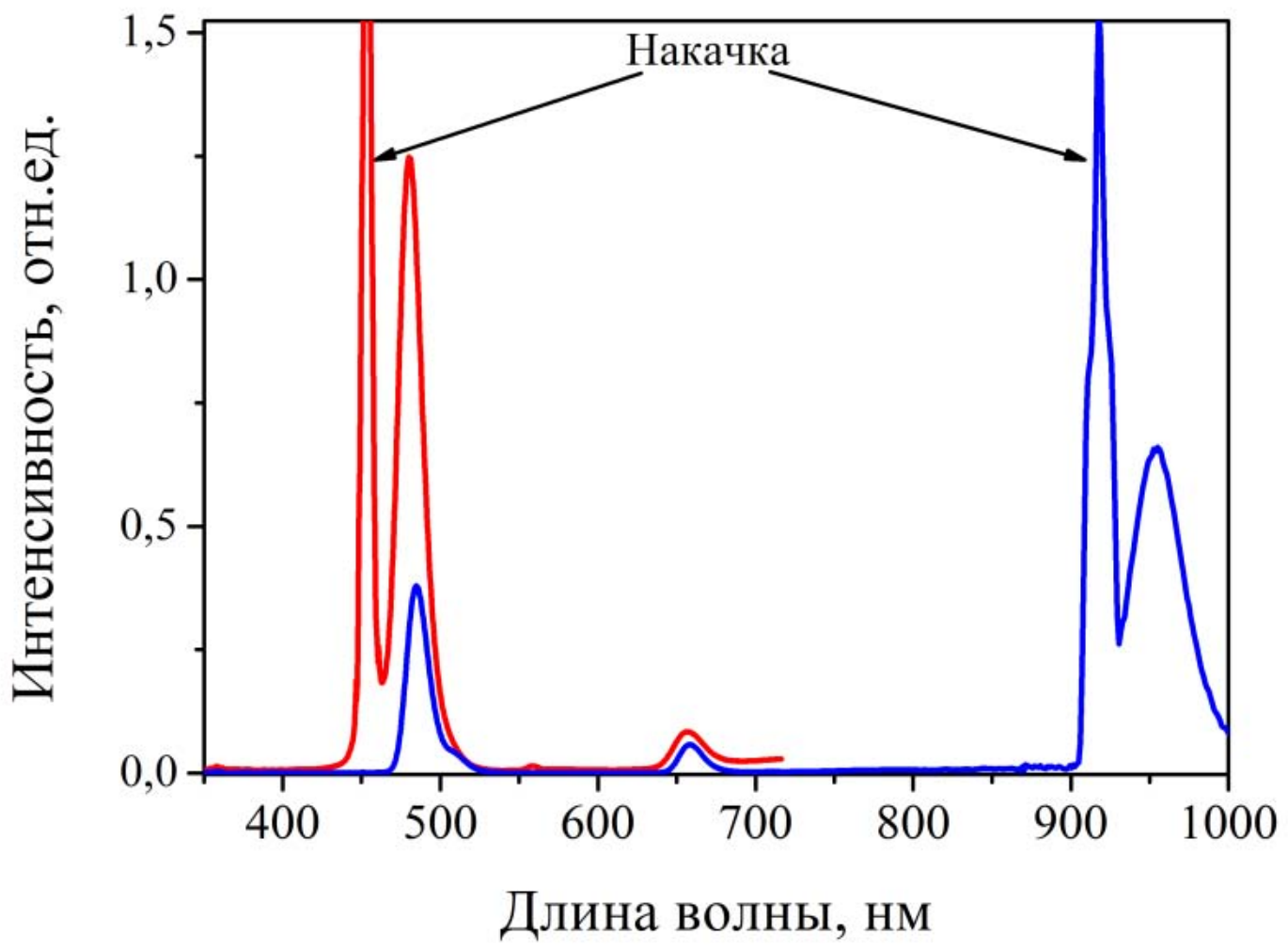

Рис. 5. 6 Спектр люминесценции GBi световода при возбуждении на 450 нм (красная линия) и 950 нм (синяя линия). Измерения выполнены при температуре световода T=77K.

Совпадение формы и положения линий люминесценции при одно- и ступенчатой двухквантовой накачке подтверждает, что в этих процессах задействованы уровни одного и того же центра.

Аналогичные результаты по ступенчатому возбуждению наблюдались и в SBi световоде при накачке на 808 нм. В этом случае также наблюдалась антистоксовая люминесценция на длине волны 420 нм и 580 нм, соответствующая переходам $E_3 \rightarrow E_0$ и $E_3 \rightarrow E_1$. Позже подобные результаты были получены в [150]. Спектры наблюдаемой в этом случае антистоксовой люминесценции представлены на Рис. 5. 7. Их сравнительно низкая интенсивность (по сравнению с аналогичными переходами в GBi световоде) может объясняться менее точным выполнением резонансного условия $2E_2 \approx E_3$ в ВАЦ-Si по сравнению с ВАЦ-Ge.

Кроме характерных полос люминесценции (ожидаемых в эксперименте), в видимой области появляется широкая полоса синей люминесценции с максимумом на 480 нм. К сожалению, полученных экспериментальных данных недостаточно для объяснения происхождения синей люминесценции.

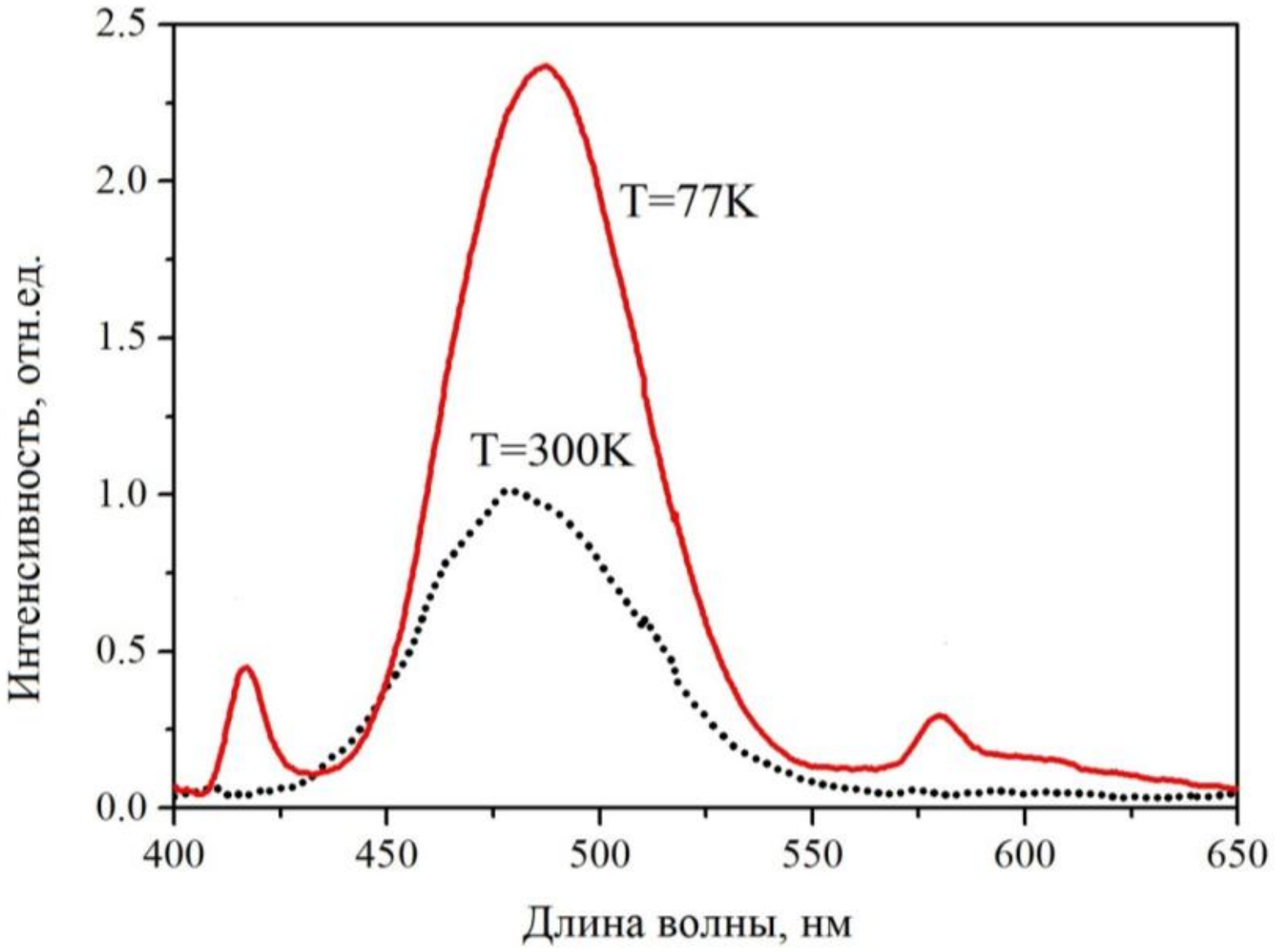

Рис. 5. 7 Спектры антистоксовой люминесценции SBi световода при возбуждении на длине волны 800 нм при комнатной и азотной температурах.

В ходе выполнения экспериментов была реализована еще одна схема возбуждения ВАЦ с использованием третьего энергетического уровня в ВАЦ-Ge. Такая схема заключалась в том, что сначала ВАЦ поглощают кванты возбуждающего излучения с энергией около 6200 см$^{-1}$ (1568 нм), осуществляя переход в первое возбужденное состояние на уровень $E_1$. Затем ВАЦ-Ge, находящиеся на уровне $E_1$, поглощают кванты излучения накачки с энергией 15200см$^{-1}$ (657 нм) и переходят на вышерасположенный возбужденный уровень $E_3$ (Рис. 5. 8 а). Для проведения данного эксперимента в качестве источников возбуждения использовались два лазера непрерывного излучения с длинами волн 657 нм (красный лазерный диод, мощность менее 1мВт) и 1568 нм (эрбиевый волоконный лазер также с мощностью ~1 мВт). Подтверждением того, что происходит возбуждение уровня $E_3$, стало появление узкой полосы синей люминесценции на 480 нм. Спектр люминесценции приведен на Рис. 5. 8 б. Данная люминесценция появлялась при T=77 К, при комнатной температуре она отсутствовала. В спектре излучения можно отчетливо наблюдать довольно узкую полосу (шириной около 25 нм) с максимумом на длине волны 480 нм и линию рассеянного излучения накачки на 657 нм.

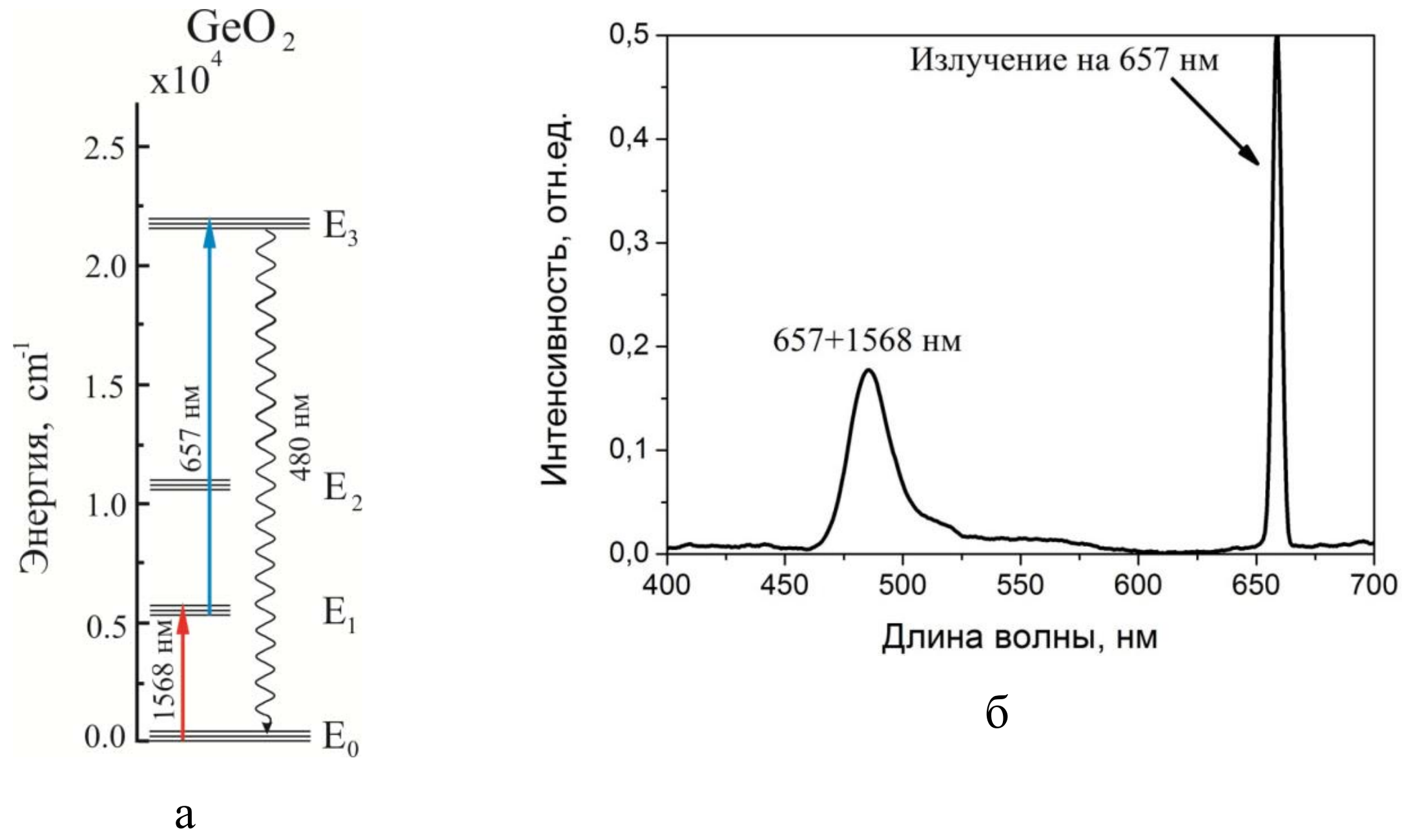

Рис. 5. 8 а) Схема возбуждения люминесценции с уровня $E_3$ ВАЦ GBi с помощью излучения двух лазеров; б) спектр синей люминесценции ВАЦ в $GeO_2$ стекле при ступенчатом возбуждении на длинах волн 657 и 1568 нм при T=77 K.

Иная ситуация в случае v-$SiO_2$ стекла с висмутом. Чтобы получить синюю люминесценцию на 420 нм, в случае кварцевого стекла для реализации аналогичной схемы с поглощением из возбужденного состояния необходимо использовать комбинацию излучений с длинами волн примерно 1400 нм и 580 нм. Но в v-$SiO_2$ длина волны возбуждения на 580 нм попадает в край полосы поглощения ионов двухвалентного висмута (которая отсутствует в GBi световоде). Они в свою очередь дают широкую полосу красной люминесценции, интенсивность которой более чем на порядок превосходит интенсивность антистоксового излучения на его длине волны, что не позволяло осуществить регистрацию синей люминесценции в SBi образце.

Таким образом, полученные результаты являются независимым подтверждением схем энергетических уровней, принадлежащих ВАЦ-Si и ВАЦ-Ge, полученных в результате экспериментов с одноквантовым возбуждении люминесценции.

## 5.3 О физической природе висмутового активного центра, излучающего в ближней ИК области спектра

Вопрос о физической природе висмутового ИК центра до сих пор остается открытым и активно обсуждается в научной литературе. Существует ряд работ, в которых выдвигаются различные гипотезы относительно строения центров люминесценции (модели ВАЦ), основные из которых обсуждались в Главе I данной работы. Однако пока ни одна из моделей не объясняет всех (или хотя бы большой части) имеющихся экспериментальных данных. Выполненные подробные исследования спектров возбуждения-эмиссии люминесценции дают новые данные, которые могут быть использованы при построение модели ВАЦ.

В данной работе показано, что в германатном, кварцевом и, возможно, фосфоросиликатном стеклах формируются висмутовые активные центры с подобной системой энергетических уровней ВАЦ. Положение каждого энергетического уровня ВАЦ-P/ВАЦ-Ge расположено выше/ниже, чем ВАЦ-Si, соответственно (Рис. 5. 9). Формирование ВАЦ-Si и ВАЦ-Ge происходит в стеклах,

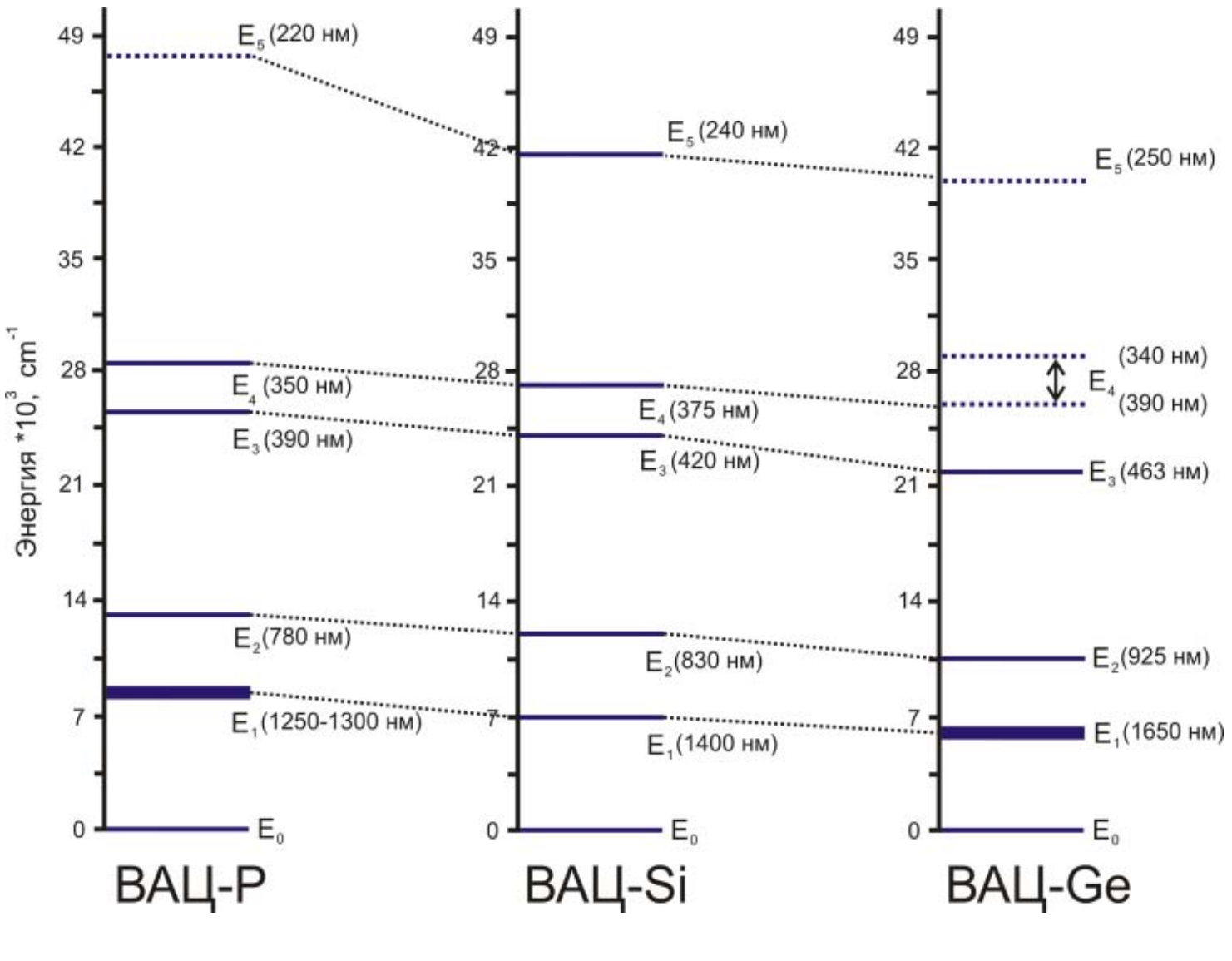

Рис. 5. 9 Схемы энергетических уровней для трех активных висмутовых центров (ВАЦ-P, ВАЦ-Si и ВАЦ-Ge), построенных по результатам данной работы.

близких по своей структуре. Тетраэдры $(SiO_4)^{4-}$ составляют основу чисто кварцевого стекла, а $(GeO_4)^{4-}$ – германатного стекла, соответственно. Размер этих тетраэдров является характерным масштабом для сетки стекла. Размер тетраэдра $(SiO_4)^{4-}$, если его оценивать, например, по длине связи Si-O, составляет 0.1605 нм [151, 152]. Размер же тетраэдра $(GeO_4)^{4-}$ в германатном стекле составляет 0.1739 нм (средняя длина связи Ge-O) [153], что на 8% больше, чем длина связи Si-O в кварцевом стекле.

Если представить себе ВАЦ в стекле как некоторую потенциальную яму, в которой находится электрон, то переходы этого электрона в потенциальной яме соответствуют переходам между уровнями энергии ВАЦ. Тогда, основываясь на подобии строения кварцевого и германатного стекла, можно предположить, что размеры соответствующих потенциальных ям будут пропорциональны характерным размерам сетки стекла, то есть размерам кислородных тетраэдров. Отсюда следует, что размер эквивалентной потенциальной ямы для электрона в германатном стекле будет примерно на 8% больше, чем в кварцевом. Поскольку величина уровней энергии ($E_n$) электрона в потенциальной яме обратно пропорциональна квадрату размера ямы ($\boldsymbol{a}$) [154]:

$$E_n = \frac{\pi^2 \hbar^2}{2 m_0 \boldsymbol{a}^2} n^2, \qquad n = 1, 2, 3, \dots$$

где $E_n$ – величина энергии $n$-го уровня, $\hbar$ – приведенная постоянная Планка, $m_0$ – масса частицы, $\boldsymbol{a}$ – размер потенциальной ямы.

Следовательно структура уровней ВАЦ в кварцевом и германатном стеклах должна быть подобна (из-за подобия структур стекол). Но уровни энергии в германатном стекле должны быть сдвинуты в область меньших энергии, примерно на 2*8 = 16 %. Что и наблюдалось в эксперименте.

В случае фосфоросиликатного стекла структура является не однородной, как в кварцевых и германатных стеклах. Впрочем, атомы фосфора, введеные в небольшом количестве (<10 мол.%), входят в сетку стекла похожим образом, находясь в центре искаженного тетраэдра. Но из-за того, что атом фосфора пятивалентен (в отличии от Si и Ge, которые четырехвалентные), один из четырех атомов кислорода в вершинах тетраэдра связан с находящимся в центре тетраэдра атома фосфора двойной связью, остальные три – одинарной. Поэтому тет-раэдер становится несколько ассиметричным, длина короткой связи P=O составляет 0.145 нм, а длинных P-O: 0.159 нм [155, 156]. При легировании висмутом фосфоросиликатного стекла наблюдаемые спектры люминесценции существенно отличаются от спектров SBi стекла, что указывает на то, что большинство атомов Bi располагаются вблизи фосфорных тетраэдров, образуя ВАЦ-P. Но поскольку размеры фосфорного тетраэдра меньше, чем кремниевого, то энергии переходов должны быть выше для ВАЦ-P по сравнению с ВАЦ-Si, что и наблюдалось в эксперименте (Рис. 5. 9). Таким образом, в случае стекол с подобным строением сетки, можно, по крайней мере, качественно предсказать изменение в положении уровней энергии ВАЦ.

Результаты данной диссертационной работы стали поводом проведения теоретических расчетов. В.О. Соколов и др. в работе [157] представили полученные расчетным путем положения энергетических уровней ВАЦ, формирующихся в кварцевом и германатном стекле в присутствии висмута. Экспериментальные, полученные в настоящей работе, и теоретические результаты показаны на Рис. 5. 10. Важно отметить, что теория описывает влияние расстояния между атомами кремния/германия на спектрально-люминесцентные свойства. Изменение данного расстояния может приводить к наблюдаемому из экспериментальных результатов сдвигу энергетических уровней ВАЦ-Ge относительно ВАЦ-Si. Хорошее совпадение теории и эксперимента, получалось только для модели, в которой висмутовый активный центр состоит из нейтрального атома $Bi^0$, находящегося вблизи нейтральной кислородной вакансии, или иона $Bi^+$ с отрицательно заряженной вакансией.

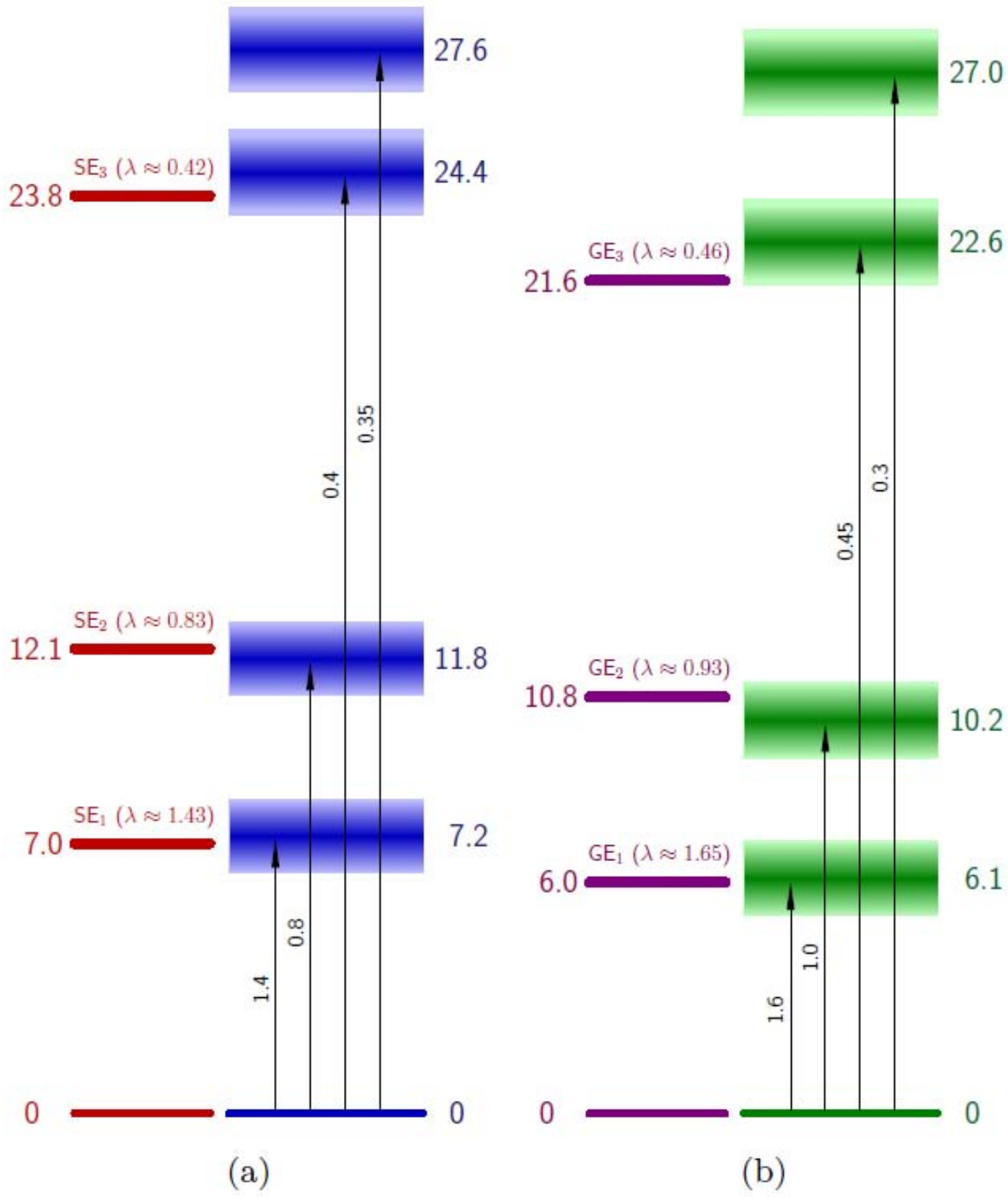

Рис. 5. 10 Экспериментальные (слева) и расчетные (справа) схемы уровней и энергии переходов, принадлежащих висмутовым активным центрам: a) v-$SiO_2$, b) v-$GeO_2$. Значения энергии уровней представлены в $10^3$ $см^{-1}$, а длины волн – в мкм [157].

Впервые модель такого активного центра – ион висмута рядом с кислородной вакансией – была предложена Е.М. Диановым в работе [149]. Данное предположение основывалось на результатах многочисленных исследований спектрально-люминесцентных и лазерных свойств кристаллов, легированных соседними с висмутом элементами (таллием и свинцом), которые находятся в одном ряду периодической системы и имеют близкие свойства.

# ЗАКЛЮЧЕНИЕ

В настоящей работе исследованы оптические свойства висмутовых активных центров в стеклообразных $SiO_2$ и $GeO_2$ и получены следующие результаты:

1. Впервые для световодов с сердцевиной из v-$SiO_2$ и v-$GeO_2$, легированных висмутом, получены спектры эмиссии и возбуждения люминесценции в широком диапазоне длин волн 240-1600 нм со спектральным разрешением 10 нм при T=77 и 300 K. Исходя из полученных данных, были определены положения 5-ти возбужденных энергетических уровней и основные излучательные переходы, принадлежащие висмутовым активным центрам. Для основных излучательных переходов ВАЦ определены времена жизни люминесценции. Установлено, что схемы энергетических уровней висмутовых активных центров, ассоциированных с кремнием (ВАЦ-Si) и германием (ВАЦ-Ge), подобны друг другу, с более низкими значениями энергии уровней ВАЦ-Ge: каждый уровень ВАЦ-Ge лежит на 10-16% по энергии ниже по сравнению с соответствующим ему уровнем в ВАЦ-Si.
2. Впервые показано, что в германосиликатных и фосфоросиликатных световодах, легированных висмутом, происходит формирование нескольких типов центров: ВАЦ-Si, ВАЦ-Ge и ВАЦ-Si, ВАЦ-P, соответственно.
3. Впервые продемонстрировано оптическое усиление и лазерная генерация в волоконном световоде из чистого $SiO_2$, легированного висмутом, на излучательном переходе $SE_1 \rightarrow SE_0$, принадлежащем ВАЦ-Si. Оптическое усиление наблюдалось в диапазоне длин волн 1410-1470 нм, лазерная генерация получена на длине волны 1460 нм при накачке на длине волны 1340 нм, КПД лазера составил 3 %.
4. Впервые измерены спектры возбуждения люминесценции на переходе между вторым возбужденным уровнем и основным уровнем энергии в висмутовых активных центрах в v-$SiO_2$ и v-$GeO_2$ при ступенчатом двухквантовом возбуждении в ИК области (1100-2000 нм).

**ЛИТЕРАТУРА**